%% file: 1_main.tex
\documentclass[11 pt]{article}

\usepackage{authblk} % author package
\usepackage{tabularx} % extra features for tabular environment
\usepackage{amsmath,amssymb}  % improve math presentation
\usepackage{mathtools} % includes \coloneqq

\usepackage{graphicx} % takes care of graphic including machinery
\usepackage{subcaption} % package for subfigure
\usepackage{float}
\usepackage[margin=1in,letterpaper]{geometry} % decreases margins
\usepackage{cite} % takes care of citations
\usepackage{hyperref}
\usepackage[nobiblatex]{xurl}
\usepackage{hyperref} % for arXiv submission
\hypersetup{colorlinks,allcolors=black} % for arXiv submission
\usepackage{lscape} % package for landscape
\usepackage{booktabs} % package for tex professional table
\usepackage{bm} % bold math 

\usepackage{setspace} % set space for text, figures, and tables 
\usepackage{natbib}
\usepackage{bibentry}
\usepackage[nottoc,numbib]{tocbibind} % add Reference to table of contents
\usepackage{sectsty} % to use \sectionfont
\usepackage{titlesec}
\titleformat{\paragraph}{\normalfont\normalsize\bfseries}{\theparagraph}{1em}{}
\titlespacing*{\paragraph}{0pt}{3.25ex plus 1ex minus .2ex}{1.5ex plus .2ex}

\usepackage{tikz} % graph in latex
\usetikzlibrary{positioning,arrows.meta} %tikz diagram
\usepackage{pgfplots}
\pgfplotsset{compat=newest} %this setting allow node to go to the correct place
\usepgfplotslibrary{fillbetween}
\usepackage{siunitx} % to use mm in tikz
\usetikzlibrary{{shapes.misc}}
\usepackage{pdflscape} %pdf landscape page

\newenvironment{chapquote}[2][2em]
  {\setlength{\@tempdima}{#1}%
   \def\chapquote@author{#2}%
   \parshape 1 \@tempdima \dimexpr\textwidth-2\@tempdima\relax%
   \itshape}
  {\par\normalfont\hfill--\ \chapquote@author\hspace*{\@tempdima}\par\bigskip}
\makeatother

\newcommand{\comment}[1]{}

\newcommand{\lnb}[1]{%
  \ln\left(#1\right)%
}

\renewcommand{\baselinestretch}{1.5} %  scales the default interline space to 1.5 its default value. 

\usepackage{setspace} % set space for text, figures, and tables 
\begin{document}
%The legibility of text can be enhanced by separating paragraphs with an amount of white space that will vary according to some design aesthetic.
\setlength{\parskip}{3pt plus1pt minus1pt}

\setlength{\abovedisplayskip}{3pt} % the space above equations 
\setlength{\belowdisplayskip}{3pt} % the space below equations 

%--------------------------------------------------------
\begin{titlepage}
\title{\textbf{Household Leverage Cycle Around \\
the Great Recession}
}

\author[1]{\textbf{Bo Li}  \thanks{ \protect\linespread{1}\protect\selectfont {\footnotesize Bo Li is with the Department of Finance, Arizona State University (boli15@asu.edu).} } }
\date{This Version: March 13th, 2024\\
\textcolor{blue}{\href{https://www.boli-finance.com/research}{\textcolor{blue}{[Click here for the latest version]}}}}
\maketitle

\vspace{-4mm}

\begin{abstract}
\fontsize{11}{13}\selectfont
\noindent This paper provides the first causal evidence that credit supply expansion caused the 1999-2010 U.S. business cycle mainly through the channel of household leverage (debt-to-income ratio). Specifically, induced by net export growth, credit expansion in private-label mortgages, rather than government-sponsored enterprise mortgages, causes a much stronger boom and bust cycle in household leverage in the high net-export-growth areas. In addition, such a stronger household leverage cycle creates a stronger boom and bust cycle in the local economy, including housing prices, residential construction investment, and house-related employment. Thus, our results are consistent with the credit-driven household demand channel \citep{mian2018finance}. Further, we show multiple pieces of evidence against the corporate channel, which is emphasized by other business cycle theories (hypotheses).

\vspace{4ex}


\end{abstract}

\thispagestyle{empty}
\end{titlepage}
%------------------------------------------------------------------

%--------------------------------------------------------
\begin{titlepage}

\centering

\textbf{\LARGE \vspace{3ex} \\
Household Leverage Cycle Around  \\
\vspace{4mm}
the Great Recession}

%\title{\textbf{Credit Expansion and Housing Cycle}}
%\author[1]{\textbf{Bo Li}  \thanks{ \protect\linespread{1}\protect\selectfont {\footnotesize Bo Li is with the Department of Finance, Arizona State University (boli15@asu.edu).} } }
%\date{Nov 16th, 2023}
%\maketitle

\vspace{16ex}

\begin{abstract}
\fontsize{11}{13}\selectfont
\noindent This paper provides the first causal evidence that credit supply expansion caused the 1999-2010 U.S. business cycle mainly through the channel of household leverage (debt-to-income ratio). Specifically, induced by net export growth, credit expansion in private-label mortgages, rather than government-sponsored enterprise mortgages, causes a much stronger boom and bust cycle in household leverage in the high net-export-growth areas. In addition, such a stronger household leverage cycle creates a stronger boom and bust cycle in the local economy, including housing prices, residential construction investment, and house-related employment. Thus, our results are consistent with the credit-driven household demand channel \citep{mian2018finance}. Further, we show multiple pieces of evidence against the corporate channel, which is emphasized by other business cycle theories (hypotheses). 

\vspace{4ex}


\end{abstract}

\thispagestyle{empty}
\end{titlepage}
%------------------------------------------------------------------

\clearpage 
\thispagestyle{empty}
\renewcommand{\baselinestretch}{1.3}
\tableofcontents
\renewcommand{\baselinestretch}{1.3}
\thispagestyle{empty}

%----------------------------------------------------------------------
% section 2: Introduction

\clearpage 
\pagenumbering{arabic}
\setcounter{page}{1}

\input{Introduction.V2}
%----------------------------------------------------------------------

%----------------------------------------------------------------------
% section 4: Research Desgin

\input{ResearchDesign}
%----------------------------------------------------------------------

%----------------------------------------------------------------------
% section 3: Data Source

\input{Data}
%----------------------------------------------------------------------

%----------------------------------------------------------------------
% section 5: Empirical Results

\input{Empirical.CreditExpansion}
%----------------------------------------------------------------------

%----------------------------------------------------------------------
% section 6: Empirical.AlternativeHypotheses

\input{Empirical.AlternativeHypotheses}
%----------------------------------------------------------------------

%----------------------------------------------------------------------
% section 7: Empirical.Robustness.tex

\input{Empirical.Robustness}
%----------------------------------------------------------------------

%----------------------------------------------------------------------
% section 8: Empirical Results: Double Difference

%\input{Empirical.DoubleDifference}
%----------------------------------------------------------------------

%----------------------------------------------------------------------
% section 9: Conclusion

\input{Conclusion}
%----------------------------------------------------------------------

%----------------------------------------------------------------------
% Section 8. bibliography
%----------------------------------------------------------------------

\clearpage

%\begingroup
%\setstretch{1.0}
%\bibliographystyle{jf}
%\printbibliography[title=References]
%\bibliography{Li_bibfile.bib}

%\endgroup

\ifx\undefined\BySame
\newcommand{\BySame}{\leavevmode\rule[.5ex]{3em}{.5pt}\ }
\fi
\ifx\undefined\textsc
\newcommand{\textsc}[1]{{\sc #1}}
\newcommand{\emph}[1]{{\em #1\/}}
\let\tmpsmall\small
\renewcommand{\small}{\tmpsmall\sc}
\fi

%----------------------------------------------------------------------
% section 9: Figures and Tables

\input{FiguresAndTables}
%----------------------------------------------------------------------

%----------------------------------------------------------------------
% section 10: Appendix 

\clearpage
\pagenumbering{arabic}% resets `page` counter to 1
\renewcommand*{\thepage}{A\arabic{page}}
% renew page numbering in the appendix 

\counterwithin{figure}{section}
\counterwithin{table}{section}
% count figures and tables within Appendix section

\appendix

\input{Appendix}
%----------------------------------------------------------------------

%----------------------------------------------------------------------
% end of document
%----------------------------------------------------------------------

\end{document}

%% file: Introduction.V2.tex
%--------------------------------------------------------------
%--------------------------------------------------------------
% This is the beginning of the entire section of Introduction
%--------------------------------------------------------------
%--------------------------------------------------------------
\clearpage

\begin{chapquote}{Atif Mian and Amir Sufi, \textit{Journal of Economic Perspectives, Summer 2018}}
\noindent ``...financial crises and a sudden collapse in credit supply are not exogenous events hitting a stable economy. As a result, we must understand the boom to make sense of the bust.”
\end{chapquote}

\vspace{-3mm}
\section{Introduction}

Understanding the cause and major mechanism of the 1999-2010 U.S. business cycle is of vital importance because the 2007-2009 Great Recession is the deepest recession since the Great Depression. The Great Recession experienced huge GDP decrease \citep{gertler2018happened}, large consumption drop \citep{mian2013household, kaplan2020non}, huge unemployment rise \citep{hoynes2012suffers, mian2014explains}, widespread mortgage defaults \citep{mayer2009rise, keys2010did}, and massive failures in the banking industry \citep{bernanke2023nobel}. Thus, identifying the cause and its major mechanisms is crucial for understanding of the economic connections among productivity, consumption, employment, housing, and credit. It is also helpful for the regulation design and macroeconomic policies that supervise the economy and avoid a similar recession. 

\comment{
In the literature there are multiple hypotheses (theories) trying to explain the business cycle. Prevalent hypotheses include speculative euphoria \citep{kindleberger1978manias,minsky1986stabilizingan}, real business cycle theory \citep{prescott1986theory}, the collateral-driven credit cycle \citep{kiyotaki1997credit}, the business uncertainty hypothesis \citep{bloom2009impact}, extrapolative expectation \citep{eusepi2011expectations}, and credit-drive housing-dominant hypothesis \citep{mian2009consequences, schularick2012credit}. Each hypothesis has found some pieces of supporting evidence. So far, however, we have not gathered enough micro-style causal evidence to distinguish which hypothesis (theory) captures the major origin of the business cycle. The empirical difficulties are twofold. First, cross-country empirical studies usually only achieve correlation because of large endogenous differences in economic development, institutions, and culture. Second, within-country studies mostly focus on the bust periods without digging into the boom period, partially because of the difficulty in finding long-term incentives that have persistent geographic divergence. Prevalent theories and ample empirical evidence support geographic convergence within a country \citep{kim2004historical}. 
}

In the literature, there is a continuing debate regarding whether firms or households play a more important role in the business cycles. First, most business cycle theories argue that firm expansion and contraction dominates the business cycle. Specifically, the real business cycle theory \citep{prescott1986theory}, the collateral-driven credit cycle theory \citep{kiyotaki1997credit}, and the business uncertainty theory \citep{bloom2009impact} only emphasize the role of firms. In addition, speculative euphoria hypothesis \citep{kindleberger1978manias,minsky1986stabilizingan} and extrapolation expectation theory \cite{eusepi2011expectations} incorporate both firms and households. The exception is the credit-driven household-demand hypothesis by \cite{mian2018finance}, which stresses the role of households over firms. Second, the empirical studies after the Great Recession with more granular data only find a correlation to support the dominant role of households. \cite{mian2010household} is the first study to show that household leverage as of 2006 is a powerful predictor of the severity of the 2007-2009 recession across U.S. counties. \cite{mian2013household,mian2014explains} emphasize the role of housing net worth in household balance sheet in explaining the consumption slump and non-tradable employment drop in the Great Recession. \cite{mian2017household} provides the first cross-country analysis on the role of household leverage in predicting short-term boom and medium-term lower economic growth and higher unemployment. However, key causal evidence is still lacking: a boom and bust cycle in household leverage causes a boom and bust cycle in consumption (particularly durable consumption) and nontradable sector employment.

To preview, by combining insights from regional economics and international economics, we design a causal framework focusing on a long-term incentive for credit that has persistent geographic divergence. This unique research design empowers us to provide micro-style causal evidence for the ``credit-driven household-demand channel" by \cite{mian2018finance}. Specifically, incentivized by net export growth, credit expansion in private-label mortgages, rather than government-sponsored enterprise mortgages, causes a stronger increase in the household leverage (debt-to-income) ratio in the high net-export-growth areas. In addition, such a stronger increase in household leverage creates a stronger boom and bust cycle in these areas in housing prices, residential construction investment, and house-related employment. Further, we provide multiple pieces of evidence against business cycle theories that emphasize the corporate channel.

%--------------------------------------------------
%\subsection{How Empirical Design Address Challenges}
%--------------------------------------------------

\noindent \textbf{Research Design} Empirical studies of the role of household leverage in business cycles by cross-country data face several challenges. First, household leverage can be driven endogenously by institutional differences, economic development, and culture. For instance, counties with higher level of economic and financial development, better legal protection of consumers, and a stronger regulatory system against financial misconduct could have higher household demand that generates higher household leverage. Alternatively, these countries could attract more international capital that results in credit expansion with lower interest rate, which induces borrowing and higher household leverage. Second, a consistent measure of household leverage does not exist due to International differences in household loan terms. For the largest part of household debt, mortgages (around 70\% in the US), the average maturity is about 15 years in Germany, 30 years in the USA, and 45 years in Sweden \citep{bernstein2021mortgage}. This fact means the same debt-to-income ratio due to the same amount of mortgage means quite high payback pressure for households in Germany but relatively low payback pressure for households in Sweden. In a similar fashion, the differences in the level and the access to social welfare programs  also present obstacles in getting a consistent measure of household leverage across countries. The above two obstacles prevent many cross-country empirical studies from providing causal evidence (e.g., \cite{mian2017household, Muller2023credit}).

Within-country empirical studies also confront with several challenges, though they can avoid the difficulties described above. First, any identification strategy requires cross-section differences in household leverage, which in turn requires that we much identify the a long-term incentive to credit expansion that has persistent geographic divergence. This requirement arises because the credit supply expansion (in either corporate debt or in home mortgages) is a long-term financial decision by lenders. In other words, we must find a persistent incentive that induces stronger credit expansion in certain areas or industries than others throughout the entire boom period preceding the bust. Short-term shocks, such as weather (e.g., rain, extreme temperature, and wildfires), are inadequate. In addition to the persistent geographic divergence, we need a good identification strategy to isolate exogenous shocks to the incentive for causal inference. Third, for completeness, the underlying economic theory (or story) must explain why the incentive can induce stronger credit expansion in the boom period rather than in other periods (such as the prior period). Such a theory (or story) requires a deep understanding of incentives to different types of market participants, the mortgage market structure, and the time-varying nature of them.

Our within-US research design tackles the above challenges in the following parts. In the first part, we follow the key idea of ``economic base theory" \citep{tiebout1962community} and construct a long-term incentive of credit expansion: metropolitan exposure to net export growth of manufacturing industries. The ``economic base" sector refers to the tradable sector that brings wealth to the local area by serving the world outside. Most of the wealth will be reused locally via a money multiplier effect by the nontradable sector. Thus, the ``economic base theory" argues that the tradable sector growth is the key driving force for the long-term local economic growth. Therefore, in response to such incentive, credit expansion in mortgages would be stronger in areas with stronger growth in tradable sector.\footnote{Please see the model by \cite{li2024credit} for more details.} Perfect measurement of the composite (share) and growth (shift) of table sector requires census-style data covering the accounting data of all tradable firms, which is non-exist in reality. Instead, following \cite{li2024credit}, we proxy the composite (share) with manufacturing employment at the industry-by-metropolitan level. In addition, we take advantage of the substantial time series change (shift) in net export growth in the U.S. International trade as a proxy for the relative growth at the industry level. Summarizing the two proxies (shift and share) can achieve a good measure of the relative growth of the tradable sector at the metropolitan level. By such construction, net export growth has enough time coverage (91-10) and enough area coverage (U.S. mainland). Its geographic variation (divergence) comes from the fact that the related manufacturing industries tend to cluster in just a few locations due to economies of scale. Internally economies of scale makes the efficiency of a single plant increase with its size while external economies of scale attracts a large number of firms in one and related industries to cluster in one location. Further, the persistence feature of net export growth comes from three dimensions. At the industry level, comparative advantages across nations are persistent and increasing return to scales likely strengthen the initial divergence. At the local level, industry clustering are shaped by long-term formation with huge fixed cost in buildings and labor migration. Changes, if any, could only happen gradually across time. At the individual level, job reallocation across different industries or locations is very costly.

\comment{
In the second part, our research design draws analysis from the model by \cite{li2024credit} to illustrate why the net export growth incentivizes mortgage credit expansion in the boom period (1999-2006) but not in the prior period (1992-1999). \cite{li2024credit} builds an economic model that incorporates the U.S. mortgage market structure and a sharp legal distinction between government-sponsored enterprise mortgages (GSEMs) and private-label mortgages (PLMs). In essence, by the legal constraint of non-discrimination, GSEMs cannot consider differences in regional economic conditions when setting up mortgage rates \citep{hurst2016regional}, where we use net export growth to capture the main differences in regional economic conditions. In contrast, PLMs are free from this constraint. Li's model explains that net export growth cannot induce house price growth in the prior period (92-99) when GSEMs dominate the market with government implicit guarantees. However, net export growth can induce a substantial house price boom and hence business boom 1999-2006 due to the sharp credit expansion in PLMs, which is documented by \cite{justiniano2022mortgage}. 
}

In the second part, our research design takes advantage of the gravity model-based instrumental variable approach by \cite{feenstra2019us} as the identification strategy.  Developing IVs from a general equilibrium model, \cite{feenstra2019us} use a very clever method to isolate the exogenous part of U.S. imports and exports. Intuitively, their IVs isolates the exogenous parts of net export growth due to (1) increasing world demand shown in US export growth, (2) increasing world supply shown in US import growth, and (3) tariff changes. They employ high-dimensional fixed effects to take out the potentially endogenous parts: (1) US industry-by-year supply shocks in exports and (2) US industry-by-year demands shocks in imports, and (3) pre-determined geographic distance between partner countries and U.S.. While they construct IVs separately for exports and imports, we combine them together as an IV for net export growth.

In the third part of our research design, we employ the insights from the model by \cite{li2024credit} to explain (1) why our framework is consistent with the fact that net export growth cannot induce credit expansion and consequent household leverage cycle in the prior period (1991-1999) and (2) the intuition for the household leverage cycle. 

\noindent \textbf{Intuition for the household leverage cycle} The basic intuition of the cross-metro differential household leverage cycle story can be illustrated in the following Figure (\ref{fig_ModelIntuition_Graphs}). Intuitively, U.S. metropolitan areas can be separated into high and low net-export growth areas. Higher net export growth causes higher household income growth, higher employment growth, and higher population growth.\footnote{For evidence, please see \cite{li2024credit}} Consequently, mortgage borrowers in the high net-export-growth area embrace at least two advantages: (1) higher foreclosure price of house given default \footnote{In the empirical literature, there is massive evidence that household income growth, employment growth, and population growth (including migration) push up housing demand and then housing price in the long term \citep{olsen1987demand}.} and (2) higher income growth that can be recoursed by lenders in the years after default.

This paragraph explains why our framework is consistent with the fact that net export growth cannot induce household leverage cycle in the prior period (1991-1999). A key legal constraint for the government-sponsored enterprise mortgages is that they cannot consider regional economic conditions (growth) \citep{hurst2016regional} in adjusting mortgage rates. But private-label mortgages can. Since all major events of the mortgage crisis, including ``global saving glut" \citep{bernanke2005global, bernanke2007global}, securitization innovation (primarily the Copula approach by \cite{li2000copula} \citep{salmon2012formula}), political lobby \citep{mian2013household}, and mortgage market deregulation \citep{di2017credit, lewis2023creditor} all occurred after 1999, private-label mortgages still had relatively high mortgage rates and a relatively small market share between 1991 and 1999. In contrast, during the same period, government-sponsored enterprise mortgages dominated the market with relatively low rates due to the government's implicit guarantee and economies of scale.\footnote{Estimates show that the spread between government-sponsored enterprise mortgages and otherwise similar jumbo loans (purchased by private issuers) are, on average, between 15-40 basis points between 1996 and 2006 (see \cite{sherlund2008jumbo} and its summary of the literature).} Consequently, without aggregate credit expansion, even high net-export-growth areas cannot undergo a household leverage boom and bust cycle before 1999.

However, there was a tremendous credit expansion in the private-label mortgages that starts in 2003 summer as documented by \cite{justiniano2022mortgage}.\footnote{\cite{justiniano2022mortgage} argues that this rate drop likely reflects mispricing, as shown in the subsequent increasing default rate.} Using granular loan-level data and a regression model, they identify a persistent and large decline in the spread between private-label mortgages and 10-year treasury yield started in 2003 summer. Given this credit expansion, we predict that private-label mortgages choose to increase strongly in high net-export-growth areas by recognizing the two above advantages in borrowers (higher foreclosure price and higher income growth by borrowers) and by being free from the legal constraint. This differentially stronger rise in private-label mortgages in high net-export-growth areas leads to a much stronger household leverage boom and bust cycle. Household leverage boom shows up in the consumption boom, especially in the durable goods consumption. Eventually, this unsustainable household leverage boom results in a bust, with consumption slump especially in the durable goods.

%------------------------------------
\begin{figure}[H]

\begin{center}

\resizebox{6in}{4in}{%
%%\resizebox{\textwidth}{!}{%
\includegraphics[]{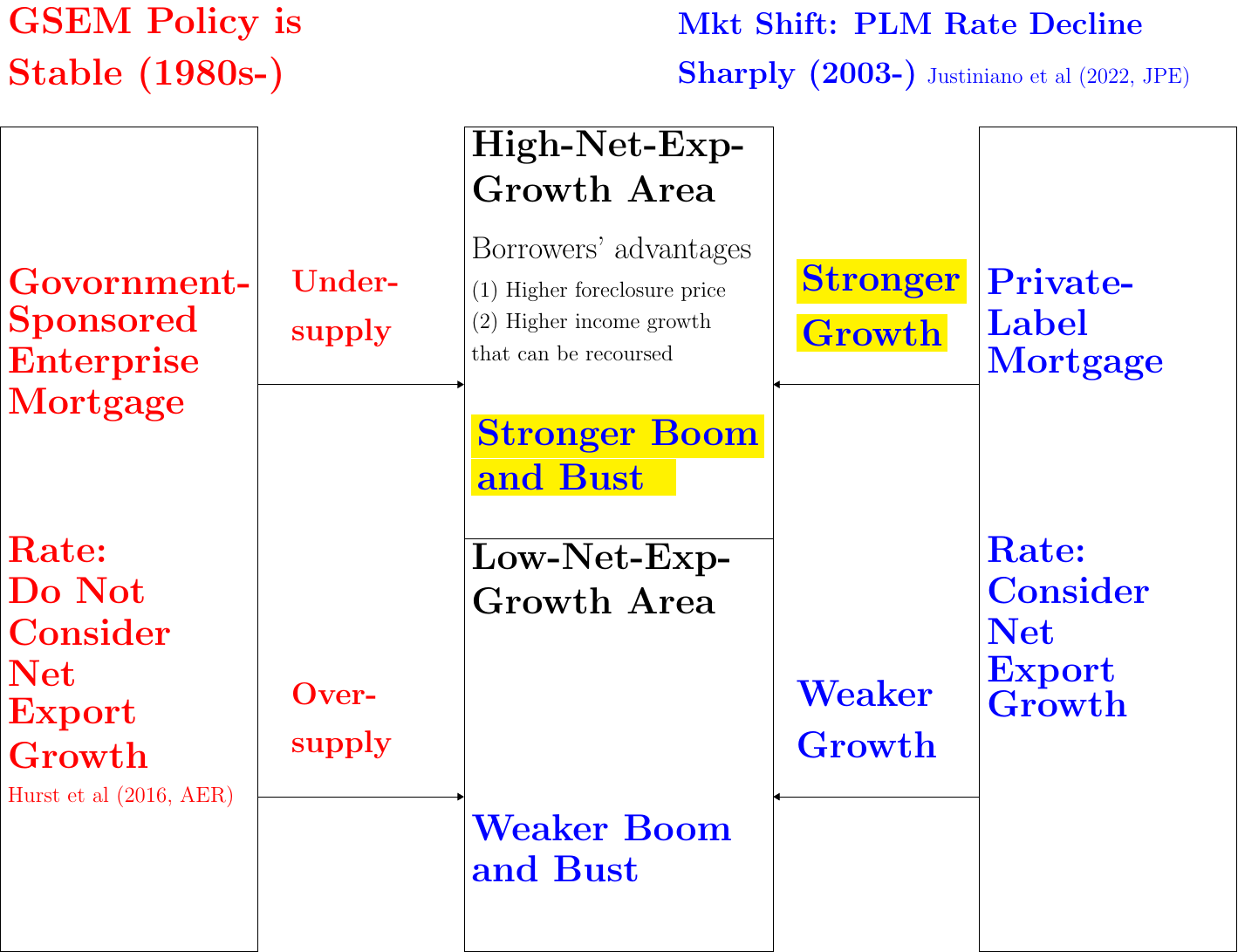}
} %end of resizebox

\end{center}

%---------------
% Figure setting: caption and label
%---------------
\caption{Model Intuition}
\label{fig_ModelIntuition_Graphs}

\end{figure}
%------------------------------------

\comment{
We explain the intuition in detail in the paragraph for the intuition of the business cycle below. A key legal requirement is that government-sponsored enterprise mortgages cannot consider regional economic conditions (growth) \citep{hurst2016regional} in setting up mortgage rates. But private-label mortgages can. Since securitization innovation (notably the Copula approach by \cite{li2000copula} \citep{salmon2012formula}), ``global saving glut" \citep{bernanke2005global, bernanke2007global}, mortgage market deregulation \citep{di2017credit, lewis2023creditor}, and political campaign \citep{mian2013household} all occurred after 1999, private-label mortgages maintained high mortgage rates between 1991 and 1999 and only had a small market share. On the contrary, government-sponsored enterprise mortgages dominated the mortgage market with low rates due to economies of scale and the government's implicit guarantee. Therefore, Without aggregate credit expansion, even high net-export-growth areas cannot undergo a business cycle.  
}

%--------------------------------------------------
%\subsection{Findings}
%--------------------------------------------------

\begin{figure}[H]

\begin{center}

\resizebox{5.4in}{!}{%
\includegraphics[width=5.4in, height=4in]{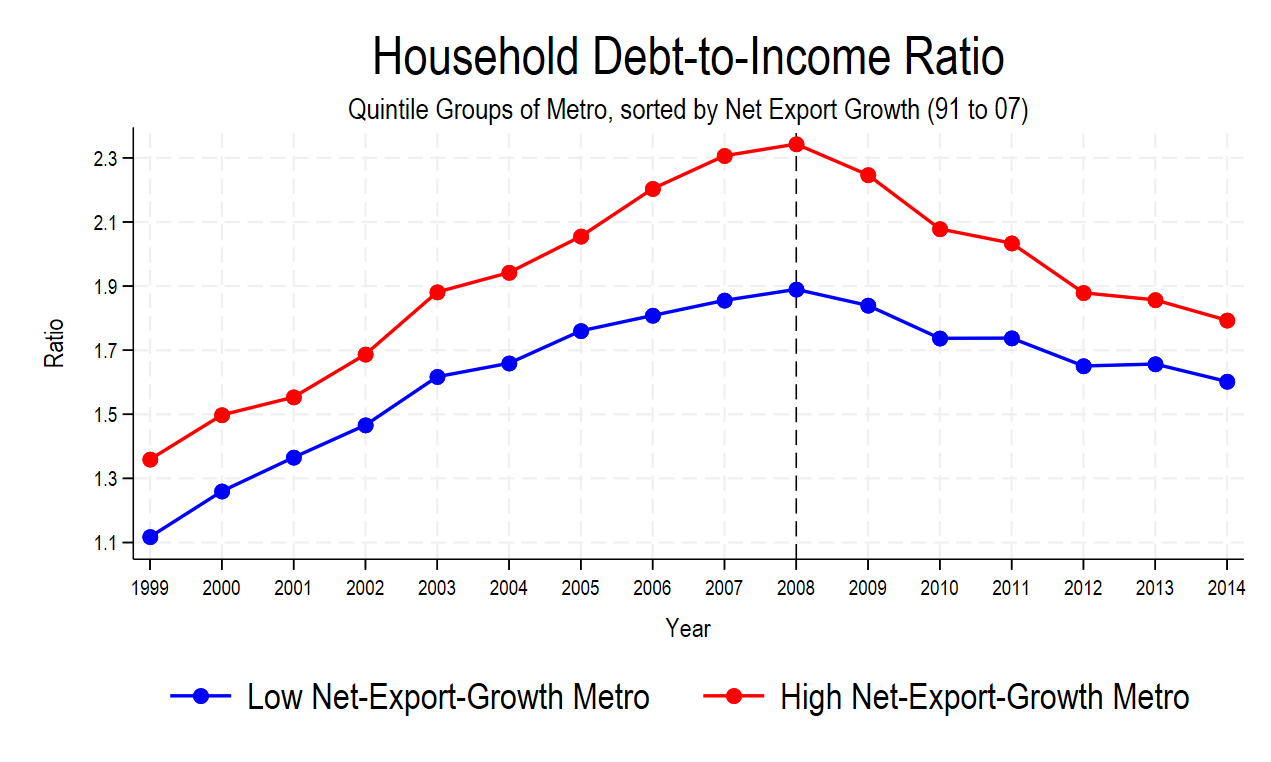}
} %end of resizebox

\end{center}

%---------------
% Figure setting: caption and label
%---------------
\caption{Household Leverage in U.S. (1999-2014)}
\label{fig_HouseholdLeverageCycle_99to14_Intro}

\end{figure}

\noindent \textbf{Findings} We illustrate our major findings in five parts. In the first part, we document a new empirical facts. There is a much stronger household leverage cycle in the high net-export-growth metropolitan areas (HNEG areas) than in the low net-export-growth areas (LNEG areas) between 1999-2014. Figure (\ref{fig_HouseholdLeverageCycle_99to14_Intro}) shows this household leverage cycle. In the boom period (1999-2005) characterized by excess credit supply in private-label mortgages \citep{justiniano2022mortgage}, the increase in household leverage (debt-to-income ratio) is much stronger in the HNEG area (0.696) than one in the LNEG area (0.643). From 2005 to 2008, driven by both credit expansion and housing price decrease, the household leverage continue to show a stronger rise in the HNEG areas (0.288) than the LNEG areas (0.130). From 2008 to 2014, however, the drop is also stronger in the HNEG area (0.550\%) than in the LNEG area (0.288\%).

%------------------------------------
%------------------------------------

In the second part, using the clever gravity model-based instrumental variable by \cite{feenstra2019us}, we provide the first causal evidence that the credit expansion in the private-label mortgages (PLMs), instead of the government-sponsored enterprise mortgages (GSEMs), causes the above household leverage boom and bust cycle to be much stronger in the high net-export-growth areas. Intuitively, the instrumental variable captures the exogenous (unexpected) parts of net export growth due to (1) rising world demand (supply) reflected in export (import) growth and (2) tariff changes after removing US industry-by-year supply (demand) shocks and pre-determined transportation cost. 

Please note that our ``causal evidence" only means that we find an incentive (net export growth) in the cross-section that induces credit expansion in private-label mortgages to be much stronger in the high net-export-growth areas. In contrast, We do not claim that we find the incentives that cause the aggregate credit expansion between 1999 and 2005. For the fundamental causes of the aggregate credit expansion, the literature has documented facts and evidence from international capital flow (``global saving glut" mainly 2003-2007 by \cite{bernanke2005global, bernanke2007global}), financial innovation in securitization (notably the Copula formula by \cite{li2000copula} \citep{salmon2012formula}),  political push (2002-2007 mortgage industry campaign contributions \citep{mian2013political}), and mortgage market deregulation (2004 preemption of national banks from state anti-predatory lending laws by the Office of the Comptroller of the Currency \citep{di2017credit}, the 2005 Bankruptcy Abuse Prevention and Consumer Protection Act \citep{lewis2023creditor}). Therefore, our boom period 1999-2005 includes all major events documented above and such a choice also matches the mortgage boom period commonly used in the literature (see \cite{griffin2021drove} for a thorough review).

In the third part, we show that induced by credit expansion, this household leverage cycle causes a differentially stronger cycle in housing prices, residential construction, and house-related industries in the high net-export-growth areas. These pieces of evidence validate the key conclusion that credit expansion causes the 1999-2010 U.S. business cycle mainly via the household leverage channel. Thus, our results are consistent with the credit-driven household demand hypothesis by \cite{mian2018finance}.

In the fourth part, we address the concern that the firm channel might play an important role. First, we show four pieces of evidence against the firm channel emphasized by the real business cycle theory \citep{prescott1986theory}, the collateral-driven credit cycle theory \citep{kiyotaki1997credit}, and the business uncertain theory \citep{bloom2009impact}. Second, we show evidence against the predictions on both the firm channel and the household channel by extrapolation expectation theory \citep{eusepi2011expectations}.

We conduct a battery of robustness tests for our major conclusion that credit expansion results in a household leverage cycle. First, we consider the state-level differences in anti-predatory lending law (APL). \cite{di2017credit} document that in January 2004 the Office of the Comptroller of the Currency (OCC) preempted national banks (instead of state-chartered banks and independent mortgage companies) from state-level anti-predatory lending laws. We find although preempted states experience a weaker boom and a weaker bust in household leverage, our main conclusion still holds across metropolitan areas. Second, we investigate state-level differences in recourse law \citep{ghent2011recourse}. We find that non-resource states experienced neither a stronger boom nor a stronger bust in household leverage, and our main conclusion still holds across metropolitan areas. Third, we account for state-level differences in the judicial requirement of foreclosure laws \citep{mian2015foreclosures}. We do not find non-judicial states experienced a stronger boom or a stronger bust in household leverage and our major conclusion still holds across metropolitan areas. Fourth, our major conclusion is robust to the inclusion of the sand-state dummy. Fifth, our main conclusion is robust after controlling the state capital gain tax rate.

%--------------------------------------------------
%\subsection{Contribution}
%--------------------------------------------------

\noindent \textbf{Contribution} Our contribution to the literature can be summarized into three parts. First, we document a new empirical fact: a stronger boom (99-08) and a stronger bust (08-14) in household leverage in the high net-export-growth metropolitan areas than in the low net-export-growth ones in the U.S..

In addition, our paper makes quite unique contributions to the literature on the causal impact of credit expansion on the U.S. business cycle via household leverage channel in four dimensions. First, our paper explains the differential household leverage cycle across metropolitan areas within the USA, a new dimension in the literature. In contrast, most empirical papers on the household (mortgage) cycle focus on the cross-country differences \cite{jorda2016great, mian2017household, Muller2023credit}. Second, our unique research design has three major advantages. The first advantage is that we capture the long-term incentive of credit expansion by operationalizing the ``economic base theory”. This theory emphasizes that the tradable sector (for which we proxy it by net export growth) determines the local economy in the long term. The second advantage is the state-of-art instrumental variable by \cite{feenstra2019us} from International Economics for causal inference. Their model-based instrumental variables have controlled many factors that cannot be removed in other methods. The third advantage is that the IV approach by \cite{feenstra2019us} can cover the entire period of the credit boom period (99-05) consisting of most major events related to the mortgage market boom. These events include international capital flow (2003-2007 ``global saving glut" by \cite{bernanke2005global, bernanke2007global}), innovation in securitization (primarily the Copula formula by \cite{li2000copula} \citep{salmon2012formula}), mortgage market deregulation (2004 preemption by the Office of the Comptroller of the Currency \citep{di2017credit}, the 2005 Bankruptcy Abuse Prevention and Consumer Protection Act \citep{lewis2023creditor}), and political lobby (2002-2007 mortgage industry campaign contributions \citep{mian2013political}). The above three advantages enable us to investigate the boom period preceding the bust period. In contrast, prior papers only focus on the bust period \citep{mian2013household,mian2014explains,kaplan2020non}. Third and most importantly, we apply our method to show how the household leverage cycle impacts the local real economy, including house prices, residential construction investment, and house-related employment. Fourth, we also employ a key legal constraint to distinguish the government-sponsored enterprise mortgages and private-label mortgages. Most other papers do not distinguish these two mortgages.\footnote{The only two exceptions that separate the role of government-sponsored enterprise mortgages and private-label mortgages are \cite{justiniano2022mortgage,mian2022credit}, though they do not show the irrelevance of government-sponsored enterprise mortgages to the differential housing cycle across metropolitan areas.} In summary, the four dimensions above distinguish our paper from other papers in the literature.

%################################################################################
%################################################################################
% This is the end of the entire section (tex file)
%################################################################################
%################################################################################

%% file: ResearchDesign.tex
%--------------------------------------------------------------------------------------
% new section
%--------------------------------------------------------------------------------------

%\clearpage

\section{Research Design}\label{sec:ResearchDesign}
This section illustrates our empirical research design, which is similar to \cite{li2024credit}, that can provide causal evidence for the household leverage cycle. We start with creating a measure of fundamental incentive for mortgage credit by operationalizing the key ideal of the ``Economic Base theory" \citep{tiebout1962community}. Specifically, we employ metropolitan exposure to net export growth of manufacturing industries (hereafter ``net export growth") as a proxy of the local tradable sector growth. To overcome the endogeneity concern of OLS, we employ the state-of-art instrumental variable approach from International Economics by \cite{feenstra2019us} as our identification strategy.

\subsection{Operationalize the ``Economic Base Theory"}
We operationalize the central idea of ``economic base theory" by using metropolitan exposure to net export growth of manufacturing industries (hereafter ``net export growth") as a proxy of the local tradable sector growth. This net export growth measure satisfies the requirement of treatment variable in the introduction: persistent geographic divergence.

Regional economists define the ``economic base sector'' (tradable sector) as business activities that a local area offers for the areas outside its boundaries \citep{tiebout1962community, nijkamp1987regional}. The tradable sector brings wealth into the local area and much of the money will be reused locally by nontradable sector via a multiplier effect. By this theory, the tradable sector growth is the most important driving force for local economic growth in the long term \citep{nijkamp1987regional,thrall2002business,ling2013real}. Therefore, the tradable sector growth can predict demand-side factors that shape the long-term business growth, including employment growth, household income growth, and population growth.\footnote{\cite{olsen1987demand} surveys on the demand factors of housing and business, including the three factors mentioned above. \cite{li2024credit} provides causal evidence that net export growth causes growth in above three elements in the long-run.} Following this logic, we expect credit expansion would be much stronger in areas with stronger tradable sector growth due to at least two factors: (1) higher foreclosure price of house given default and (2) higher income growth in the future that can be recoursed by lenders after default.

Perfect measurement of local tradable sector growth mandates census-style data to include the accounting data of all tradable firms, which is not exist. As an alternative, we follow\cite{li2024credit} and use the local employment composite as a measure for the composite (share) of local tradable sector. In addition, we use the cross-time change in U.S. net export growth in manufacturing industries as a measure for the relative growth (shift) at the industry level. Summarizing the above two proxies (share and shift) can generate a good measure of the relative growth of the local treatable sector across metropolitan areas.\footnote{We admit that manufacturing sector does not include several other elements in the tradable sector: college town, retirement community, other tradable goods industries (e.g., natural resource), and tradable services (e.g., information technology and medical sector). But in most regression analysis, we control these elements by their value in the starting year in our period. } Considering the work commuting within local areas, we summarize net export growth at the metropolitan level (2003 CBSA code). 

Specifically, we implement the above method in two steps in the data. First, in industry $g$ in year $t$, net export measure is defined as $ \text{NetExp}_{g,t}= \frac{Export_{g,t} - Import_{g,t}}{Y_{g,91}} $, where $Export_{g,t}$, $Import_{g,t}$, and $Y_{g,91}$ are US export, import and domestic production value in industry $g$ in year $t$ (or 1991), respectively. All values are converted to 2007 US dollars. US domestic production in 1991 serves as the scaling factor. 1991 is intentionally used to avoid potential response of domestic production to trade in later periods.\footnote{The same choice of domestic production in 1991 as the denominator is used by other papers like \cite{barrot2022import}.} Second, we employ local employment composite (share) to aggregate net export growth at the metropolitan level across period: 
\begin{equation}\label{equ:NEG_m}
    \triangle_{t_{1},t_{2}}\text{NetExp}_{m} = \sum_{g} \big[  (L_{m,g,t_{0}}/L_{m.t_{0}}) * (\text{NetExp}_{g,t_{2}} - \text{NetExp}_{g,t_{1}}) \big]
\end{equation}
where $L_{m,g,t_{0}}$ and $L_{m.t_{0}}$ are the employment of industry $g$ and total employment, respectively, in metropolitan area $m$ in year $t_{0}$. We choose $t_{0} = t_{1}-1$ to make sure the employment share is pre-determined to the trade measure so that all cross-time variation in net export growth ($\triangle_{t_{1},t_{2}}\text{NetExp}_{m}$) is driven by changes in trade measures rather than employment composite. The time horizon between $t_{1}$ and $t_{2}$ is the period of interest.

The above net export growth satisfies the requirement of persistent geographic divergence. First, its geographic divergence arises from the empirical fact that the local tradable sector tends to cluster within several related industries due to economies of scale. Internal economies of scale enlarges the size of local manufacturing firms \citep{worldbank2009ch4} and external economies of scale draws firms in the same and related industries to the local clustering \citep{krugman2018internationalCh7,worldbank2009ch4}.Internal economies of scale consist fixed cost of plant operating, input purchase at a volume discount, and learning in operation. External economies of scale include labor market pooling, specialization of suppliers, and knowledge spillovers. Second, the long-term decline in transportation costs and increased labor mobility in the last century served to enforce the tradable sector clustering. Third, many global events in the 1980s and 1990s promoting international trade also magnified the local industry clustering at the global level. These events include huge tariff reduction by international negotiation, reforms and opening in many emerging countries, the Dissolution of the Soviet Union in 1991, the 1994 establishment of North American Free Trade Agreement (NAFTA), and the 1995 creation of World Trade Organization (WTO). See \cite{worldtradereport2007} for a review. Fourth, the persistence feature of net export growth at the metropolitan level comes from three dimensions. At the industry level, both comparative advantages (due to technology level or natural endowment) and horizontal specialization (due to economies of scale) across nations make trade patterns persistent across time. At the local level, formation of industry clusters is associated with huge costs and is unlikely to change rapidly in a short period. At the individual level, human capital build-up and job reallocation across industries are very time- and effort-consumption locally and remotely.

\subsection{OLS and Its Bias}
This subsection illustrates the potential bias from OLS regression. we start with OLS specification that relates the household leverage growth to the growth of private-label mortgages (non-jumbo) (PLMNJ) at the county level:
\begin{equation}\label{eq:OLS_HHDTI_PLMNJ}
\begin{split}
     \triangle_{99,05} HHDTI_{c} = \beta * \triangle_{99,05} Ln(\text{Private-Mort}_{c}) + \gamma* Controls_{c}  + \epsilon_{c} 
\end{split}
\end{equation}

\noindent Here, the dependent variable $\triangle_{99,05} HHDTI_{c}$ is the change of the household leverage (debt-to-income ratio) at county $c$ 99-05. The independent variable $\triangle_{99,05} Ln(PLMNJ_{c})$ is the growth rate of the dollar amount (07USD) of private-label mortgage (non-jumbo) (PLMNJ) at county $c$ 99-05.

\noindent \textbf{Omitted Variable Problems} 
Potential omitted variables could bias $\beta$. For example, the rapid net export growth and, hence, mortgage growth in 1999-2005 has been anticipated by employees in Silicon Valley so that house price and household leverage increase before 1999 to reflect such expectation. In this case, $\beta$ could be biased downward because some of the effect of the mortgage shows up in household leverage increase in earlier period, reducing the household leverage increase between 1999 and 2005.

\subsection{Gravity Model-based Instrumental Variable}\label{subsec:GIV_exports}
To overcome the endogenous concern on OLS specification, we employ the gravity model-based instrumental variable by \cite{feenstra2019us} for net export growth. Whereas they construct IVs for exports and imports separately, we combine both as a single IV for net export growth. For illustration, we detail the model of exports and leave the model of imports in Appendix Section \ref{subsec:GIV_imports}. 

To develop an instrumental variable for US exports at the industry-by-year level, \cite{feenstra2019us} build on the idea that eight other high-income countries' exports can instrument for the US exports because they are both related to the world's rising demand \citep{autor2013china}. Additionally, \cite{feenstra2019us} incorporate tariff changes, which are plausibly exogenous to firms, employment, and households. Moreover, their model controls for the supply-side shocks in the home country by employing a fixed effect to remove them.

To predict US exports, the gravity model begins with a symmetric constant-elasticity equation by \cite{romalis2007nafta} for exports: 
\begin{equation}{\label{eq:gravity_export}}
    \frac{X^{US,j}_{s,v,t}}{X^{i,j}_{s,v,t}} = \Bigg( \frac{w^{US}_{s,t}*d^{US,j}*\tau^{US,j}_{s,t}}{w^{i}_{s,t}*d^{i,j} * \tau^{i,j}_{s,t}} \Bigg) ^{1-\sigma}
\end{equation}
Here $X^{US,j}_{s,v,t}$ is US exports to country $j$ in industry $s$ in product variant $v$ in year $t$. Likewise, $X^{i,j}_{s,v,t}$ represents exports from country $i$ to $j$. $w^{US}_{s,t}$ and $w^{i}_{s,t}$ stand for the relative marginal cost of production in the industry $s$ in the US and country $i$, respectively. $\tau^{US,j}_{s,t}$ and $\tau^{i,j}_{s,t}$ represent the \textit{ad valorem} import tariff imposed by country $j$ on exports by the US and country $i$, respectively. $d^{US,j}$ and $d^{i,j}$ stand for the bilateral distance and other fixed trade costs from US to country $j$ and from country $i$ to country $j$, respectively. Finally, $\sigma$ is the constant elasticity of substitution ($\sigma > 1$). 

The basic intuition of this gravity model is quite straightforward. Competing with country $i$, US exports to country $j$ are decreasing with the ratio of bilateral distance, with the ratio of relative marginal cost, and with the ratio of \textit{ad valorem} total import tariff. 

Assume $N^{i}_{s,t}$ identical product varieties in exports by country $i$ to the country $j$ in the industry $s$ and year $t$, \cite{feenstra2019us} re-arrange this equation, multiply both sides by $N^{i}_{s,t}$, and sum over countries $i$ ($i \neq US$):
\begin{equation*}
    X^{US,j}_{s,v,t}*\sum_{i\neq US} \big[ N^{i}_{s,t}(w^{i}_{s,t} d^{i,j})^{1-\sigma} \big] = (w^{US}_{s,t}d^{US,j}\tau^{US,j}_{s,t})^{1-\sigma} * \sum_{i\neq US} \big[N^{i}_{s,t}X^{i,j}_{s,v,t}  (\tau^{i,j}_{s,t})^{\sigma-1}\big]
\end{equation*}

Since the above equation holds for any country $i$ except for the U.S., one can choose a list of countries that have a similar level of economic development (so that they are close competitors of US exports) to make the prediction more accurate. \cite{feenstra2019us} choose the eight high-income countries as \cite{autor2013china}. 

Then, one can multiple $N^{US}_{s,t}$ (number of variants of products in US exports) on both sides and denote the sectoral exports $X^{US,j}_{s,t} \equiv X^{US,j}_{s,v,t}N^{US}_{s,t}$ and $X^{i,j}_{s,t} \equiv X^{i,j}_{s,v,t}N^{i}_{s,t}$. This step achieves the following result
\begin{equation*}
    X^{US,j}_{s,t}*\sum_{i\neq US} \big[ N^{i}_{s,t}*(w^{i}_{s,t} d^{i,j})^{1-\sigma} \big] = N^{US}_{s,t}*(w^{US}_{s,t}d^{US,j}\tau^{US,j}_{s,t})^{1-\sigma} * \sum_{i\neq US} \big[X^{i,j}_{s,t} * (\tau^{i,j}_{s,t})^{\sigma-1}\big]
\end{equation*}

Just with a few re-arrangements, one can derive the formula for $ X^{US,j}_{s,t}$:
\begin{equation}
    X^{US,j}_{s,t} =   \frac{N^{i}_{s,t}*(w^{US}_{s,t}d^{US,j}\tau^{US,j}_{s,t})^{1-\sigma}}{\sum_{i\neq US} \big[ N^{i}_{s,t}*(w^{i}_{s,t} d^{i,j})^{1-\sigma} \big]}  * \bigg( \sum_{k\neq US} X^{k,j}_{s,t} \bigg)  * \sum_{i\neq US} \bigg[  \frac{ X^{i,j}_{s,t} }{\sum_{k\neq US} X^{k,j}_{s,t}}  * (\tau^{i,j}_{s,t})^{\sigma-1} \bigg]
\end{equation}
 
In the above formula, we both multiply and divide by $\sum_{k\neq US} X^{k,j}_{s,t}$ to achieve the regression specification. By taking the natural logarithms for the above equation, we reach a regression-style formula:
\vspace{-1mm}
\begin{equation} \label{eq:exp_gravityRegression}
\resizebox{0.92\textwidth}{!}{%
\begin{math}
\begin{aligned}
\lnb{X^{US,j}_{s,t}} & = \underbrace{\lnb{\sum_{k\neq US}X^{k,j}_{s,t}} }_{\text{Term 0}} +  \underbrace{\lnb{N^{US}_{s,t}(w^{US}_{s,t})^{1-\sigma}}}_{\text{Ind-Year FE: } \alpha^{US}_{s,t}} + \underbrace{(1-\sigma)\lnb{d^{US,j}}}_{\text{Importing-country FE: } \delta^{US,j}}  \\
& + \underbrace{ (1 - \sigma) \lnb{\tau^{US,j}_{s,t}} }_{\text{Term 1}} + \underbrace{ (\sigma-1) \lnb{ \Bigg\{  \sum_{i\neq US} \bigg[ \frac{ X^{i,j}_{s,t} }{\sum_{k\neq US} X^{k,j}_{s,t}} (\tau^{i,j}_{s,t})^{\sigma -1} \bigg] \Bigg\} ^{\frac{1}{\sigma-1}}  }}_{\text{Term 2:} (\sigma -1)\lnb{T^{j}_{s,t}} } + \epsilon^{j}_{s,t} \\ 
\end{aligned} 
\end{math}
} %end of \scalemath \resizebos
\end{equation}
From the above formula, We can see that US exports to the country $j$ in the industry $s$ year $t$ can be separated into six terms. ``Term 0'' is the exports from eight other high-income countries to the country $j$, which reflects the world demand. The second term $\alpha^{US}_{s,t}$, which represents the US supply shocks at the industry-by-year level, is potentially endogenous. We remove this term by the US industry-by-year fixed effects. The third term $\delta^{US,j}$ reflects the predetermined distance from the US to the destination market $j$, including all other industry- and year-invariant trade costs. We remove it by the importing-country fixed effects. ``Term 1" is the absolute tariffs on US exports imposed by country $j$, which is out of control by US firms, employment, and households. I retain this term to capture the shocks from tariffs. ``Term 2" is the weighted average tariffs on non-US exports charged by destination country $j$, which is arguably exogenous to US firms, employment, and households. Intuitively, as this weighted average tariffs on non-US exports rise, country $j$ will import more US goods as substitutions. I keep this term to reflect this substitution effect. The last term $\epsilon^{j}_{s,t} = - \lnb{ \sum_{k\neq US} [ N^{i}_{s,t}(w^{i}_{s,t}d^{i,j})^{1-\sigma} ] } $ is unobserved and remains in the regression error term. 

By the above regression, we can construct predicted US exports that exclude supply-side shocks and the predetermined distance:
\begin{equation} \label{eq:gravityPreUSExp}
\lnb{\widehat{X^{US,j}_{s,t}}} = \lnb{\sum_{k\neq US}X^{k,j}_{s,t}} + \hat{\beta_1} *\lnb{\tau^{US,j}_{s,t}} + \hat{\beta_2}* \lnb{T^{j}_{s,t}}
\end{equation}

\subsection{Data Implementation}\label{subsec:Data_Implementation}
This subsection illustrates four steps in data to get net export and its GIV at the metropolitan level across periods. First, in order to derive predicted US export by Eq (\ref{eq:gravityPreUSExp}), we estimate Eq (\ref{eq:exp_gravityRegression}) at the 6-digit HS industry level (5673 industries). Second, we summarize predicted US exports over importing countries and crosswalk the 6-digit HS code (5673 industries) to the 4-digit revised SIC code (392 manufacturing industries) via the crosswalk file with weights by \cite{acemoglu2016import}. such steps give us predicted US exports to the world at the industry $g$ year $t$ level. We perform this summary by $ \widehat{X^{US}_{g,t}} = \sum_{s\in g}\sum_{j} \widehat{ X^{US,j}_{s,t} }$. Similarly, we get predicted US imports from the world $ \widehat{M^{US}_{g,t}}$. Third, we calculate the gravity model-based instrumental variable for net export at the industry-by-year level 
\begin{equation}\label{equ:givNEP_gt}
\text{givNetExp}_{g,t}^{US}= \frac{\widehat{X^{US}_{g,t}} - \widehat{M^{US}_{g,t}} }{ Y_{g,91} }
\end{equation}
Here $Y_{g,91}$ is US domestic production in the year 1991. Fourth, we use employment data as local industry share to aggregate $givNEP$ at the metropolitan level across periods. 
\begin{equation}\label{equ:delta_givNEP_m}
    \triangle_{t_{1},t_{2}}\text{givNetExp}_{m} = \sum_{g} \big[  (L_{m,g,t^{\prime}_{0}}/L_{m.t^{\prime}_{0}}) * (\text{givNetExp}_{g,t_{2}} - \text{givNetExp}_{g,t_{1}}) \big]
\end{equation}
where $L_{m,g,t^{\prime}_{0}}$ and $L_{m.t^{\prime}_{0}}$ are the counts of employment  in industry $g$ and total employment, respectively, in metropolitan area $m$ in year $t^{\prime}_{0}$. Inspired by \cite{acemoglu2016import}, we use $t^{\prime}_{0} = t_{1}-3$ to avoid potential covariance rising from data error between the dependent variable and the independent variable.

\noindent \textbf{Relevance Condition} To sum up, the gravity model-based IV by \cite{feenstra2019us} captures the exogenous part of net export growth due to (1) increasing world demand in net export growth by other eight high-income counties and (2) tariff changes, after accounting for the US industry-year supply shocks and predetermined bilateral distance. The relevance condition is satisfied because it starts from the general equilibrium model by \cite{romalis2007nafta} and is derived from specific decomposition above. We will test this condition in regression by first-stage Kleibergen-Paap (2006) robust (clustered) F-statistics \citep{kleibergen2006generalized} and Montiel Olea-Pflueger (2013) efficient F-statistics \citep{olea2013robust}.

\noindent \textbf{Exclusion Restriction} We have exclusion restrictions at two levels by nature. The first level refers to the gravity model-based instruments by \cite{feenstra2019us}. They have already removed supply-side shocks via industry-by-year fixed-effect in predicted US exports and demand-side shocks via industry-by-year fixed-effect in predicted US imports. Thus, exclusion restriction holds for the gravity model-based IV. The second level refers to our regression specification Eq (\ref{eq:OLS_HHDTI_PLMNJ}), where the exclusion restriction means that net export growth can only affect house prices via private-label mortgages.  Section \ref{sec:Irrelevance_GSEM} provides evidence supporting this claim by showing government-sponsored enterprise mortgages are unrelated with household leverage cycle. For other evidence supporting such an exclusion restriction, refer to Section 4.3 Exclusion Restriction in \cite{li2024credit}. 
 
Just like all instrumental variable estimates, our 2SLS estimates capture the local average treatment effects on compilers \citep{imbens1994identification}. In our setting, compilers are metropolitan-by-period observations that experience more US net export growth to the world following increases in gravity model-based predicted US net export growth described above.
 
%-------------------------------------------------------
%-------------------------------------------------------

%################################################################################
%################################################################################
% This is the end of the entire section (tex file)
%################################################################################
%################################################################################

%% file: Data.tex
%--------------------------------------------------------------------------------------------------------------------------------------------------------------------------

%\clearpage
\section{Data Sources}\label{sec:data}
%--------------------------------------------------------------------------------------------------------------------------------------------------------------------------

We combine several datasets for to study the central role of household leverage in the business cycle. The International trade and tariff data and the household leverage data are new to the literature.

%--------------------------------------------------------------------------------------------------------------------------------------------------------------------------
\subsection{Data for Household Leverage}
%---------------------------------------------------------------------------------------------------------------------------
We obtain household leverage (debt-to-income) ratio at the county and state level in annual frequency from 1999 to 2023 from the Board of Governors of the Federal Reserve System.\footnote{The website of household debt-to-income ratio is available here: \url{https://www.federalreserve.gov/releases/z1/dataviz/household_debt/}.} Though the data table only provides upper bound and lower bound of the household debt-to-income ratio, the map does provide the true value.\footnote{We are grateful for an anonymous staff who replied to our email and gave us such important information.} We webscrape the map to obtain the true values of household debt-to-income ratio.

%--------------------------------------------------------------------------------------------------------------------------------------------------------------------------
\subsection{Data for Net Export Growth and Tariff}
%---------------------------------------------------------------------------------------------------------------------------

%---------------------------------------------------------------------------------------
\noindent \textbf{Trade Flow Data} We get International trade flow data from the United Nations Comtrade Database.\footnote{The website of UN Comtrade Database is \url{https://comtrade.un.org/data/}.} This database maintains bilateral exports and imports data for detailed products documented under the six-digit Harmonized Commodity Description and Coding System (HS code). To deflate trade value to 2007 USD dollar, I employ the Personal Consumption Expenditures Chain-type Price Index from Federal Reserve in St. Louis.\footnote{Federal Reserve in St. Louis offers Personal Consumption Expenditures Chain-type Price Index at \url{https://fred.stlouisfed.org/series/PCEPI}}. To convert these trade data from a six-digit HS system to a four-digit SIC system, we adopt the crosswalk file and revised SIC system (392 manufacturing industries) in \cite{acemoglu2016import}.\footnote{This crosswalk file is available from Prof. David Dorn's website: \url{https://www.ddorn.net/data.htm}. The further refined SIC system (392 manufacturing industries) and crosswalk file are from \cite{acemoglu2016import}.}  \footnote{To double check our calculation is correct, we calculate China's exports to the US and eight other high-income countries at the industry-by-year level from 1991 to 2007. We compare our calculation to data maintained at David Dorn's website: section [D] Industry Trade Exposure at \url{https://www.ddorn.net/data.htm}. Correlations between our calculation and his data are 0.9983 for China's exports to the US and 0.9973 for China's exports to eight other high-income countries.}

%---------------------------------------------------------------------------------------
\noindent \textbf{Tariff Data} We obtain Bilateral tariff schedule data at five-digit SITC product level between 1984 to 2011 in \cite{feenstra2014international}.\footnote{Their original data are collected from the TRAINS, IDB databases, and multiple other resources via several cleaning steps and filling in missing values with other resources. The detailed data work is described in Appendix C in \cite{feenstra2014international}.} To convert tariff data from a five-digit SITC system to a six-digit HS system, we follow the methods in \cite{feenstra2019us}. Specifically, we first convert the HS 2007 version to the HS 2002 version by the crosswalk files from the Trade Statistics Branch (TSB) of the United Nations Statistics Division.\footnote{The crosswalks files between different HS versions are maintained by the UN Comtrade database: \url{https://unstats.un.org/unsd/trade/classifications/correspondence\%2Dtables.asp}.} Then we match each six-digit HS code to one five-digit SITC2 via a crosswalk from \cite{feenstra2005world}. When one six-digit HS code is matched to multiple SITC2 codes, we follow \cite{feenstra2019us} and use the one having the highest value share.

%---------------------------------------------------------------------------------------
\noindent \textbf{Manufacturing Production Data} To scale the trade value in calculating growth rate, we use the US 4-digit SIC manufacturing industry total domestic production (vship) in 1991 as the denominator. Such data are from NBER-Center for Economics Studies (NBER-CES) Manufacturing Industry Database. We choose 1991 because it is the first year in analysis so production is unlikely to respond to trade change afterward.\footnote{This choice of scaling is also used by \cite{barrot2018import}.}

%---------------------------------------------------------------------------------------
\noindent \textbf{Manufacturing Employment Data} We get employment data at the county-by-year-by-industry level in the U.S. from County Business Patterns (CBP) Database in U.S. Census.\footnote{The website of County Business Patterns Database is \url{https://www.census.gov/programs-surveys/cbp/data/datasets.html}.} Following \cite{acemoglu2016import}, we use manufacturing employment data to aggregate net export growth and its instrumental variable at the metropolitan areas across periods. Section \ref{subsec:Data_Implementation} illustrate more details.

%--------------------------------------------------------------------------------------------------------------------------------------------------------------------------
\subsection{Data for Mortgages and House Prices}
%--------------------------------------------------------------------------------------------------------------------------------------------------------------------------

%---------------------------------------------------------------------------------------
\noindent \textbf{Mortgage Data} We get detailed loan-level mortgage data from the Home Mortgage Disclosure Act (HMDA) database.\footnote{The Consumer Financial Protection Bureau (CFPB) maintains 2007-2017 HMDA data"  \url{https://www.consumerfinance.gov/data\%2Dresearch/hmda/historic\%2Ddata/}. The Federal Financial Institutions Examination Council (FFIEC) discloses 2017-2021 HMDA data at \url{https://ffiec.cfpb.gov/data\%2Dpublication/2021}. CFPB providess links of 1990-2006 HMDA to the National Archives at \url{https://github.com/cfpb/HMDA_Data_Science_Kit/blob/master/hmda_data_links.md}.} Congress enacted HMDA to improve public reporting and monitoring of mortgage loans in 1975. Any financial institution is mandated to report HMDA data to its regulator once it meets certain standard, such as a threshold for assets and if the institution has a home office or branch in a Metropolitan Statistical Area (MSA). This database contains information on lender identifiers, borrower demographics, loan applications, and loan specifics such as location, purpose, and amount. The HMDA database offers near-universal coverage of the mortgage market. \cite{avery20102008} confirm that in 2008, commercial banks filing HMDA carried 93\% of the total mortgage dollars outstanding on commercial bank portfolios.\footnote{Though lenders with offices only in non-metropolitan areas are not required to file HMDA, 83.2\% of the population lived in metropolitan areas as of 2006 \citep{dell2012credit}.}

We use the following filtering criteria. First, we keep originated mortgages and delete applications that are denied, withdrawn, or not accepted. Second, for mortgage types, we keep conventional and Federal Housing Administration-insured (FHA-insured) mortgages and delete ones insured by the Veterans Administration, Farm Service Agency, and Rural Housing Service. Third, for purposes, we mainly use home purchase mortgages for most empirical tests. Fourth, for occupancy types, we keep owner-occupied both non-owner-occupied mortgages and treat ``not applicable" as owner-occupied.\footnote{Based on the HMDA manual (\url{https://www.ffiec.gov/hmda/pdf/1998guide.pdf}),  ``not applicable" occupancy likely refers to a multifamily dwelling where the borrower lives in. In terms of loan numbers, this ``not applicable" occupancy is only around 3.5\% of non-owner-occupied loans and 0.59\% of owner-occupied loans as of 2007. Thus, this filtering choice cannot affect conclusions in large magnitude.}

We employ the HMDA database to construct loan volume (number and dollar amount) at the county-by-year level for government-sponsored enterprise mortgages (GSEM) and private-label mortgages (PLM). Based on the HMDA examination procedures, an institution is mandated to report the type of entities that purchase the loans in the same calendar year.\footnote{See ``Home Mortgage Disclosure Act Examination Procedures" at \url{https://www.federalreserve.gov/boarddocs/caletters/2009/0910/09\%2D10_attachment.pdf}. This reporting requirement implies that the potential under-estimates of GSEMs and PLMs because the mortgages originated near the end of a year need some time to be sold. Nonetheless, this potential underestimation can only bias my results to zero.} I employ the categorization of PLMs and GSEMs by \cite{mian2022credit}.\footnote{\cite{mian2022credit} group five categories as PLMs if a mortgage is sold: (1) into private securitization, (2) to a commercial bank, savings bank, or savings affiliation affiliate, (3) to a life insurance company, credit union, mortgage bank, or finance company, (4) to an affiliate institution, and (5) to other types of purchasers.}

%---------------------------------------------------------------------------------------
\noindent \textbf{Conforming Loan Limits Data} We employ conforming loan limits (CCLs) by county and year from Federal Housing Finance Agency. Before and in 2007, conforming loan limits are set at the national level.\footnote{CCLs before 2007 are available here: \url{https://www.fhfa.gov/AboutUs/Policies/Documents/Conforming\%2DLoan\%2DLimits/loanlimitshistory07.pdf}.} From 2008 onward, conforming loan limits are set by county and year.\footnote{County-by-year CCLs are available here: \url{https://www.fhfa.gov/DataTools/Downloads/Pages/Conforming\%2DLoan\%2DLimit.aspx}.} In general, conforming loan limits are different for 1-unit, 2-unit, 3-unit, and 4-unit dwellings in each year (and county). Since the HMDA data does not disclose information on the number of units in a home between 1991 and 2009, we only use the 1-unit conforming loan limit for all mortgages. Our conservative measure of non-jumbo mortgages can help avoid a potentially upward bias in results.

%---------------------------------------------------------------------------------------
\noindent \textbf{Consistent Counties Covered by HMDA} Following the suggestion by \cite{avery2007opportunities}, we focus on counties that are consistently covered by HMDA based on the coverage of metropolitan areas defined and updated by the U.S. Office of Management and Budget. Detailed steps are the same in \cite{li2024credit}. Consequently, We derive 800 ``HMDA consistent counties after 1996" and 712 ``HMDA consistent counties after 1990".

%---------------------------------------------------------------------------------------
\noindent \textbf{U.S. House Price Data} We get the U.S. house price index data based on repeated sales at the county level from the Federal Housing Finance Agency.\footnote{The data are available at \url{https://www.fhfa.gov/DataTools/Downloads}.FHFA working paper \cite{bogin2019missing} details the construction of the index and shows its accuracy via various methods. }

%---------------------------------------------------------------------------------------
\noindent \textbf{Merge House Price and Mortgage Data} We keep counties covered by both mortgage data and house price data for both figure and regression analysis including house prices. The merged data set contains fewer counties compared to sole mortgage data because house price data covers fewer counties.

%--------------------------------------------------------------------------------------------------------------------------------------------------------------------------
\subsection{Data for Employment and Business}
%--------------------------------------------------------------------------------------------------------------------------------------------------------------------------
%---------------------------------------------------------------------------------------
\noindent \textbf{Aggregate Debts for Sectors} We acquire annual debt data for households and nonprofit organizations, business (corporate and non-corporate), and government (federal and local) from the Federal Reserve.\footnote{The debt data are available here: \url{https://www.federalreserve.gov/releases/z1/dataviz/z1/nonfinancial_debt/table/}.} Such data also contains subcategories for debt of households and nonprofit organizations: mortgages, consumer credit, and other liability.

%---------------------------------------------------------------------------------------
\noindent \textbf{BEA Employment Data} We obtain annual employment data at the county level from the U.S. Bureau of Economic Analysis (BEA).\footnote{The BEA employment data is available at \url{https://apps.bea.gov/regional/downloadzip.cfm}. In the category ``Personal Income (State and Local)", "CAEMP25S" contains data from 1969 to 2000, while "CAEMP25N" contains data from 2001 and onward.} This coverage is better than County Business Pattern employment data, which does not contain self-employment (proprietor employment) not working in establishments.

%---------------------------------------------------------------------------------------
\noindent \textbf{CBP Employment Data} We acquire employment data in detailed industries at the county-by-year level from the County Business Pattern (CBP) database from the U.S. Census.\footnote{County Business Patterns Database is available at: \url{https://www.census.gov/programs-surveys/cbp/data/datasets.html}.} To derive the accurate number from ranges reported in CBP, we obtain the carefully imputed CBP data from \cite{eckert2020imputing}.\footnote{\cite{eckert2020imputing} provide final data, code, and detailed documentation of their methodology in imputing CBP data at \url{https://fpeckert.me/cbp/}.} For industry classification, we follow \cite{goukasian2010reaction} and \cite{mian2014explains}.

%---------------------------------------------------------------------------------------
\noindent \textbf{New Residential Unit Permits} I get county-by-year new residential unit permit data from the U.S. Census.\footnote{New residential unit permits data is available at: \url{https://www.census.gov/construction/bps/index.html}.} To avoid reduced sample size due to missing observations because of non-survey years for some counties, I use the Census-imputed permit data.

%--------------------------------------------------------------------------------------------------------------------------------------------------------------------------
\subsection{Local Economic Conditions}
%--------------------------------------------------------------------------------------------------------------------------------------------------------------------------

%---------------------------------------------------------------------------------------
\noindent \textbf{IRS Household Income Data} We acquire household income data at the county-by-year level from the U.S. Internal Revenue Service (IRS).\footnote{For 1989 to 2018, the data is available at \url{https://www.irs.gov/statistics/soi\%2Dtax\%2Dstats\%2Dcounty\%2Ddata}.} The average household income at the county level is the adjusted gross income divided by the number of returns (households).

%---------------------------------------------------------------------------------------
\noindent \textbf{Local Control Variables} We obtain detailed control variables at county level from U.S. Decennial Census Summary Files. Control variables in 1989 at the county level are from 1990 (March) Census Summary File 1C and 3C.\footnote{The 1990 U.S. Census Summary File 3 is available at \url{https://www.census.gov/data/datasets/1990/dec/summary-file-3.html}. } Control variables in 1999 at the county level level are from both 2000 (March) Census Summary File 1 and 3.\footnote{The 2000 U.S. Census Files are available at: \url{https://www.census.gov/programs-surveys/decennial-census/guidance/2000.html}.}

%--------------------------------------------------------------------------------------------------------------------------------------------------------------------------
\subsection{Counties Severely Affected by 2005 Hurricanes}\label{subsec:2005Hurricanes}
%--------------------------------------------------------------------------------------------------------------------------------------------------------------------------
Following \cite{li2024credit}, we remove twelve ``deeply affected counties by 2005 Hurricanes'' since they experienced unusual growth in mortgages due to hurricane damage and subsequent government subsidies.\footnote{We try our best to present the most robust results. Since outliers only largely affect results in regression but not the illustration in figures, we include these twelve counties in the figures but remove them from regressions and summary statistics.} In 2005, three Category 5 hurricanes (Katrina, Rita, and Wilma) caused enormous fatalities and damage (estimated \$125 billion).\footnote{These ``deeply affected counties'' include Monroe County (FL, 12087), Cameron Parish (LA, 22023), Jefferson Parish (LA, 22051), Orleans Parish (LA, 22071), Plaquemines Parish (LA, 22075), St. Bernard Parish (LA, 22087), St. Tammany Parish (LA, 22103), Vermilion Parish (LA, 22113), Hancock County (MS, 28045), Harrison County (MS, 28047), Jackson County (MS, 28059), Stone County (MS, 28131). }

\subsection{Summary Statistics and Figures}

%################################################################################
%################################################################################
% This is the end of the entire section (tex file)
%################################################################################
%################################################################################

%% file: Empirical.CreditExpansion.tex
%------------------------------------------------------------
%------------------------------------------------------------
%\clearpage
%------------------------------------------------------------
%------------------------------------------------------------
\section{Empirical: Credit-Induced Household Leverage Cycle}
Our first set of empirical tests provide direct evidence that credit expansion in private-label mortgages (PLMs) rather than government-sponsored enterprise mortgages (GSEMs) cause the 1999-2010 household leverage cycle. Then our second set of tests show the impact of household leverage cycle on real business activities, including housing price, residential construction, and employment. 

To compare these two types of mortgages under the same criteria, we employ the conforming loan limits to get the non-jumbo category of private-label mortgage (PLMNJ) and delete the jumbo ones. Thus, our tests focus on the comparison between private-label mortgages (non-jumbo) (PLMNJs) and the government-sponsored enterprise mortgages (GSEMs).\footnote{First, the majority of jumbo loan borrowers are not low-income households so credit expansion study shall primarily focus on non-jumbo loans. Second, the absolute number of jumbo private-label mortgages is much smaller than the non-jumbo ones. Third, including jumbo ones only strengthens (rather than weakening) my results on the impact of private-label mortgages on the dependent variables across metropolitan areas.}

%------------------------------------------------------------
%------------------------------------------------------------
%------------------------------------------------------------
\subsection{Household Leverage Boom (99-05) and Bust (08-14)}\label{subsec:causal_householdLeverage_boom_bust}
%------------------------------------------------------------
%------------------------------------------------------------
%------------------------------------------------------------

%------------------------------------------------------------
\subsubsection{Private-label Mortgages}
%------------------------------------------------------------

Based on the intuition described in the instruction, we predict that induced by net export growth, private-label mortgages (non-jumbo) expand more in the high net-export-growth areas, thus causing a much stronger boom and a subsequent stronger bust of household leverage in these areas.

We test the household leverage boom and bust by the following stacked regression 
\begin{equation}\label{eq:HHLeverageBoomBust_on_PLMNJ}
\resizebox{0.92\textwidth}{!}{$
\begin{aligned}
\triangle_{99,05} \& \triangle_{08,14} HHDTI_{c} & = \beta_{99,05} * \triangle_{99,05} Ln(PLMNJ_{c}) \times Dum_{99,05} + \beta_{08,14} * \triangle_{99,05} Ln(PLMNJ_{c}) \times Dum_{08,14} \\
& + \gamma_{99,05}* \bm{Controls_{c}} \times Dum_{99,05} + \gamma_{08,14}* \bm{Controls_{c}} \times Dum_{08,14} + \epsilon_{period, c}
\end{aligned}
$} %end of \resizebox
\end{equation}
The dependent variable $\triangle_{99,05} \& \triangle_{08,14} HHDTI_{c}$ is the stacked increase in household leverage (debt-to-income ratio) at county $c$ 99-05 and 08-14. The key independent variable $\triangle_{99,05} Ln(PLMNJ_{c})$ is the growth rate of the dollar amount (deflated to 2007 USD) of private-label mortgages (non-jumbo) at county $c$ 99-05. We use the gravity model-based instrumental variable $\triangle_{99,05}\text{givNetExp}_{m}$ as the IV for $\triangle_{99,05}Ln(PLMNJ_{c})$.

We include control variables at county $c$ only in 1999, which are used to neutralize factors that may affect credit expansion for reasons unrelated to the net export growth. Our basic controls include the number of households, average household income, and the fraction of the labor force in population at the county $c$. Our housing controls include the number of house units, housing supply elasticity \citep{saiz2010geographic}, house vacancy rate, and fraction of renters in the occupied house units. Demographic controls include the fraction of population holding a Bachelor's degree and above, the percentage of the white population, and the count of immigrants entering the U.S. between 1990 and 1999. Industry controls include the ratio of the population that are in the art, entertainment, and recreation industries, that are in the health industries, that are in the tradable service industries, and that are college students. The industry controls capture the phenomena of retirement towns, medical centers, and college towns and the effect of the tradable service sector. Each regression is weighted by the natural logarithm of the number of house units in 1999 in each county. Logarithm rather than the absolute number of house units is chosen to keep results from being dominated by a few super-populous counties. To account that households might commute to work across counties within a metropolitan area, we measure net export growth at the metropolitan level.\footnote{According to US Census, ``the general concept of a metropolitan statistical area is that of a core area containing a substantial population nucleus, together with adjacent communities having a high degree of economic and social integration with that core." (\url{https://www.census.gov/programs-surveys/metro-micro/about.html})} Furthermore, we cluster standard errors at the metropolitan level (2003 CBSA code).

Table (\ref{table_HHDTI.D99t05vsD08t14.PLMNJ.4Reg}) reports OLS, reduced-form, second stage, and the first stage of the stacked regression of household leverage rise in the boom period (99-05) and the bust period (08-14). First, the OLS coefficients in panel A show that the impact of PLMNJ growth (99-05) is positive and marginally significant in the boom period (99-05) and significantly negative at 1\% in the bust period (08-14). The weaker significance in the boom period is consistent with our prediction in the research design that some of the net export growth can be anticipated by households and local mortgage managers. In contrast, all coefficients are significant at 1\% in the reduced-form regressions in panel B and 2SLS regressions in panel C. This comparison indicates that our instrumental variable approach captures the unexpected shocks successfully. First-stage estimates in panel D show the stable and strong positive correlation between the PLMNJ growth and gravity model-based IV for Net Export Growth (GIV-NEG), with large enough first-stage F-Statistics (clustered Kleibergen-Paap F-statistic is 11.43, and the Montiel Olea-Pflueger Efficient F-Statistic is 11.37). Therefore, our results are free from weak IV concerns. As for the coefficient equality test of the impact of PLMNJ growth (99-05) in the boom (99-05) and bust (08-14) periods in Table (\ref{table_HHDTI.D99t05vsD08t14.PLMNJ.2SLS.wide}), the chi-square statistics are large, and p-values are below 0.01, meaning the two coefficients are statistically different. To sum up, induced by net export growth, the growth of private-label mortgages (non-jumbo) (99-05) causes household leverage to experience a larger boom (99-05) and a larger bust (08-14) in the high net-export-growth areas than in the low net-export-growth areas.

In terms of economic meaning, one standard deviation in six-year PLMNJ growth (99-05) causes household leverage to rise $8.137\% \times 0.940 \times 6 = 45.891\%$ 1999-2005 and to drop $ 8.137\% \times 1.302 \times 6 = 63.564\%$ 2008-2014. For household leverage change, one standard deviation is $12.922\% \times 6 = 77.529\%$ 1999-2005 and $10.818\% \times 6 = 64.909\%$ 2008-2014. The two results mean that one standard deviation in six-year PLMNJ growth (99-05) can explain $45.891\% / 77.529\% = 59.19\%$ of one standard deviation in six-year household leverage boom (99-05) and $63.564\% / 64.909\% = 97.93\%$ of one standard deviation in six-year household leverage bust (08-14). The higher explanatory power in the bust is consistent with findings in many other settings and topics. In contrast, the 2SLS estimates in column (5) in Table (\ref{table_HHDTI.D99t05vsD08t14.PLMNJ.2SLS.wide}) show that housing supply elasticity does not impact the household leverage boom or bust directly after controlling other factors, particularly the private-label mortgages. The explanatory magnitude is also low. One standard deviation in housing supply elasticity can only explain $1.217 \time 0.005 \time 6 / 77.529\% = 4.71\%$ household leverage growth 1999-2005 and $1.217 \time 0.007 \time 6 / 64.909\% = 7.87\%$ household leverage drop 2008-2014.

\comment{
We also show that the above results are not driven by extreme values or outliers. Specifically, we show our results are robust when household leverage are winsoried at 1\% and 99\% in Appendix Section .
}

%------------------------------------------------------------

%------------------------------------------------------------
\subsubsection{Leverage Transition Period (2005-2008)}
%------------------------------------------------------------
This section shows evidence that the 2005-2008 household leverage rise is caused by growth in private-label mortgages (1999-2005) induced by net export growth (1999-2005)

Figure \ref{fig_HouseholdLeverageCycle_99to14_Intro} shows that household leverage continues to rise from 2005 to 2008. Two major reasons contribute to this trend. The first one is the declining but still high private-label mortgage amount in 2005, shown in Figure \ref{fig_GSEMvsPLMNJ_91t11_combine}. From 2006 and onward, the private-label mortgages declined sharply. The second reason is the declining housing price from 2007 to 2008 caused by foreclosures, which is ultimately caused by unsustainable credit expansion earlier on. Thus, we expect that we can attribute the 2005-2008 continuing rise in household leverage to the growth in private-label mortgages 1999-2005.

We present causal evidence for the above prediction in Table (\ref{table_HHDTI.D05t08.PLMNJ.D99t05.4Reg}), which reports OLS, reduced-form, second stage, and the first stage of the regression of household leverage rise between 2005 and 2008. First, the OLS coefficients in panel A show that the impact of PLMNJ growth (99-05) is positive and significant at 1\% level. Similarly, all coefficients are significant in the reduced-form regressions in panel B and 2SLS regressions in panel C. The marginal significance in column (5) likely reflect the fact that 2005-2008 is a short period with reduced variation when comparing to 1999-2005 and 2008-2014. As before, first-stage estimates in panel D show the stable and strong positive correlation between the PLMNJ growth and gravity model-based IV for Net Export Growth (GIV-NEG), meaning our results are free from weak IV concerns. To sum up, induced by net export growth, the growth of private-label mortgages (non-jumbo) (99-05) causes household leverage to show a stronger rise between 2005 and 2008 in the high net-export-growth areas than in the low net-export-growth areas.

In 2009 and forward, credit kept at low level and a lot of government policies were carried out for debt relief, foreclosure prevention, and default prevention. Therefore, household leverage decline gradually after 2008.\footnote{For a complete review of U.S. government policies for the Great Recession, please see the summary from Yale School of Management at \url{https://ypfs.som.yale.edu/us-government-crisis-response}.}

\subsubsection{Irrelevance of Government-Sponsored Enterprise Mortgages}\label{sec:Irrelevance_GSEM}
%------------------------------------------------------------

We also perform a placebo test to show that government-sponsored enterprise mortgages (GSEMs) do not experience an expansion and cannot explain the household leverage increase across metropolitan areas. 

We use the same regression specification as in Equation (\ref{eq:HHLeverageBoomBust_on_PLMNJ}) except that we use government-sponsored enterprise mortgages instead. Table (\ref{table_HHDTI.D99t05vsD08t14.GSEM.4Reg}) reports OLS, reduced-form, second stage, and the first stage of the stacked regression of household leverage rise in the boom period (99-05) and the bust period (08-14). First, the OLS estimates in panel A show that the impact of GSEM growth is neither positively significant in the boom period (99-05) and nor negatively significant in the bust period (08-14). Likewise, 2SLS estimates are not significant in panel C. Further, first-stage estimates in panel D are insignificant, with too small first-stage F-Statistics (clustered Kleibergen-Paap F-statistic is 0.773, and the Montiel Olea-Pflueger Efficient F-Statistic is 0.777 in column (5)). As for the coefficient equality test in Table (\ref{table_HHDTI.D99t05vsD08t14.GSEM.2SLS.wide}), the chi-square statistics are small, and p-values are larger than 0.05.  In sharp contrast, all coefficients are significant at 1\% in the reduced-form regressions in panel B. This sharp comparison indicates that government-sponsored enterprise mortgages do not respond to the geographic divergence in net export growth, thus not relating to the stronger household leverage cycle in the high net-export-growth areas.

%------------------------------------------------------------
\subsubsection{Causal Evidence on Housing Net Worth}\label{sec:HousingNetWorth}
%------------------------------------------------------------

We also show that, induced by net export growth (99-05), growth in private-label mortgages (non-jumbo) (9-05) causes the massive reduction in housing net worth (07-09) calculated in \cite{mian2013household, mian2014explains}. The housing net worth is designed as a measure of how house price drops (07-09) deteriorate the household balance sheet at the county level.

We conduct tests by the following regression specification
\begin{equation}\label{eq:HousingNetWorthChangeD07t09_on_PLMNJD99t05}
\resizebox{0.92\textwidth}{!}{$
\triangle_{07,09} \text{Housing Net Worth}_{c} = \beta * \triangle_{99,05} Ln(PLMNJ_{c})  + \gamma* \bm{Controls_{c}} + \alpha + \epsilon_{c}
$} %end of \resizebox
\end{equation}
The left-hand-side dependent variable $\triangle_{07,09} \text{Housing Net Worth}_{c}$ is the housing net worth change at county $c$ 07-09. The key independent variable, instrumental variable, controls, weight, and clustered standard errors are the same as in Equation (\ref{eq:HHLeverageBoomBust_on_PLMNJ}).

Table (\ref{table_HousingNetWorth.D07t09.PLMNJ.D99t05.4Reg}) reports OLS, reduced-form, second stage, and the first stage of the regression of housing net worth change (07-09) on PLMNJ growth (99-05). First, the OLS coefficients in panel A show that the PLMNJ growth (99-05) is negatively significant at 1\%. Likewise, negative coefficients are significant in the reduced-form regressions in panel B and in 2SLS regressions in panel C. First-stage estimates in panel D show the stable and strong positive correlation between the gravity model-based IV for Net Export Growth (99-05)) and PLMNJ growth (99-05) at 1\% level. The smaller first-stage F-statistics are due to the smaller sample caused by the data availability of housing net worth change (07-09) because Table (\ref{table_HHDTI.D99t05vsD08t14.PLMNJ.4Reg}) panel D show large enough F-statistics when the sample size is large enough. Though with reduced sample, the first-stage coefficient 11.844 in column (5) in panel D in Table (\ref{table_HousingNetWorth.D07t09.PLMNJ.D99t05.4Reg}) is quite close to 11.716 in column (5) in panel D in Table (\ref{table_HHDTI.D99t05vsD08t14.PLMNJ.4Reg}) with a larger sample. Therefore, we still interpret our results as being free from the weak IV concern. To sum up, induced by net export growth, the growth of private-label mortgages (non-jumbo) (99-05) causes a stronger negative change in housing net worth in the bust period (07-09) in the high net-export-growth areas than in the low net-export-growth areas. To clearly see the predicting power of net export growth (99-05) on housing net worth (07-09), we present scatter plot and added-variable plot of reduced-form regression in Figure \ref{fig_ReducedForm_HousingNetWorth_on_GIV}. Both plots show a very strong correlation between the gravity model-based instrument variable of net export growth (99-05) and housing net worth change (07-09).

In terms of economic meaning, one standard deviation in six-year PLMNJ growth (99-05) causes housing net worth to change $8.078\% \times 0.817 \times 2 = 13.199\%$ 2007-2009. For housing net worth change, one standard deviation is $4.361\% \times 2 = 8.723\%$ 2007-2009. The two results mean that one standard deviation in six-year PLMNJ growth (99-05) can explain $13.199\% / 8.723\% = 151.32\%$ of one standard deviation in two-year change in housing net worth (07-09). In contrast, the 2SLS estimates in column (6) in Table (\ref{table_HousingNetWorth.D07t09.PLMNJ.D99t05.2SLS}) show that housing supply elasticity does not impact the housing net worth change (07-09) directly after controlling other factors, particularly the growth in private-label mortgages.

%------------------------------------------------------------
%------------------------------------------------------------
%------------------------------------------------------------
\subsection{Impact on Real Economy}\label{subsec:HousingIndustryChannel}
%------------------------------------------------------------
%------------------------------------------------------------
%------------------------------------------------------------
This section provides causal evidence that household leverage rise 1999-2005 causes boom and bust in many economic variables, including house price, residential construction planning, and house-related employment

\subsubsection{House Price Boom (99-05) and Bust (07-09)}
Based on the intuition described in the instruction and the evidence from Section \ref{subsec:causal_householdLeverage_boom_bust}, we predict that induced by net export growth, household leverage rises more in the high net-export-growth areas, thus causing a much stronger boom and a subsequent stronger bust in house price in these areas.

We test the house price boom and bust by the following stacked regression 
\begin{equation}\label{eq:HPI.BoomBust_on_HHDTI}
\resizebox{0.92\textwidth}{!}{$
\begin{aligned}
\triangle_{99,05} \& \triangle_{07,09} Ln(HPI_{c}) & = \beta_{99,05} * \triangle_{99,05} HHDTI_{c} \times Dum_{99,05} + \beta_{07,09} * \triangle_{99,05} HHDTI_{c} \times Dum_{07,09} \\
& + \gamma_{99,05}* \bm{Controls_{c}} \times Dum_{99,05} + \gamma_{07,09}* \bm{Controls_{c}} \times Dum_{07,09} + \epsilon_{period, c}
\end{aligned}
$} %end of \resizebox
\end{equation}
The left-hand-side dependent variable $\triangle_{99,05} \& \triangle_{07,09} Ln(HPI_{c})$ is the stacked growth rate of the house price index (deflated to 2007) at county $c$ 99-05 and 07-09. The key independent variable $\triangle_{99,05} HHDTI_{c}$ is the rise in household leverage (debt-to-income ratio) at county $c$ 99-05. We use the gravity model-based instrumental variable $\triangle_{99,05}\text{givNetExp}_{m}$ as the IV for $\triangle_{99,05} HHDTI_{c}$. Controls, weight, and standard errors are the same as Equation (\ref{eq:HHLeverageBoomBust_on_PLMNJ}).

Table (\ref{table_HPI.D99t05vsD07t09.HHDTI.4Reg}) reports OLS, reduced-form, second stage, and the first stage of the stacked regression of house price growth in the boom period (99-05) and the bust period (07-09) . First, the OLS coefficients in panel A shows that the impact of household leverage rise (99-05) are positively significant at 1\% in the boom period (99-05) and negatively significant at 1\% in the bust period (07-09). The similar trends apply to the reduced-form regressions in panel B and 2SLS regressions in panel C. First-stage estimates in panel D show a stable and strong positive correlation between the household leverage change and gravity model-based IV for Net Export Growth (GIV-NEG), with acceptable first-stage F-Statistics (clustered Kleibergen-Paap F-statistic is 8.736, and the Montiel Olea-Pflueger Efficient F-Statistic is 7.959).\footnote{Since we have shown the causal relationship between net export growth and household leverage change in Section \ref{subsec:causal_householdLeverage_boom_bust}, we believe the first-stage F statistics here are OK as they are close to or above eight.} Therefore, our results are free from weak IV concerns. As for the coefficient equality test of impact of household leverage rise (99-05) in the boom (99-05) and bust (07-09) periods in Table (\ref{table_HHDTI.D99t05vsD08t14.PLMNJ.2SLS.wide}), the chi-square statistics are large, and p-values are below 0.01, meaning the impact on difference periods are statistically different. To sum up, induced by net export growth, the rise in household leverage (99-05) causes house price to experience a stronger boom (99-05) and a stronger bust (07-09) in the high net-export-growth areas than in the low net-export-growth areas.

In terms of economic meaning, one standard deviation in six-year household leverage rise (99-05) causes house price to rise $12.961\% \times 0.445  \times 6 = 34.608\%$ 1999-2005 and to drop $ 12.961\% \times 0.304 \times 2 = 7.881\%$ 2007-2009. For house price growth, one standard deviation of six-year boom (99-05) is $3.421\% \times 6 = 20.528\% $ and one standard deviation of two-year bust (07-09) is $2.052\% \times 2 =4.103 \%$. The two results mean that one standard deviation in six-year household leverage rise (99-05) can explain $34.608\% / 20.528\% = 168.59\%$ of one standard deviation in six-year house price boom (99-05) and $7.881\% / 4.103\% = 192.06\%$ of one standard deviation in two-year house price bust (07-09). The higher explanatory power in the bust period is consistent with findings in many other settings. However, the 2SLS estimates in column (5) in Table (\ref{table_HPI.D99t05vsD07t09.HHDTI.2SLS.wide}) show that housing supply elasticity does not impact the house price boom or bust directly after controlling other factors, including the household leverage rise. The explanatory magnitude is also low. One standard deviation in housing supply elasticity can only explain $1.217 \time 0.003 \time 6 / 20.528\% = 10.67\%$ house price growth 1999-2005 and $1.217 \time 0.001 \time 2 / 4.103\% = 5.92\%$ house price drop 2007-2009.

%------------------------------------------------------------
%------------------------------------------------------------
\subsubsection{Residential Construction Investment Boom (99-05) and Bust (05-09)}
%------------------------------------------------------------
%------------------------------------------------------------
In addition, we also predict that the household leverage cycle causes a much stronger boom and bust cycle in residential construction investment (building permit value) in the high net-export-growth areas.

We test the residential construction investment (building permit value) boom and bust by the following stacked regression 
\begin{equation}\label{eq:Permit.BoomBust_on_HHDTI}
\resizebox{0.92\textwidth}{!}{$
\begin{aligned}
\triangle_{99,05} \& \triangle_{05,09} Ln(\text{PermitValue}_{c}) & = \beta_{99,05} * \triangle_{99,05} HHDTI_{c} \times Dum_{99,05} + \beta_{05,09} * \triangle_{99,05} HHDTI_{c} \times Dum_{05,09} \\
& + \gamma_{99,05}* \bm{Controls_{c}} \times Dum_{99,05} + \gamma_{05,09}* \bm{Controls_{c}} \times Dum_{05,09} + \epsilon_{period, c}
\end{aligned}
$} %end of \resizebox
\end{equation}
The left-hand-side dependent variable $\triangle_{99,05} \& \triangle_{05,09} Ln(\text{PermitValue}_{c})$ is the stacked growth rate of the residential construction investment (building permit value) (deflated to 2007) at county $c$ 99-05 and 05-09. The key independent variable $\triangle_{99,05} HHDTI_{c}$ is the rise in household leverage (debt-to-income ratio) at county $c$ 99-05. We use the gravity model-based instrumental variable $\triangle_{99,05}\text{givNetExp}_{m}$ as the IV for $\triangle_{99,05} HHDTI_{c}$. Controls, weight, and standard errors are the same as Equation (\ref{eq:HHLeverageBoomBust_on_PLMNJ}).

Table (\ref{table_Permit.D99t05vsD05t09.HHDTI.4Reg}) reports OLS, reduced-form, second stage, and the first stage of the stacked regression of residential construction investment growth in the boom period (99-05) and the bust period (05-09) . First, the OLS coefficients in panel A shows that the impact of household leverage rise (99-05) are positive and only significant in column (1)-(3) in the boom period (99-05) and negatively significant at 1\% for all columns in the bust period (07-09). The relatively weaker impact in the boom period in column (4) and (5) likely reflect the fact that some household leverage growth is expected. However, results are all significant in both the boom and bust periods in reduced-form regressions in panel B and 2SLS regressions in panel C. First-stage estimates in panel D show a stable and strong positive correlation between the household leverage change and gravity model-based IV for Net Export Growth (GIV-NEG), with acceptable first-stage F-Statistics (clustered Kleibergen-Paap F-statistic is 9.267, and the Montiel Olea-Pflueger Efficient F-Statistic is 8.726).\footnote{Since we have already shown the causal relationship between net export growth and household leverage rise in Section \ref{subsec:causal_householdLeverage_boom_bust}, we believe the first-stage F statistics are OK as they are above eight.} Thus, our results are free from weak IV concerns. As for the coefficient equality test in Table (\ref{table_Permit.D99t05vsD05t09.HHDTI.2SLS.wide}), the chi-square statistics are large, and p-values are all below 0.02, meaning the impact on difference periods are statistically different. To sum up, induced by net export growth, the rise in household leverage (99-05) causes residential construction investment to experience a stronger boom (99-05) and bust (05-09) in the high net-export-growth areas than in the low net-export-growth areas.

In terms of economic meaning, one standard deviation in six-year household leverage rise (99-05) causes residential construction investment to rise $12.877\% \times 0.749  \times 6 = 57.867\%$ 1999-2005 and to drop $ 12.877\% \times -1.839 \times 4 = 97.720\%$ 2005-2009. For residential construction investment growth, one standard deviation of six-year boom (99-05) is $9.425\% \times 6 = 56.550\% $ and one standard deviation of four-year bust (05-09) is $16.906\% \times 4 =67.623 \%$. The two results mean that one standard deviation in six-year household leverage rise (99-05) can explain $57.867\% / 56.550\% = 102.33\%$ of one standard deviation in six-year residential construction investment boom (99-05) and $97.720\% / 67.623\% = 140.07\%$ of one standard deviation in four-year residential construction investment bust (07-09). The higher explanatory power in the bust period is consistent with findings in many other settings. However, the 2SLS estimates in column (5) in Table (\ref{table_Permit.D99t05vsD05t09.HHDTI.2SLS.wide}) show that housing supply elasticity does not impact the residential construction investment boom or bust directly after controlling other factors, including the household leverage rise.

%------------------------------------------------------------
%------------------------------------------------------------
\subsubsection{Refined House Employment (00-06) and Bust (07-10)}
%------------------------------------------------------------
%------------------------------------------------------------

Further, we predict that the household leverage cycle causes a much stronger boom and bust cycle in house-related employment in the high net-export-growth metropolitan areas. We focus on refined house employment that covers employment in residential construction industries, other industries, and mortgage banker industries defined in \cite{goukasian2010reaction}. We list detailed industries in 1987 SIC code in Table (\ref{table_EmploymentIndustryClassification}).

We test the above prediction with the following stacked regression 
\begin{equation}\label{eq:RefineHouseEmp.BoomBust_on_HHDTI}
\resizebox{0.92\textwidth}{!}{$
\begin{aligned}
\triangle_{00,06} \& \triangle_{07,10} RefinedHouseEmpShr_{c} & = \beta_{00,06} * \triangle_{99,05} HHDTI_{c} \times Dum_{00,06} + \beta_{07,10} * \triangle_{99,05} HHDTI_{c} \times Dum_{07,10} \\
& + \gamma_{00,06}* \bm{Controls_{c}} \times Dum_{00,06} + \gamma_{07,10}* \bm{Controls_{c}} \times Dum_{07,10} + \epsilon_{period, c}
\end{aligned}
$} %end of \resizebox
\end{equation}
The left-hand-side dependent variable $\triangle_{00,06} \& \triangle_{07,10} RefinedHouseEmpShr_{c}$ is is the change of the refined house employment share in working-age population at county $c$ 00-06 and 07-10. The key independent variable $\triangle_{99,05} HHDTI_{c}$ is the rise in household leverage (debt-to-income ratio) at county $c$ 99-05. We use the gravity model-based instrumental variable $\triangle_{99,05}\text{givNetExp}_{m}$ as the IV for $\triangle_{99,05} HHDTI_{c}$. To reduce the impact of outliers resulted from data imputation in County Business Pattern data, the dependent variable is winsorized at 5\% and 95\% levels in each period. Controls, weight, and standard errors are the same as Equation (\ref{eq:HHLeverageBoomBust_on_PLMNJ}).

Table (\ref{table_RefineHouse.D00t06vsD07t10.HHDTI.4Reg}) reports OLS, reduced-form, second stage, and the first stage of the stacked regression of refined house employment share change in the boom period (00-06) and the bust period (07-10) . First, the OLS coefficients in panel A shows that the impact of household leverage rise (99-05) are positively significant in the boom period (00-06) and negatively significant in the bust period (07-10), except for the last column. However, all key coefficients are significant in reduced-form regressions in panel B and 2SLS regressions in panel C. First-stage estimates in panel D display a stable and strong positive correlation between the gravity model-based IV for Net Export Growth (GIV-NEG) and household leverage rise, with acceptable first-stage F-Statistics (clustered Kleibergen-Paap F-statistic is 10.34, and the Montiel Olea-Pflueger Efficient F-Statistic is 9.921).\footnote{Since we have shown the robust causal relationship between net export growth and household leverage rise in Section \ref{subsec:causal_householdLeverage_boom_bust}, we think the first-stage F statistics here are good as they are close to or above eight.} Consequently, our results are free from weak IV concerns. As for the coefficient equality test in Table (\ref{table_HHDTI.D99t05vsD08t14.PLMNJ.2SLS.wide}), the chi-square statistics are large, and p-values are below 0.05, meaning the impact on difference periods are statistically different. To sum up, induced by net export growth, the rise in household leverage (99-05) causes refined house employment to experience a stronger boom (00-06) and a stronger bust (07-10) in the high net-export-growth areas than in the low net-export-growth ones.

In terms of economic meaning, one standard deviation in six-year household leverage rise (99-05) causes refined house employment share to rise $12.945\% \times 0.421  \times 6/100 = 0.327\%$ 1999-2005 and to drop $ 12.945\% \times -0.696 \times 3 /100 = -0.270\%$ 2007-2010. For refined house employment share change, one standard deviation of six-year boom (99-05) is $0.041\% \times 6/100 = 0.249\% $ and one standard deviation of three-year bust (07-10) is $0.074\% \times 3/100 = 0.222\%$. The two results mean that one standard deviation in six-year household leverage rise (99-05) can explain $0.327\% / 0.249\% = 131.36\%$ of one standard deviation in six-year refined house employment share boom (99-05) and $0.270\% / 0.222\% = 121.80\%$ of one standard deviation in three-year refined house employment share bust (07-10). However, the 2SLS estimates in column (5) in Table (\ref{table_RefineHouse.D00t06vsD07t10.HHDTI.2SLS.wide}) show that housing supply elasticity does not impact the refined house employment boom or bust directly after controlling other factors, notably the household leverage rise.

%------------------------------------------------------------
%------------------------------------------------------------
\subsubsection{BEA Construction Employment (00-06) and Bust (07-10)}
%------------------------------------------------------------
%------------------------------------------------------------

Moreover, we expect that the household leverage cycle causes a much stronger boom and bust cycle in broader construction sector defined by the U.S. Bureau of Economic Analysis in the high net-export-growth metropolitan areas.

We test the above prediction with the same regression in Equation (\ref{eq:RefineHouseEmp.BoomBust_on_HHDTI}) except that the dependent variable is now the change of the BEA construction employment share in working-age population at county $c$ 00-06 and 07-10. To reduce the impact of outliers, the dependent variable is winsorized at 2\% and 98\% levels in each period.

Table (\ref{table_BEAConstEmp.D00t06vsD07t10.HHDTI.4Reg}) reports OLS, reduced-form, second stage, and the first stage of the stacked regression of BEA construction employment share change in the boom period (00-06) and the bust period (07-10) . First, the OLS coefficients in panel A are positively significant but not stable in the boom period (00-06) and negatively significant in the bust period (07-10). However, most key coefficients are significant in reduced-form regressions and 2SLS regressions. First-stage estimates display a stable and strong positive correlation between the gravity model-based IV for Net Export Growth (GIV-NEG) and household leverage rise. Thus, our results are free from weak IV concerns. As for the coefficient equality test in Table (\ref{table_BEAConstEmp.D00t06vsD07t10.HHDTI.2SLS.wide}), the chi-square statistics are large, and p-values are below 0.03 (except for being 0.062 in column (3)), meaning the impact on difference periods are statistically different. To sum up, induced by net export growth, the rise in household leverage (99-05) causes BEA construction employment to experience a stronger boom (00-06) and a stronger bust (07-10) in the high net-export-growth areas than in the low net-export-growth ones.

In terms of economic meaning, one standard deviation in six-year household leverage rise (99-05) causes BEA construction employment share to rise $12.581\% \times 0.019  \times 6 = 1.478\%$ 1999-2005 and to drop $ 12.581\% \times -0.025 \times 3  = -0.972\%$ 2007-2010. For BEA construction employment share change, one standard deviation of six-year boom (99-05) is $0.199\% \times 6 = 1.192\% $ and one standard deviation of three-year bust (07-10) is $0.323\% \times 3= 0.969\%$. The two results mean that one standard deviation in six-year household leverage rise (99-05) can explain $1.478\% / 1.192\% = 124.01\%$ of one standard deviation in six-year BEA construction employment share boom (99-05) and $0.972\% / 0.969\% = 100.33\%$ of one standard deviation in three-year BEA construction employment share bust (07-10). However, the 2SLS estimates in column (4) in Table (\ref{table_BEAConstEmp.D00t06vsD07t10.HHDTI.2SLS.wide}) show that housing supply elasticity does not impact the BEA construction employment boom or bust directly after controlling other factors, particularly the household leverage rise.

%################################################################################
%################################################################################
% This is the end of the entire section (tex file)
%################################################################################
%################################################################################

%% file: Empirical.AlternativeHypotheses.tex
%------------------------------------------------------------
%------------------------------------------------------------
%\clearpage
%------------------------------------------------------------
%------------------------------------------------------------
\section{Empirical: Addressing Alternative Hypotheses}
This section provides evidence against alternative hypotheses regarding the major channel leading to the Great Recession. 

%------------------------------------------------------
%------------------------------------------------------
\subsection{Addressing Hypotheses Predicting The Firm Channel}

Many theories (hypotheses) regarding the business cycles or financial crises predict that firms rather than households play the dominant role in driving the boom and the bust. These theories (hypotheses) include real business cycle theories by \cite{prescott1986theory}, the collateral-driven credit cycle theory by \cite{kiyotaki1997credit}, and the business uncertainty theory by \cite{bloom2009impact}. 

We provide four pieces of evidence against the above theories. First, the aggregate level of debt-to-GDP ratio in household and nonprofit organizations rather than the non-financial corporate experienced a persistent boom between 1999 and 2007 and a persistent bust between 2008 and 2014 in Subfigure (a) in Figure \ref{fig_DebtToGDPRatio_HHBusiGov_HHSubCategories}. Additionally, Subfigure (b) shows that the mortgages dominated the boom and bust cycle in debt of households and nonprofit organizations while little changes happened in consumer credit and other debts. Such empirical facts at the aggregate level are directly against the there theories above.

Second, we show that the tradable employment experiences a stronger growth in both the boom (00-06) and bust (07-10) periods in the high net-export-growth areas. Table \ref{table_Tradable.D00t06vsD07t10.NEP.D99t05.2SLS.wide} and \ref{table_Tradable.D00t06vsD07t10.NEP.D99t05.4Reg} report the corresponding regression results. This piece of evidence is against the following two theories. If the real business cycle theory were to explain the 1999-2010 U.S. business cycle, then there shall be some technology shocks or natural disasters disproportionately affect the high net-export-growth areas and result in a stronger bust. This prediction is in contradiction of the above empirical evidence because tradable employment enjoy a differently stronger growth in the bust period (2007-2010). If the business uncertain theory were to explain the business cycle, then there shall be a stronger bust in employment in all sectors in the high net-export-growth areas, including the tradable sector. On the contrary, this prediction is directly disproved by the above empirical evidence.

Third, we present evidence that commercial employment experiences neither a stronger boom nor a stronger bust in the high net-export-growth areas. Table \ref{table_ComConstEmp.D00t06vsD07t10.NEP.D99t05.2SLS.wide} and \ref{table_ComConstEmp.D00t06vsD07t10.NEP.D99t05.4Reg} report the regression results. This evidence is against the following three theories. If the real business cycle theory is correct, commercial construction employment shall experience a stronger boom and a stronger bust in the high net-export-growth areas due to a stronger boom and bust cycle in the local economy. On the contrary, this prediction is directly disputed by the above empirical result. In addition, if the collateral-driven credit cycle theory is correct, due to a stronger house price cycle in the high net-export-growth areas, commercial construction employment shall experience a stronger cycle as well resulting from collateral-driven debt by firms. However, such a cycle in commercial construction employment is not supported by the above empirical evidence. Further, if the business uncertain theory is correct, the uncertainty must be higher in the high net-export-growth areas in the bust period (07-10) in order to explain the stronger bust in local economy. This prediction implies that a stronger bust in employment in many sectors. However, we do not see such a stronger bust in the commericial construction sector.

\comment{
The real business cycle theory cannot explain why the rise and fall of total productivity fact of corporations do not result in boom and bust in commercial construction employment. The collateral-driven credit cycle theory cannot explain why house price cycle-induced business cycle do not lead to a similar cycle in commercial construction sector. The business uncertainty theory cannot explain why the predicted higher uncertainty in the high net-export-growth areas do not result in a stronger bust in the commercial construction employment. 
}

Fourth, in Figure \ref{fig_PLMNJ_vs_HousePriceIndex}, we show that at aggregate level private-label mortgages increases at a similar pace with house prices in the boom (1999-2005) but declining before house prices crash in the bust (2007-2009). This evidence, particularly in the bust period, is directly against the prediction by the collateral-driven credit cycle theory.

%------------------------------------------------------
%------------------------------------------------------
\subsection{Addressing Hypotheses Predicting Both Firm and Household Channels}

The extrapolative expectation theory by \cite{eusepi2011expectations} predict that both firms and households play important roles in business cycle. We show two pieces of evidence against this theory. First, this theory predicts that the initial economic growth (likely driven by net export growth) triggers extrapolative expectation in both firms and households. Then debt in both parts shall experience a boom and bust cycle. However, Figure \ref{fig_DebtToGDPRatio_HHBusiGov_HHSubCategories} demonstrates that debt of households and nonprofit organizations instead of non-financial corporations experienced a strong boom and bust cycle. Second, if the extrapolative expectation is the key driving force and is stronger in the high net-export-growth areas, then both private-label mortgages (non-jumb) by the relatively low-credit households and government-sponsored enterprise mortgages by the credit-qualified households shall experience boom and bust cycle. However, Table \ref{table_HHDTI.D99t05vsD08t14.GSEM.4Reg} shows that government-sponsored enterprise mortgages did not experience a stronger boom (99-05) or a stronger bust (05-08) in the high net-export-growth areas.

The speculative euphoria hypothesis by \cite{kindleberger1978manias, minsky1986stabilizingan} predict that both firms and households speculate with bank-expanded credit, thus causing a boom and bust cycle in asset prices. If speculation were the most important driving force of the 1999-2010 U.S. business cycle, this theory must explain why the aggregate data in Figure \ref{fig_DebtToGDPRatio_HHBusiGov_HHSubCategories} shows possible speculation (debt cycle) in the households and nonprofit organizations but not in the non-financial corporations. Second, for detailed evidence against the speculation in housing market, which is out of the scope of this paper, please refer to \cite{li2024credit}.

%% file: Empirical.Robustness.tex
%------------------------------------------------------------
%------------------------------------------------------------
\section{Empirical: Robustness}\label{subsec:Empirical.Robustness}
%------------------------------------------------------------
%------------------------------------------------------------

In this subsection, we perform robustness tests to support the main conclusion that, induced by net export growth, credit expansion in private-label mortgages (non-jumbo) causes the household leverage boom (9-05) and bust (08-14). This household leverage cycle eventually results in the house price boom (99-05) and bust (07-09). We show that the main conclusion is quite robust to state-level distinctions in anti-predatory lending laws \citep{di2017credit}, recourse laws \citep{ghent2011recourse}, judicial requirement in house foreclosure \citep{mian2015foreclosures}, category of sand states \citep{choi2016sand}, and state capital gain tax \cite{gao2020economic}.

%------------------------------------------------------------
\subsection{Anti-Predatory-Lending States vs. Other States}
%------------------------------------------------------------

This subsection shows that our main conclusion still holds after controlling state-level differences in the anti-predatory lending law. \cite{di2017credit} show that a dozen states had already implemented the anti-predatory law to protect borrowers before 2004. Nonetheless, on January 7th, 2004, the Office of the Comptroller of the Currency (OCC) preempted national banks (rather than state-chartered depository institutions or independent mortgage companies) from state-level anti-predatory lending law (APL law). They document that such deregulation caused the credit expansion in national banks (relative to state-regulated institutions), house price rise, and nontradable employment growth in 2004-2006, and a dramatic decline subsequently in these states. Appendix Table 1 in \cite{di2017credit} summarizes the list of APL states before 2004. As HMDA only contains annual data, we restrict our APL states to those that adopted APL at least half a year before 2004, ending up with eleven APL states.\footnote{Using the information in \cite{di2017credit}, our sample of eleven APL states are California, Connecticut, District of Columbia, Georgia, Maryland, Michigan, Minnesota, New York, North Carolina, Texas, and West Virginia.}

Our stacked 2SLS regression is 
\begin{equation}\label{eq:HPI.BoomBustonPLMNJ_APLvsNone}
\resizebox{0.92\textwidth}{!}{$
\begin{aligned}
\triangle_{99,05} \& \triangle_{07,09} Ln(HPI_{c}) & = \beta_{Boom} * \triangle_{99,05} HHDTI_{c} \times Dum_{99,05} + \beta_{Bust} * \triangle_{99,05} HHDTI_{c} \times Dum_{07,09} \\
 & + \beta_{APL, Boom} * \triangle_{99,05} HHDTI_{c} \times Dum_{99,05} \times Dum_{APL} + \beta_{APL, Bust} * \triangle_{99,05} HHDTI_{c} \times Dum_{07,09} \times Dum_{APL} \\
 & + \gamma_{Boom} * \bm{Controls_{c}} \times Dum_{99,05} + \gamma_{Bust} * \bm{Controls_{c}} \times Dum_{07,09} + \epsilon_{c}
\end{aligned}
$} %end of \resizebox
\end{equation}
Controls, weight, and standard errors are the same as Eq (\ref{eq:HPI.BoomBust_on_HHDTI}).

Please note that we have two endogenous variables in the above specification in either boom or bust period. $\triangle_{99,05} HHDTI_{c}$ is instrumented by $\triangle_{99,05}\text{givNetExp}_{m}$ and $\triangle_{99,05} HHDTI_{c} \times Dum_{APL}$ is instrumented by $\triangle_{99,05}\text{givNetExp}_{m} \times Dum_{APL}$. For each of the two first-stage regression F-tests, we employ Sanderson-Windmeijer robust (clustered) F-statistics \citep{sanderson2016weak}. The SW F-statistics is 9.283 for household leverage rise and 22.78 for the interaction between household leverage rise and a dummy of APL-states in column (5), meaning each F-stage regression is significant, and each instrument is strong for its endogenous variable. To evaluate the overall strength of the two instruments, we use the p-value of robust (clustered) Kleibergen-Paap test statistic developed by \citep{windmeijer2021testing}. The p-value is 0.0110 in column (5), meaning the two instruments are jointly strong for the two endogenous variables.

Table (\ref{table_Robust.APLvsNone.HPI.D99t05.D07t09.HHDTI}) reports results of the above 2SLS regression and displays two key conclusions. First, after accounting for the potential differential trend in APL-states, for all metropolitan areas, the household leverage rise 1999-2005 leads to a stronger house price boom (99-05) and a stronger bust (07-09) in the high net-export-growth areas. Second, compared to non-APL states, APL-states seems to experience a differentially weaker house price boom (99-05) or a differentially weaker bust (07-09) caused by household leverage rise in the high net-export-growth metropolitan areas. Though the second result seems to be odd with the conclusion by \cite{di2017credit}, further exploration in this direction goes beyond our paper. We have achieved our purpose of testing the robustness of the main conclusion.

%------------------------------------------------------------
\subsection{Non-Recourse vs. Recourse States}
%------------------------------------------------------------

In this subsection, we show that the main conclusion is robust to the difference between non-recourse and recourse states. \cite{ghent2011recourse} show that mortgages are recourse loans in 39 states and Washington, D.C. in U.S..\footnote{The 39 recourse states are in Table 1 in \cite{ghent2011recourse}. The eleven non-recourse states are Alaska, Arizona, California, Iowa, Minnesota, Montana, North Carolina, North Dakota, Oregon, Washington, and Wisconsin.} In such states, lenders could go after the borrower's other assets to recover the mortgage loss not covered by the foreclosure sale by obtaining a deficiency judgment. They document that in recourse states, borrowers are less responsive to negative equity, and defaults are more likely to proceed via a lender-friendly procedure. We worry that the non-recourse law may induce a weaker credit expansion in mortgages and induce borrowers to be more willing to default amid falling house price. 

We employ the same regression as Eq (\ref{eq:HPI.BoomBustonPLMNJ_APLvsNone}) except that the dummy variable is for non-recourse states. Table (\ref{table_Robust.NRCvsRC.HPI.D99t05.D07t09.HHDTI}) reports the above 2SLS results and shows two key conclusions. First, after accounting for the potentially different trend in the non-recourse states, the household leverage rise 1999-2005 leads to a stronger house price boom (99-05) and a stronger bust (07-09) in the high net-export-growth metropolitan areas. Second, Compared with recourse states, non-recourse states did not go through a stronger house price bust (07-09) or a stronger bust (07-09). The SW F-statistics and p-value of robust (clustered) Kleibergen-Paap test statistics together show that instruments are separately and jointly valid for the two endogenous variables.

%------------------------------------------------------------
\subsection{Non-Judicial vs Judicial States}
%------------------------------------------------------------

In this subsection, we show that our main conclusion is robust to the state-level difference in judicial requirement in foreclosure. \cite{mian2015foreclosures} show that the foreclosure of a delinquent property needs judicial judgement in 20 states of the U.S..\footnote{20 judicial states are summarized in Figure 2 in \cite{mian2015foreclosures} and \url{https://www.realtytrac.com/real-estate-guides/foreclosure-laws/}. The twenty judicial states are Connecticut, Delaware, Florida, Illinois, Indiana, Kansas, Kentucky, Louisiana, Maine, Maryland, Massachusetts, Nebraska, New Jersey, New Mexico, New York, North Dakota, Ohio, Pennsylvania, South Carolina, and Vermont.} In these states, in order to sell a delinquent property through foreclosure to recover loss, lenders are mandated by law to file a notice with a judge to provide evidence of the delinquency and get court approval. In comparison, in non-judicial states, the foreclosure procedure is much easier and does not require court approval. For more details, see \cite{mian2015foreclosures}. They document that lenders are twice as likely to foreclose on delinquent property in non-judicial states. We worry that non-judicial states might dominate our results by encouraging lenders to expand mortgages more aggressive, resulting in a stronger house price boom (99-05) and a stronger bust (07-09) subsequently.

We conduct the same regression test as in Eq (\ref{eq:HPI.BoomBustonPLMNJ_APLvsNone}) except that dummy variable is for non-judicial states. Table (\ref{table_Robust.NJDvsJD.HPI.D99t05.D07t09.HHDTI}) reports the above 2SLS results and shows two major conclusions. First, after accounting for the potentially different trend in non-recourse states, the household leverage rise 1999-2005 led to both a stronger house price boom (99-05) and a stronger bust (07-09) in the high net-export-growth areas. Second, In comparison to judicial states, non-judicial states did not experience either a stronger boom (99-05) or a stronger bust (07-09) in house price, caused by the household leverage rise 1999-2005. The SW F-statistics and p-value of robust (clustered) Kleibergen-Paap test statistics combined show that instruments are both separately and jointly strong for the two endogenous variables.

%------------------------------------------------------------
\subsection{Sand vs Other States}
%------------------------------------------------------------

In this subsection, we show that our main conclusion is still valid after controlling the difference between the sand and non-sand states. In addition, the household leverage rise (99-05) led to a stronger bust in house prices in sand states. Many studies document that sand states (Arizona, California, Florida, and Nevada) experienced an amplified housing cycle than the rest of the United States \citep{choi2016sand}. We worry that the sand states might dominate the results for our major conclusion.

We add an interaction of dummy variable for sand states and period dummy to Eq (\ref{eq:HPI.BoomBust_on_HHDTI}). We do not use an interaction of three terms as limited number of metropolitan counties (only 73 metro counties in sand states) present a weak IV concern. Since our major concern is the differential housing boom and bust of sand states, a dummy variable can help us address this concern. 

Table (\ref{table_Robust.SandvsNone.HPI.D99t05.D07t09.HHDTI}) reports the above 2SLS results and displays two key results. First, compared to other states, sand states experienced a stronger bust (07-09) not a stronger boom (99-09). Second, after accounting for the differential trend in the sand states, we only use the within-sand-states and within-other-states differences across metropolitan areas. Since the cross-group difference between sand states and other states is removed by the interaction terms of sand dummy and period dummy, our 2SLS result shall be interpreted as evidence strongly supporting our main conclusion: induced by net export growth, private-label mortgage caused a stronger boom (99-05) in household leverage, which eventually resulted in a stronger boom (99-05) and a stronger bust (07-09) in the house price in the high net-export-growth areas. In addition, the reduced cross-metro variation after controlling for the sand-state dummy reduces the F-statistics: the kleibergen-Paap (2006) robust (clustered) statistics is 8.736 and Montiel Olea-Pflueger (2013) efficient statistics is 7.959. The reduced cross-metro variation might also explain the weaker but still marginal significant coefficient of household leverage rise in the column (5) in the boom period.

%------------------------------------------------------------
\subsection{State Capital Gain Tax}
%------------------------------------------------------------

In this section, we show that our major conclusion is valid after inclusion of state capital gain tax as a control variable. \cite{gao2020economic} document that speculation (measured by non-owner-occupied purchase mortgages) is discouraged by the state capital gain tax and such speculation contributes to the housing boom and bust. Based on their findings, we worry the main results might be dominated by metropolitan areas in states with low capital gain tax since state capital gain tax discourages housing boom and bust. 

For the regression test, we add an interaction of state capital gain tax rate and period dummy to Eq (\ref{eq:HPI.BoomBust_on_HHDTI}). Table (\ref{table_Robust.StCapGainTax.HPI.D99t05.D07t09.HHDTI}) displays the above 2SLS results and presents two key conclusions. First, after controlling for state capital gain tax rate, our major conclusion holds. Second, consistent with the conclusion in \cite{gao2020economic}, state capital gain tax rate weakens the house price boom (99-05) and weakens the bust (07-09).

%################################################################################
%################################################################################
% This is the end of the entire section (tex file)
%################################################################################
%################################################################################

%% file: Conclusion.tex
%----------------------------------------------------------------------------
%\clearpage

\section{Conclusion}

Understanding the cause and the dominant mechanism of the 1999-2010 U.S. business cycle is important because the U.S. economy experienced the worst recession since the Great Depression. The literature on business cycles, however, has a continuing debate on the role between firms and households. First, most theories emphasize the role of firms while more recent empirical studies uncover the dominant role of households. Second, cross-country empirical studies usually cannot achieve causal inference whereas within-country papers mostly focus on the bust period. 

First, this paper documents a new empirical fact: the household leverage cycle is much stronger in the high net-export-growth metropolitan areas than in the low ones. Second, by a unique research design, this paper provides the first causal evidence that credit supply expansion caused the 1999-2010 U.S. business cycle mainly through the channel of household leverage (debt-to-income ratio). Specifically, induced by net export growth, credit expansion in private-label mortgages (non-jumbo), rather than the government-sponsored enterprise mortgages causes a much stronger boom and bust cycle in household leverage in the high net-export-growth areas. Third, such a stronger household leverage cycle leads to a stronger boom and bust cycle in the local economy in these areas, including housing prices, residential construction investment, and house-related employment. Thus, our results are consistent with the credit-driven household demand channel by \cite{mian2018finance}. Fourth, we provide multiple pieces of evidence against the firm channel predicted by the real business cycle theory \citep{prescott1986theory}, the collateral-driven credit cycle theory \citep{kiyotaki1997credit}
, the business uncertainty theory \citep{bloom2009impact}, and the extrapolative expectation theory \citep{eusepi2011expectations}.

\comment{
(To-do list) In addition, the household leverage-induces boom and bust cycle is much stronger in durable goods consumption than in nondurable goods consumption. Further, in the spirit of \cite{li2024credit}, we provide evidence of ``double differences": the differential higher growth rate of consumption in low-income than high-income ZIP codes within the same metropolitan area is significantly caused by net export growth across metropolitan areas. Evidence of ``double differences" further validates the ``credit expansion" view while against the ``demand-based" view. Moreover, by comparison of inferences with different sets of data, we demonstrate the importance of using data in various sectors (nontradable vs. tradable employment, nondurable vs. durable consumption) across metropolitan areas in disciplining macroeconomic models of business cycles.
}

%################################################################################
%################################################################################
% This is the end of the entire section (tex file)
%################################################################################
%################################################################################

%% file: FiguresAndTables.tex
%------------------------------------------------------------
%------------------------------------------------------------
\pagebreak
%------------------------------------------------------------
%------------------------------------------------------------
\section{Figures and Tables}

%------------------------------------------------------------
% figure 1: fig_USMetroCty_NetExpGrowth

\input{Figure_HH/fig_USMetroCty_NetExpGrowth}

\pagebreak
%------------------------------------------------------------
% fig_HouseholdLeverage_99To14
%------------------------------------------------------------
% figure 2: 

\input{Figure_HH/fig_HouseholdLeverage_99To14}

\pagebreak
%------------------------------------------------------------
%------------------------------------------------------------
% figure 3: fig_GSEMvsPLMNJ_91t11_combine

\input{Figure_HH/fig_GSEMvsPLMNJ_91t11_combine}

\pagebreak
%------------------------------------------------------------
%------------------------------------------------------------
% figure 4: fig_ReducedForm_HousingNetWorth_on_GIV

\input{Figure_HH/fig_ReducedForm_HousingNetWorth_on_GIV}

\pagebreak
%------------------------------------------------------------
%------------------------------------------------------------
% figure 5: fig_PLMNJ_vs_HousePriceIndex

\input{Figure_HH/fig_PLMNJ_vs_HousePriceIndex}

%\pagebreak
%------------------------------------------------------------
%------------------------------------------------------------
% figure 6: fig_DebtToGDPRatio_HHBusiGov_HHSubCategories

\input{Figure_HH/fig_DebtToGDPRatio_HHBusiGov_HHSubCategories
}

\pagebreak
%------------------------------------------------------------
% table 0: table_EmploymentIndustryClassification

%\input{Table_HH/table_EmploymentIndustryClassification}

\pagebreak
%------------------------------------------------------------
% table 0: table_SumStat1

%\input{Table_HH/table_SumStat1}

\pagebreak
%------------------------------------------------------------
% table 0: table_SumStat2

%\input{Table_HH/table_SumStat2}

\pagebreak
%------------------------------------------------------------
% table_EmploymentIndustryClassification

\input{Table_HH/table_EmploymentIndustryClassification}

%---------------------------------------------------------------
%---------------------------------------------------------------
% Empirical: Main Tests 1.1
% Household Leverage Cycle
%---------------------------------------------------------------
%---------------------------------------------------------------

\pagebreak
%-----------------------------------------------------------------

%%%%%%%%%%%%%%%%%%%%%%%%%%%%%%%%%%%%
% table_HHDTI.D99t05vsD08t14.PLMNJ.2SLS.wide
%%%%%%%%%%%%%%%%%%%%%%%%%%%%%%%%%%%%

\input{Table_HH/table_HHDTI.D99t05vsD08t14.PLMNJ.2SLS.wide}

\pagebreak 
%---------------------------------------------------------------

%%%%%%%%%%%%%%%%%%%%%%%%%%%%%%%%%%%%%%%%%%%%%%%%
% table_HHDTI.D99t05vsD08t14.PLMNJ.4Reg
%%%%%%%%%%%%%%%%%%%%%%%%%%%%%%%%%%%%%%%%%%%%%%%%

\input{Table_HH/table_HHDTI.D99t05vsD08t14.PLMNJ.4Reg}

%---------------------------------------------------------------
%---------------------------------------------------------------
% Empirical: Main Tests 1.2
% Household Leverage Cycle.
% Placebo: GSEM 
%---------------------------------------------------------------
%---------------------------------------------------------------

\pagebreak
%-----------------------------------------------------------------

%%%%%%%%%%%%%%%%%%%%%%%%%%%%%%%%%%%%
% table_HHDTI.D99t05vsD08t14.GSEM.2SLS.wide
%%%%%%%%%%%%%%%%%%%%%%%%%%%%%%%%%%%%

\input{Table_HH/table_HHDTI.D99t05vsD08t14.GSEM.2SLS.wide}

\pagebreak 
%---------------------------------------------------------------

%%%%%%%%%%%%%%%%%%%%%%%%%%%%%%%%%%%%%%%%%%%%%%%%
% table_HHDTI.D99t05vsD08t14.GSEM.4Reg
%%%%%%%%%%%%%%%%%%%%%%%%%%%%%%%%%%%%%%%%%%%%%%%%

\input{Table_HH/table_HHDTI.D99t05vsD08t14.GSEM.4Reg}

%---------------------------------------------------------------
%---------------------------------------------------------------
% Empirical: Main Tests 1.3
% Household Leverage Cycle 
% Transition Period (2005-2008)
%---------------------------------------------------------------
%---------------------------------------------------------------

\pagebreak
%-----------------------------------------------------------------

%%%%%%%%%%%%%%%%%%%%%%%%%%%%%%%%%%%%
% table_HHDTI.D05t08.PLMNJ.D99t05.4Reg
%%%%%%%%%%%%%%%%%%%%%%%%%%%%%%%%%%%%

\input{Table_HH/table_HHDTI.D05t08.PLMNJ.D99t05.4Reg}

%---------------------------------------------------------------
%---------------------------------------------------------------
% Empirical: Main Tests 1.4
% Impact on Housing Net Worth Change (07-09)
%---------------------------------------------------------------
%---------------------------------------------------------------

\pagebreak
%-----------------------------------------------------------------

%%%%%%%%%%%%%%%%%%%%%%%%%%%%%%%%%%%%
% table_HousingNetWorth.D07t09.PLMNJ.D99t05.2SLS
%%%%%%%%%%%%%%%%%%%%%%%%%%%%%%%%%%%%

\input{Table_HH/table_HousingNetWorth.D07t09.PLMNJ.D99t05.2SLS}

\pagebreak 
%---------------------------------------------------------------

%%%%%%%%%%%%%%%%%%%%%%%%%%%%%%%%%%%%%%%%%%%%%%%%
% table_HousingNetWorth.D07t09.PLMNJ.D99t05.4Reg
%%%%%%%%%%%%%%%%%%%%%%%%%%%%%%%%%%%%%%%%%%%%%%%%

\input{Table_HH/table_HousingNetWorth.D07t09.PLMNJ.D99t05.4Reg}

%---------------------------------------------------------------
%---------------------------------------------------------------
% Empirical: Main Tests 2.1 Real Economy
% The Impact of Household Leverage on House Price Cycle
%---------------------------------------------------------------
%---------------------------------------------------------------

\pagebreak
%-----------------------------------------------------------------

%%%%%%%%%%%%%%%%%%%%%%%%%%%%%%%%%%%%
% table_HPI.D99t05vsD07t09.HHDTI.2SLS.wide
%%%%%%%%%%%%%%%%%%%%%%%%%%%%%%%%%%%%

\input{Table_HH/table_HPI.D99t05vsD07t09.HHDTI.2SLS.wide}

\pagebreak 
%---------------------------------------------------------------

%%%%%%%%%%%%%%%%%%%%%%%%%%%%%%%%%%%%%%%%%%%%%%%%
% table_HPI.D99t05vsD07t09.HHDTI.4Reg
%%%%%%%%%%%%%%%%%%%%%%%%%%%%%%%%%%%%%%%%%%%%%%%%

\input{Table_HH/table_HPI.D99t05vsD07t09.HHDTI.4Reg}

%---------------------------------------------------------------
%---------------------------------------------------------------
% Empirical: Main Tests 2.2
% The Impact of Household Leverage on Residential Permit Cycle
%---------------------------------------------------------------
%---------------------------------------------------------------

\pagebreak
%-----------------------------------------------------------------

%%%%%%%%%%%%%%%%%%%%%%%%%%%%%%%%%%%%
% table_Permit.D99t05vsD05t09.HHDTI.2SLS.wide
%%%%%%%%%%%%%%%%%%%%%%%%%%%%%%%%%%%%

\input{Table_HH/table_Permit.D99t05vsD05t09.HHDTI.2SLS.wide}

\pagebreak 
%---------------------------------------------------------------

%%%%%%%%%%%%%%%%%%%%%%%%%%%%%%%%%%%%%%%%%%%%%%%%
% table_Permit.D99t05vsD05t09.HHDTI.4Reg
%%%%%%%%%%%%%%%%%%%%%%%%%%%%%%%%%%%%%%%%%%%%%%%%

\input{Table_HH/table_Permit.D99t05vsD05t09.HHDTI.4Reg}

%---------------------------------------------------------------
%---------------------------------------------------------------
% Empirical: Main Tests 2.3
% The Impact of Household Leverage on Refined House Employment
%---------------------------------------------------------------
%---------------------------------------------------------------

\pagebreak
%-----------------------------------------------------------------

%%%%%%%%%%%%%%%%%%%%%%%%%%%%%%%%%%%%
% table_RefineHouse.D00t06vsD07t10.HHDTI.2SLS.wide
%%%%%%%%%%%%%%%%%%%%%%%%%%%%%%%%%%%%

\input{Table_HH/table_RefineHouse.D00t06vsD07t10.HHDTI.2SLS.wide}

\pagebreak 
%---------------------------------------------------------------

%%%%%%%%%%%%%%%%%%%%%%%%%%%%%%%%%%%%%%%%%%%%%%%%
% table_RefineHouse.D00t06vsD07t10.HHDTI.4Reg
%%%%%%%%%%%%%%%%%%%%%%%%%%%%%%%%%%%%%%%%%%%%%%%%

\input{Table_HH/table_RefineHouse.D00t06vsD07t10.HHDTI.4Reg}

%---------------------------------------------------------------
%---------------------------------------------------------------
% Empirical: Main Tests 2.4
% The Impact of Household Leverage on BEA Construction Employment 
%---------------------------------------------------------------
%---------------------------------------------------------------

\pagebreak
%-----------------------------------------------------------------

%%%%%%%%%%%%%%%%%%%%%%%%%%%%%%%%%%%%
% table_BEAConstEmp.D00t06vsD07t10.HHDTI.2SLS.wide
%%%%%%%%%%%%%%%%%%%%%%%%%%%%%%%%%%%%

\input{Table_HH/table_BEAConstEmp.D00t06vsD07t10.HHDTI.2SLS.wide}

\pagebreak 
%---------------------------------------------------------------

%%%%%%%%%%%%%%%%%%%%%%%%%%%%%%%%%%%%%%%%%%%%%%%%
% table_BEAConstEmp.D00t06vsD07t10.HHDTI.4Reg
%%%%%%%%%%%%%%%%%%%%%%%%%%%%%%%%%%%%%%%%%%%%%%%%

\input{Table_HH/table_BEAConstEmp.D00t06vsD07t10.HHDTI.4Reg}

%%%%%%%%%%%%%%%%%%%%%%%%%%%%%%%%%%%%%%%%%%%%%%%%%%%%%%%%%%%%%%%%%%%%%%%%%%%
%%%%%%%%%%%%%%%%%%%%%%%%%%%%%%%%%%%%%%%%%%%%%%%%%%%%%%%%%%%%%%%%%%%%%%%%%%%
% Main Test 2. Address Alternative Channel: Corporate Channel

%%%%%%%%%%%%%%%%%%%%%%%%%%%%%%%%%%%%%%%%%%%%%%%%%%%%%%%%%%%%%%%%%%%%%%%%%%%
%%%%%%%%%%%%%%%%%%%%%%%%%%%%%%%%%%%%%%%%%%%%%%%%%%%%%%%%%%%%%%%%%%%%%%%%%%%

%---------------------------------------------------------------
%---------------------------------------------------------------
% Main Test 2. Address Alternative Channel: Corporate Channel
% % Main Test 2.1. 
% Tradable Sector Employment Growth in Boom (00-06) and Bust (07-10)
%---------------------------------------------------------------
%---------------------------------------------------------------

\pagebreak
%-----------------------------------------------------------------

%%%%%%%%%%%%%%%%%%%%%%%%%%%%%%%%%%%%
% table_Tradable.D00t06vsD07t10.NEP.D99t05.2SLS.wide
%%%%%%%%%%%%%%%%%%%%%%%%%%%%%%%%%%%%

\input{Table_HH/table_Tradable.D00t06vsD07t10.NEP.D99t05.2SLS.wide}

\pagebreak 
%---------------------------------------------------------------

%%%%%%%%%%%%%%%%%%%%%%%%%%%%%%%%%%%%%%%%%%%%%%%%
% table_Tradable.D00t06vsD07t10.NEP.D99t05.4Reg
%%%%%%%%%%%%%%%%%%%%%%%%%%%%%%%%%%%%%%%%%%%%%%%%

\input{Table_HH/table_Tradable.D00t06vsD07t10.NEP.D99t05.4Reg}

%---------------------------------------------------------------
%---------------------------------------------------------------
% Main Test 2. Address Alternative Channel: Corporate Channel
% % Main Test 2.2. 
% Commercial Construction Employment in Boom (00-06) and Bust (07-10)
%---------------------------------------------------------------
%---------------------------------------------------------------

\pagebreak
%-----------------------------------------------------------------

%%%%%%%%%%%%%%%%%%%%%%%%%%%%%%%%%%%%
% table_ComConstEmp.D00t06vsD07t10.NEP.D99t05.2SLS.wide
%%%%%%%%%%%%%%%%%%%%%%%%%%%%%%%%%%%%

\input{Table_HH/table_ComConstEmp.D00t06vsD07t10.NEP.D99t05.2SLS.wide}

\pagebreak 
%---------------------------------------------------------------

%%%%%%%%%%%%%%%%%%%%%%%%%%%%%%%%%%%%%%%%%%%%%%%%
% table_ComConstEmp.D00t06vsD07t10.NEP.D99t05.4Reg
%%%%%%%%%%%%%%%%%%%%%%%%%%%%%%%%%%%%%%%%%%%%%%%%

\input{Table_HH/table_ComConstEmp.D00t06vsD07t10.NEP.D99t05.4Reg}

%%%%%%%%%%%%%%%%%%%%%%%%%%%%%%%%%%%%%%%%%%%%%%%%%%%%%%%%%%%%%%%%%%%%%%%%%%%
%%%%%%%%%%%%%%%%%%%%%%%%%%%%%%%%%%%%%%%%%%%%%%%%%%%%%%%%%%%%%%%%%%%%%%%%%%%
% Main Test 3. Consumption Cycle
% 
% 
%%%%%%%%%%%%%%%%%%%%%%%%%%%%%%%%%%%%%%%%%%%%%%%%%%%%%%%%%%%%%%%%%%%%%%%%%%%
%%%%%%%%%%%%%%%%%%%%%%%%%%%%%%%%%%%%%%%%%%%%%%%%%%%%%%%%%%%%%%%%%%%%%%%%%%%

\pagebreak
%-----------------------------------------------------------------

%%%%%%%%%%%%%%%%%%%%%%%%%%%%%%%%%%%%
% 
%%%%%%%%%%%%%%%%%%%%%%%%%%%%%%%%%%%%

%\input{Table_HH/}

\pagebreak 
%---------------------------------------------------------------

%%%%%%%%%%%%%%%%%%%%%%%%%%%%%%%%%%%%%%%%%%%%%%%%
% 
%%%%%%%%%%%%%%%%%%%%%%%%%%%%%%%%%%%%%%%%%%%%%%%%

%\input{Table_HH/}

%################################################################################
%################################################################################
% This is the end of the entire section (tex file)
%################################################################################
%################################################################################

%% file: Figure_HH/fig_USMetroCty_NetExpGrowth.tex
%------------------------------------------------------------
% fig_USMetroCty_NetExpGrowth

%------------------------------------
\begin{figure}[h!] 
    \centering
    \includegraphics[width=16cm, height=12cm]{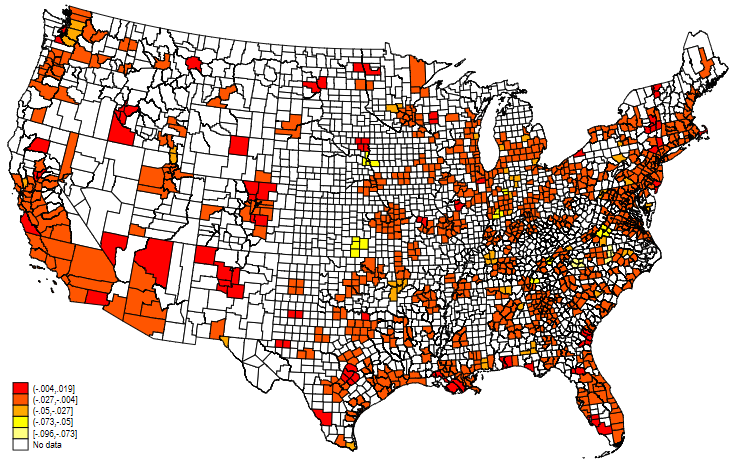}
    \caption{\textbf{U.S. Mainland Metropolitan Heat Map of Net Export Growth: 1999-2005}  \smallskip  \newline 
    {\footnotesize This figure displays the U.S. mainland metropolitan heat map of net export growth measure of period 19990-2005. The net export growth measure at the metropolitan level for a period is defined in equation (\ref{equ:NEG_m}). In the above figures, each small area with a boundary is a county. Counties with white color are non-metropolitan areas in the 2003 CBSA version (1085 counties in metropolitan areas in the U.S. mainland). Metropolitan counties are painted with colors ranging from yellow (for low net export growth) to red (for high net export growth) in five categories. 
        } %end of small font
    } % end of caption
    \label{fig_USMetroCty_NetExpGrowth}
    
\end{figure} 
%------------------------------------

%% file: Figure_HH/fig_HouseholdLeverage_99To14.tex
%------------------------------------------------------------
% figure 1: fig_HouseholdLeverage_99To14
%------------------------------------------------------------

\begin{figure}[h!] 
    \centering
    \includegraphics[width=16cm, height=12cm]{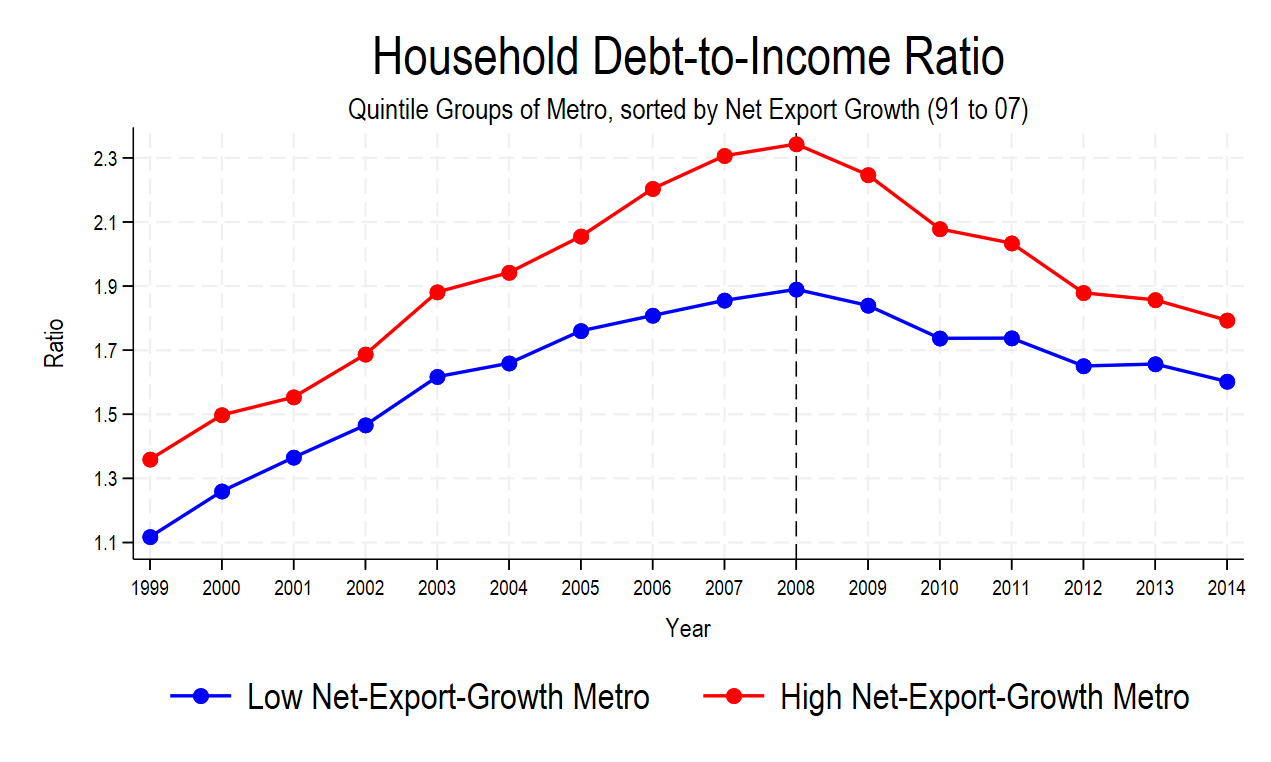}
    \caption{\textbf{Household Leverage (99-14) in Metro Areas, High vs. Low Quintile of Net Export Growth (91 to 07)} \smallskip \newline 
    {\footnotesize This figure displays the time series of weighted average household leverage (debt-to-income ratio) for high and low quintile groups of Metropolitan Areas (MSAs) from 1999 to 2014. For the entire period, the quintile groups are sorted by net export growth (1991 to 2007) at the metropolitan level (CBSA code 2003 version). The sample includes 319 metropolitan areas (800 counties) that are consistently covered by the HMDA sample after 1996. The household leverage data from the Board of Governors of the Federal Reserve System covers all 800 counties from 1999 to 2014. The red line represents the high quintile group, while the blue line represents the low quintile group. The low quintile group comprises 64 metros (103 counties), and the high quintile group comprises 63 metros (129 counties) throughout the entire period. The household leverage is weighted by the county-level number of households within each group in each year. Throughout the entire period, the time series of weighted-average household leverage of each group is divided by their 1991 values, ensuring that both groups start at a value of 1 in 1991.
    } %end of small font
    } % end of caption
    \label{fig_HouseholdLeverage_99To14}
    % note that \label is given after \caption.
\end{figure}

%% file: Figure_HH/fig_GSEMvsPLMNJ_91t11_combine.tex
%------------------------------------------------------------
% figure 2: fig_GSEMvsPLMNJ_91t11_combine

\begin{figure}[h!] 
    \centering
    \includegraphics[width=16cm, height=12cm]{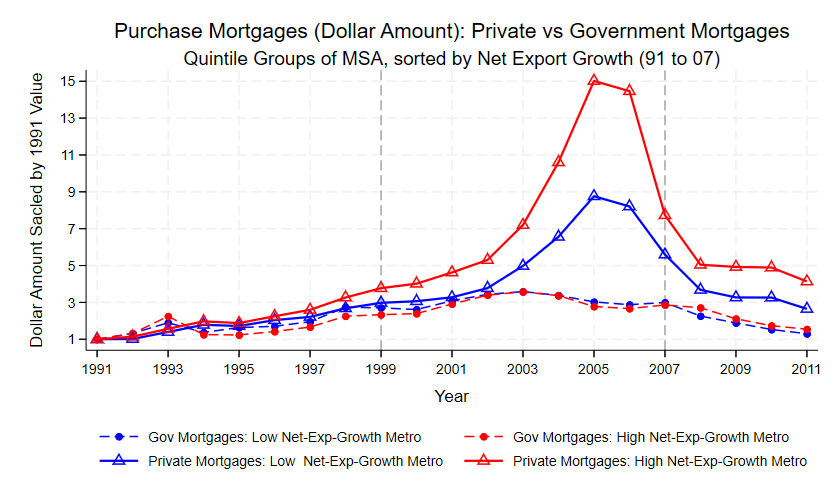}
    \caption{\textbf{Mortgage Growth (91-11) across Metropolitan Areas: GSEM vs. PLMNJ, High vs. Low Quintile Sorted by Net Export Growth (91-07).}  \smallskip \newline 
    {\footnotesize This figure displays the time series of the weighted-average dollar amount of Government-Sponsored Enterprise Mortgages (GSEM) (in dash lines with dots) and Private-Label Mortgages (Non-Jumbo) (PLMNJ) (in solid lines with triangles) for high and low quintile groups of metropolitan statistical areas (MSA) from 1991 to 2011. Both types of mortgages only include purchase loans. For the entire period, the quintile groups are sorted by net export growth (1991 to 2007) at the MSA level. The whole sample includes 301 MSA (712 counties) that are consistently covered by the HMDA sample after 1990 due to the smaller coverage of metropolitan areas in the early years. The low quintile group comprises 61 MSA (94 counties), and the high quintile group comprises 60 MSA (112 counties) throughout the entire period. The number of loans is weighted by the county-level housing units within each group in each year. Throughout the entire period, the time series of the weighted-average number of loans of each group are divided by their 1991 values, ensuring that both groups start at a value of 1 in 1991. The red lines represent the high quintile group, while the blue lines represent the low quintile group.
        } %end of small font
    } % end of caption
    \label{fig_GSEMvsPLMNJ_91t11_combine}
    % note that \label is given after \caption.
\end{figure}

%% file: Figure_HH/fig_ReducedForm_HousingNetWorth_on_GIV.tex
%------------------------------------------------------------
% fig_ReducedForm_HousingNetWorth_on_GIV

%------------------------------------
\begin{figure}[h!] 
    \centering
    \begin{subfigure}[t]{0.9\textwidth}
        \centering
        \includegraphics[height=7.5cm]{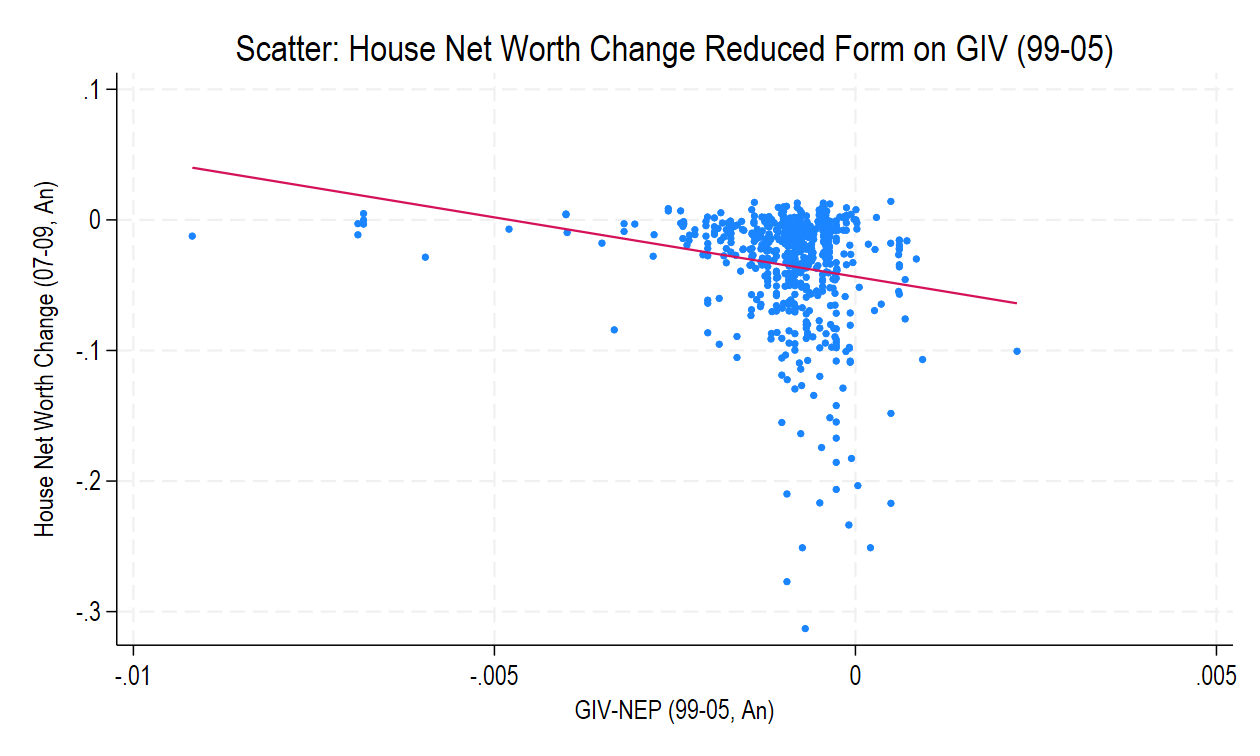}
        \caption{Scatter Plot}
    \end{subfigure}%
    \hfill 
    \begin{subfigure}[t]{0.9\textwidth}
        \centering
        \includegraphics[height=7.5cm]{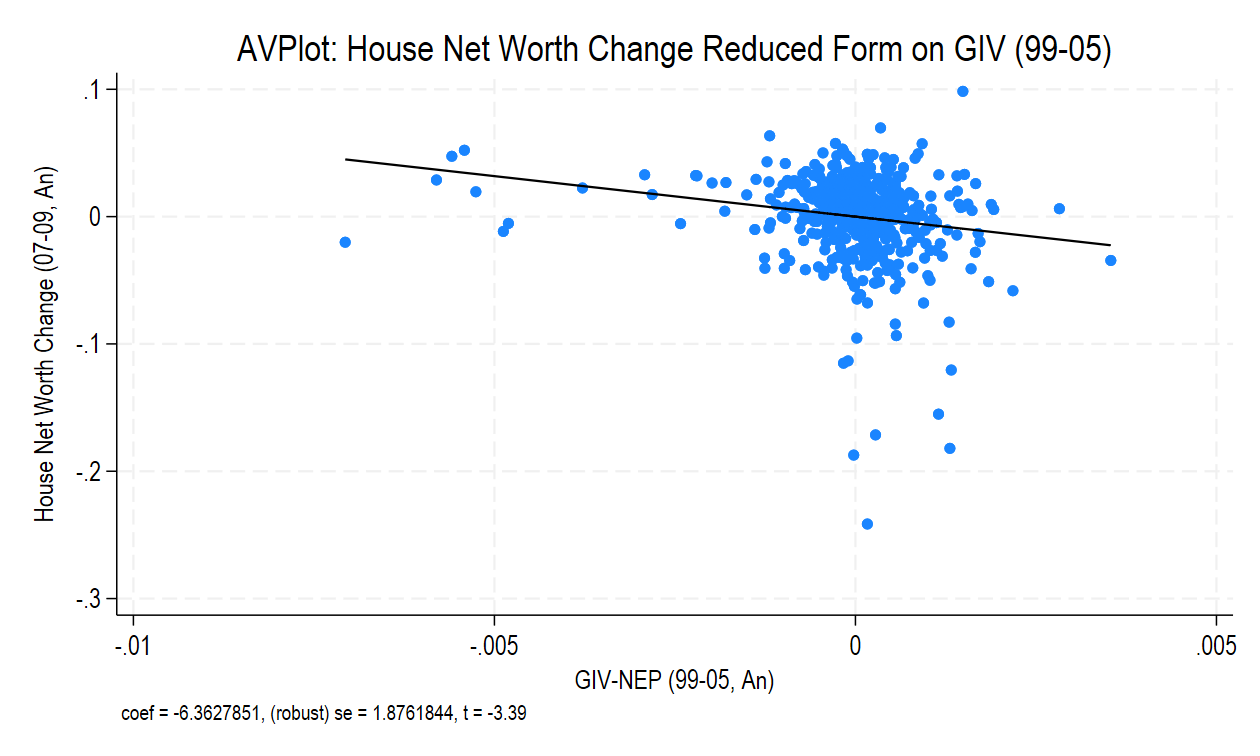}
        \caption{Added-Variable Plot}
    \end{subfigure}
    \caption{\textbf{Plots of Reduced-Form Regression of Housing Net Worth Change (07-09)}  \smallskip  \newline 
    {\footnotesize This figure displays the scatter plot (subfigure a) and added-variable plot (subfigure b) of reduced-form regression of housing net worth change (07-09) \citep{mian2013household} on gravity model-based instrumental variable of net export growth (99-05).
        } %end of small font
    } % end of caption
    \label{fig_ReducedForm_HousingNetWorth_on_GIV}
    
\end{figure} 
%------------------------------------

%% file: Figure_HH/fig_PLMNJ_vs_HousePriceIndex.tex
%------------------------------------------------------------
% fig_PLMNJ_vs_HousePriceIndex

\begin{figure}[h!] 
    \centering
    \includegraphics[width=16cm, height=12cm]{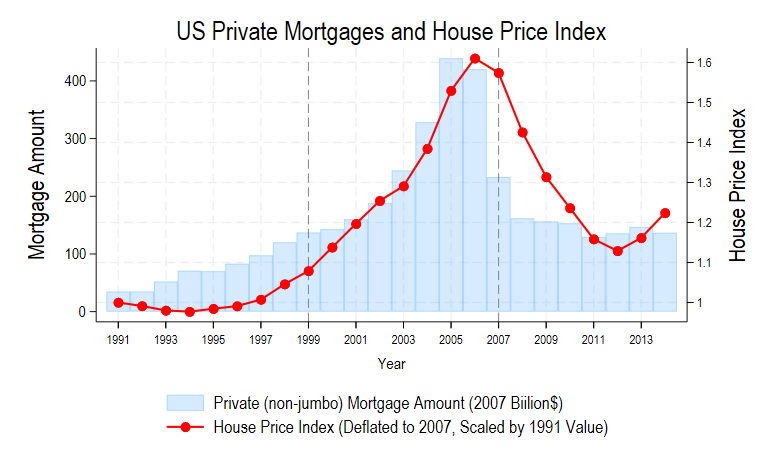}
    \caption{\textbf{Credit Expansion vs. House Price Index in USA (1991-2014).}  \smallskip \newline 
    {\footnotesize This figure displays the time series of dollar amount of private-label mortgages (non-jumbo) (in blue bars) and house price index (in red line) in USA from 1991 to 2014. The whole sample includes 301 MSA (679 counties) that are consistently covered by the HMDA sample after 1990 and by the house price index from Federal Housing Financing Agency (FHFA). The dollar amount of mortgages in billions is deflated to the 2007 USD by the Personal Consumption Expenditures Chain-type Price Index (PCEPI) from Federal Reserve Bank of St. Louis. The housing price index is first deflated by PCEPI to 2007 and further scaled by its 1991 value.
        } %end of small font
    } % end of caption
    \label{fig_PLMNJ_vs_HousePriceIndex}
    % note that \label is given after \caption.
\end{figure}

%% file: Figure_HH/fig_DebtToGDPRatio_HHBusiGov_HHSubCategories.tex
%------------------------------------------------------------
% fig_DebtToGDPRatio_HHBusiGov_HHSubCategories

%------------------------------------
\begin{figure}[h!] 
    \centering
    \begin{subfigure}[t]{0.9\textwidth}
        \centering
        \includegraphics[height=7.5cm]{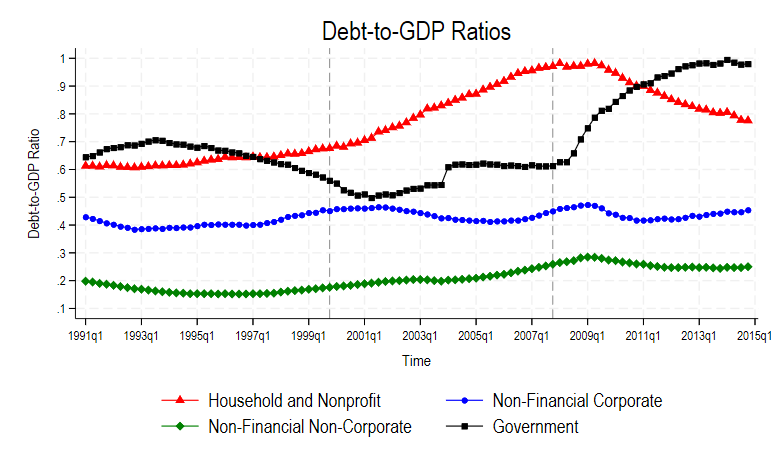}
        \caption{Households, Business, and Government}
    \end{subfigure}%
    \hfill 
    \begin{subfigure}[t]{0.9\textwidth}
        \centering
        \includegraphics[height=7.5cm]{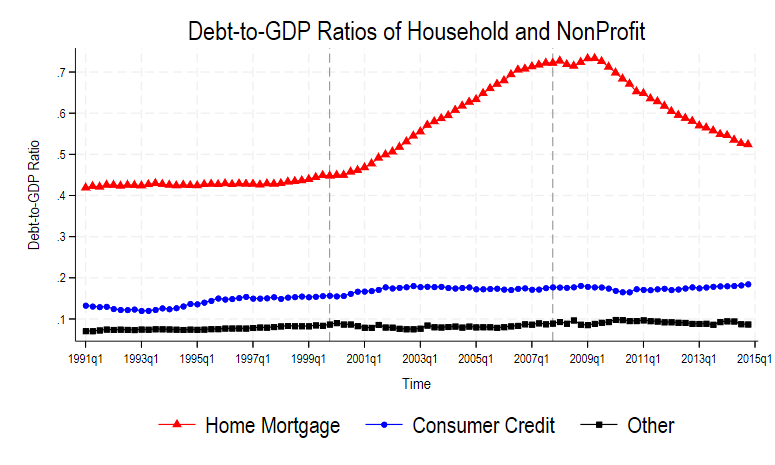}
        \caption{Households: Mortgages, Consumer Credit, and Other Liability}
    \end{subfigure}
    \caption{\textbf{Debt-to-GDP Ratios: Households Subcategories, Business, and Government}  \smallskip  \newline 
    {\footnotesize This figure displays the time series of debt-to-GDP ratios of households and nonprofit organizations, business organizations, and government in the USA from 1991 to 2014. Subfigure (a) includes debt-to-GDP ratios of three groups: households and nonprofit organizations, business organizations, and government. Subfigure (b) includes debt-to-GDP ratios of three sub-categories within the households and nonprofit organizations: mortgages, consumer credit and other liabilities. 
        } %end of small font
    } % end of caption
    \label{fig_DebtToGDPRatio_HHBusiGov_HHSubCategories}
    
\end{figure} 
%------------------------------------

%% file: Table_HH/table_EmploymentIndustryClassification.tex
%---------------------------------------------------------------

%%%%%%%%%%%%%%%%%%%%%%%%%%%%%%%%%%%%%%%%%%%%%%%%
% table_EmploymentIndustryClassification
%%%%%%%%%%%%%%%%%%%%%%%%%%%%%%%%%%%%%%%%%%%%%%%%

\noindent 

\begin{table}[h!]
\centering
\caption{
\textbf{Employment Industry Classification} \smallskip \newline
{\scriptsize
This table reports industry classification for employment data.
} % end of small font size
} % end of caption
\label{table_EmploymentIndustryClassification}
\resizebox{\columnwidth}{!}{%
\begin{tabular}{lllll}
\toprule 
\addlinespace
\multicolumn{5}{l}{\begin{tabular}[c]{@{}l@{}}\textbf{Industry Classification by \cite{goukasian2010reaction} in County} \\ \textbf{Business Pattern Data}\end{tabular}}                                                                                                                            \\
\textbf{1987 SIC Code}     & \multicolumn{4}{c}{\textbf{Industry Description}}                                                                                                                                                          \\
\midrule
\addlinespace

\multicolumn{1}{c}{}                                          & \multicolumn{4}{c}{Residential Construction}                                                                                                                                                      \\
1521                                                          & \multicolumn{4}{l}{General contractors—single-family houses}                                                                                                                                      \\
1522                                                          & \multicolumn{4}{l}{General contractors—residential buildings, other than single-family}                                                                                                           \\
1531                                                          & \multicolumn{4}{l}{Operative builders}                                                                                                                                                            \\
6552                                                          & \multicolumn{4}{l}{Land subdividers and developers, except cemeteries}                                                                                                                            \\
                                                              & \multicolumn{4}{c}{Other Employment}                                                                                                                                                              \\
1741                                                          & \multicolumn{4}{l}{Masonry, stone setting and other stone work}                                                                                                                                   \\
1771                                                          & \multicolumn{4}{l}{Concrete work}                                                                                                                                                                 \\
1791                                                          & \multicolumn{4}{l}{Structural steel erection}                                                                                                                                                     \\
1742                                                          & \multicolumn{4}{l}{Plastering, drywall, acoustical, and insulation work}                                                                                                                          \\
1761                                                          & \multicolumn{4}{l}{Roofing, siding, and sheet metal work}                                                                                                                                         \\
1731                                                          & \multicolumn{4}{l}{Electrical work}                                                                                                                                                               \\
                                                              & \multicolumn{4}{c}{Mortgage Employment}                                                                                                                                                           \\
6162                                                          & \multicolumn{4}{l}{Mortgage bankers and loan correspondents}                                                                                                                                      \\
\bottomrule
\addlinespace
\multicolumn{5}{l}{\begin{tabular}[c]{@{}l@{}}\textbf{Goukasian and Majbouri (2010) House Employment} includes all industries in Table 1 \\ in  \cite{goukasian2010reaction}\end{tabular}} \\
\multicolumn{5}{l}{\begin{tabular}[c]{@{}l@{}}Our \textbf{Refined House Employment} includes Residential Construction, Other Employment, \\ and Mortgage Employment\end{tabular}}                          \\
\addlinespace
                                                                                              
\end{tabular}

} % end of resize box

\end{table}

%% file: Table_HH/table_HHDTI.D99t05vsD08t14.PLMNJ.2SLS.wide.tex
%-----------------------------------------------------------------------
%%%%%%%%%%%%%%%%%%%%%%%%%%%%%%%%%%%%
% table_HHDTI.D99t05vsD08t14.PLMNJ.2SLS.wide
%%%%%%%%%%%%%%%%%%%%%%%%%%%%%%%%%%%%

\noindent 

\begin{table}[h!]
\centering
\caption{
\textbf{2SLS Stacked Regression of Household Leverage Increase in Boom (99-05) and Bust (08-14) Periods on PLMNJ Growth (99-05)} \smallskip \newline
{\scriptsize
This table reports 2SLS regression $\triangle_{99,05} \& \triangle_{08,14} HHDTI_{c} = \beta_{99,05} * \triangle_{99,05} Ln(PLMNJ_{c}) \times Dum_{99,05} + \beta_{08,14} * \triangle_{99,05} Ln(PLMNJ_{c}) \times Dum_{08,14} + \gamma_{99,05}* \bm{Controls_{c}} \times Dum_{99,05} + \gamma_{08,14}* \bm{Controls_{c}} \times Dum_{08,14} + \epsilon_{period, c}$. The left-hand-side dependent variable $\triangle_{99,05} \& \triangle_{08,14} HHDTI_{c}$ is the stacked increase in household leverage (debt-to-income ratio) at county $c$ 99-05 and 08-14. The key independent variable $\triangle_{99,05} Ln(PLMNJ_{c})$ is the growth rate of the dollar amount (deflated to 2007 USD) of private-label mortgages (non-jumbo) at county $c$ 99-05. $Controls_{c}$ indicates control variables at county $c$ in the period start year 1999. We use the gravity model-based instrumental variable $\triangle_{99,05}\text{givNetExp}_{m}$ as the IV for $\triangle_{99,05}Ln(PLMNJ_{c})$. Regression is weighted by the natural logarithm of housing units in 1999.  For the first-stage F-test of two non-stacked samples, we report Kleibergen-Paap (2006) robust (clustered) statistics and Montiel Olea-Pflueger (2013) efficient statistics. Standard errors are clustered at the CBSA level. ***, **, and * indicate significance at the 1\%, 5\%, and 10\% levels, respectively.
\smallskip
} % end of small font size
} % end of caption
\label{table_HHDTI.D99t05vsD08t14.PLMNJ.2SLS.wide}

\vspace{-2mm}

\resizebox{\columnwidth}{!}{%
\begin{tabular}{l*{10}{c}}
\toprule
\textbf{TSLS estimates}            &\multicolumn{10}{c}{Household Leverage Increase (99-05 and 08-14, Annualized)} \\
            \cmidrule{2-11} 
            &\multicolumn{2}{c}{(1)}&\multicolumn{2}{c}{(2)}&\multicolumn{2}{c}{(3)}&\multicolumn{2}{c}{(4)}&\multicolumn{2}{c}{(5)}\\
            
\midrule
PLMNJ Growth (99-05, An) x Dum99t05&    0.779***&  (0.246)&    0.796***&  (0.258)&    0.793** &  (0.312)&    0.912***&  (0.305)&    0.940***&  (0.342)\\ 
\addlinespace
PLMNJ Growth (99-05, An) x Dum08t14&   -1.099***&  (0.289)&   -1.071***&  (0.323)&   -1.174***&  (0.410)&   -1.216***&  (0.401)&   -1.302***&  (0.433)\\ \addlinespace
Dum99t05       &    0.013   &  (0.041)&   -0.919***&  (0.258)&   -0.418   &  (0.338)&   -0.334   &  (0.566)&   -0.311   &  (0.549)\\ \addlinespace
Dum08t14       &    0.100** &  (0.048)&    0.913***&  (0.275)&    0.762** &  (0.371)&    1.203** &  (0.571)&    1.146*  &  (0.607)\\ \addlinespace
Ln(Num of HH, 99) x Dum99t05&            &         &   -0.038***&  (0.008)&    0.034   &  (0.062)&   -0.031   &  (0.056)&   -0.035   &  (0.057)\\ \addlinespace
Ln(Num of HH, 99) x Dum08t14&            &         &    0.015** &  (0.006)&    0.030   &  (0.067)&    0.101   &  (0.067)&    0.082   &  (0.068)\\ \addlinespace
Ln(HH Income, 99) x Dum99t05&            &         &    0.119***&  (0.031)&    0.067*  &  (0.040)&    0.065   &  (0.062)&    0.064   &  (0.060)\\ \addlinespace
Ln(HH Income, 99) x Dum08t14&            &         &   -0.100***&  (0.029)&   -0.077** &  (0.039)&   -0.127** &  (0.055)&   -0.123** &  (0.057)\\ \addlinespace
Ratio of Labor Force (1999) x Dum99t05&            &         &    0.097   &  (0.122)&    0.179   &  (0.192)&    0.309*  &  (0.167)&    0.290*  &  (0.160)\\ \addlinespace
Ratio of Labor Force (1999) x Dum08t14&            &         &    0.149   &  (0.119)&    0.055   &  (0.149)&   -0.100   &  (0.148)&   -0.058   &  (0.149)\\ \addlinespace
Ln(Num of HU, 99) x Dum99t05&            &         &            &         &   -0.064   &  (0.067)&   -0.006   &  (0.061)&   -0.006   &  (0.062)\\ \addlinespace
Ln(Num of HU, 99) x Dum08t14&            &         &            &         &   -0.019   &  (0.067)&   -0.085   &  (0.067)&   -0.065   &  (0.069)\\ \addlinespace
Housing supply elasticity x Dum99t05&            &         &            &         &   -0.007   &  (0.006)&   -0.004   &  (0.006)&   -0.005   &  (0.006)\\ \addlinespace
Housing supply elasticity x Dum08t14&            &         &            &         &   -0.004   &  (0.007)&   -0.006   &  (0.007)&   -0.007   &  (0.007)\\ \addlinespace
House Vacancy Rate (1999) x Dum99t05&            &         &            &         &    0.140   &  (0.195)&    0.055   &  (0.190)&    0.125   &  (0.190)\\ \addlinespace
House Vacancy Rate (1999) x Dum08t14&            &         &            &         &    0.021   &  (0.184)&    0.082   &  (0.188)&    0.009   &  (0.174)\\ \addlinespace
Ratio of Renters (1999) x Dum99t05&            &         &            &         &   -0.260** &  (0.128)&   -0.438***&  (0.118)&   -0.358***&  (0.109)\\ \addlinespace
Ratio of Renters (1999) x Dum08t14&            &         &            &         &    0.132   &  (0.109)&    0.266***&  (0.098)&    0.137   &  (0.109)\\ \addlinespace
Ratio of Bachelor Educated (1999) x Dum99t05&            &         &            &         &            &         &   -0.120   &  (0.141)&   -0.147   &  (0.180)\\ \addlinespace
Ratio of Bachelor Educated (1999) x Dum08t14&            &         &            &         &            &         &    0.261*  &  (0.157)&    0.166   &  (0.187)\\ \addlinespace
Ratio of White Race (1999) x Dum99t05&            &         &            &         &            &         &   -0.043   &  (0.074)&   -0.023   &  (0.074)\\ \addlinespace
Ratio of White Race (1999) x Dum08t14&            &         &            &         &            &         &    0.087   &  (0.055)&    0.044   &  (0.055)\\ \addlinespace
Ratio of Immigration (90-00) x Dum99t05&            &         &            &         &            &         &    1.199***&  (0.287)&    1.133***&  (0.305)\\ \addlinespace
Ratio of Immigration (90-00) x Dum08t14&            &         &            &         &            &         &   -0.854***&  (0.301)&   -0.608*  &  (0.319)\\ \addlinespace
Ratio of Art, Enter, and Recre Emp (1999) x Dum99t05&            &         &            &         &            &         &            &         &   -2.183** &  (0.909)\\ \addlinespace
Ratio of Art, Enter, and Recre Emp (1999) x Dum08t14&            &         &            &         &            &         &            &         &    2.909** &  (1.369)\\ \addlinespace
Ratio of Health Emp (1999) x Dum99t05&            &         &            &         &            &         &            &         &   -0.127   &  (0.687)\\ \addlinespace
Ratio of Health Emp (1999) x Dum08t14&            &         &            &         &            &         &            &         &    1.347*  &  (0.689)\\ \addlinespace
Ratio of Tradable Service Emp (1999) x Dum99t05&            &         &            &         &            &         &            &         &    0.409   &  (0.449)\\ \addlinespace
Ratio of Tradable Service Emp (1999) x Dum08t14&            &         &            &         &            &         &            &         &   -0.171   &  (0.481)\\ \addlinespace
Ratio of College Students (1999) x Dum99t05&            &         &            &         &            &         &            &         &   -0.423** &  (0.213)\\ \addlinespace
Ratio of College Students (1999) x Dum08t14&            &         &            &         &            &         &            &         &    0.236   &  (0.250)\\ \addlinespace
\midrule
Obs            &     1584   &         &     1584   &         &     1402   &         &     1402   &         &     1402   &         \\
Cluster SE     &     CBSA   &         &     CBSA   &         &     CBSA   &         &     CBSA   &         &     CBSA   &         \\ 
Weight         & Ln(HU99)   &         & Ln(HU99)   &         & Ln(HU99)   &         & Ln(HU99)   &         & Ln(HU99)   &         \\ 
KP F-Stat (99-05, non-stack sample)       &    23.43   &         &    21.04   &         &    13.97   &         &    11.74   &         &    11.43   &         \\
MOP F-Stat (99-05, non-stack sample)  &    22.30   &         &    20.29   &         &    14.03   &         &    11.80   &         &    11.37   &         \\
CoefEqual\_Chi2 &   15.947   &         &   12.191   &         &    9.512   &         &   11.555   &         &   10.476   &         \\
CoefEqual\_PValue &    0.000   &         &    0.000   &         &    0.002   &         &    0.001   &         &    0.001   &         \\
\bottomrule

\end{tabular}

} % end of resize box

\end{table}

%% file: Table_HH/table_HHDTI.D99t05vsD08t14.PLMNJ.4Reg.tex
%---------------------------------------------------------------

%%%%%%%%%%%%%%%%%%%%%%%%%%%%%%%%%%%%%%%%%%%%%%%%
% table_HHDTI.D99t05vsD08t14.PLMNJ.4Reg
%%%%%%%%%%%%%%%%%%%%%%%%%%%%%%%%%%%%%%%%%%%%%%%%

\noindent 

\begin{table}[h!]
\centering
\caption{
\textbf{Four Stacked Regressions of Household Leverage Increase in Boom (99-05) and Bust (08-14) Periods on PLMNJ Growth (99-05)} \smallskip \newline
{\scriptsize
This table reports OLS, reduced-form, first stage, and second stages of stacked 2SLS regression $\triangle_{99,05} \& \triangle_{08,14} HHDTI_{c} = \beta_{99,05} * \triangle_{99,05} Ln(PLMNJ_{c}) \times Dum_{99,05} + \beta_{08,14} * \triangle_{99,05} Ln(PLMNJ_{c}) \times Dum_{08,14} + \gamma_{99,05}* \bm{Controls_{c}} \times Dum_{99,05} + \gamma_{08,14}* \bm{Controls_{c}} \times Dum_{08,14} + \epsilon_{period, c}$. The left-hand-side dependent variable $\triangle_{99,05} \& \triangle_{08,14} HHDTI_{c}$ is the stacked increase in household leverage (debt-to-income ratio) at county $c$ 99-05 and 08-14. The key independent variable $\triangle_{99,05} Ln(PLMNJ_{c})$ is the growth rate of the dollar amount (deflated to 2007 USD) of private-label mortgages (non-jumbo) at county $c$ 99-05. $Controls_{c}$ indicates control variables at county $c$ in the period start year 1999. We use the gravity model-based instrumental variable $\triangle_{99,05}\text{givNetExp}_{m}$ as the IV for $\triangle_{99,05}Ln(PLMNJ_{c})$. Regression is weighted by the natural logarithm of housing units in 1999.  For the first-stage F-test of two non-stacked samples, we report Kleibergen-Paap (2006) robust (clustered) statistics and Montiel Olea-Pflueger (2013) efficient statistics. Standard errors are clustered at the CBSA level. ***, **, and * indicate significance at the 1\%, 5\%, and 10\% levels, respectively.
} % end of small font size
} % end of caption
\label{table_HHDTI.D99t05vsD08t14.PLMNJ.4Reg}
\resizebox{\columnwidth}{!}{%
\begin{tabular}{l*{5}{c}}
\toprule
Dep Var (Panel A, B, and C)                      &\multicolumn{5}{c}{Household Leverage Increase (99-05 \& 08-14, annualized)} \\
            \cmidrule{2-6} 
            &\multicolumn{1}{c}{(1)}&\multicolumn{1}{c}{(2)}&\multicolumn{1}{c}{(3)}&\multicolumn{1}{c}{(4)}&\multicolumn{1}{c}{(5)}\\

\midrule
\multicolumn{6}{l}{\textbf{Panel A. OLS estimates}} \\
PLMNJ Growth (99-05, An) x Dum99t05&    0.213***&    0.253***&    0.130*  &    0.113*  &    0.123*  \\
               &  (0.056)   &  (0.055)   &  (0.066)   &  (0.063)   &  (0.064)   \\
\addlinespace
PLMNJ Growth (99-05, An) x Dum08t14&   -0.288***&   -0.288***&   -0.225***&   -0.204***&   -0.213***\\
               &  (0.056)   &  (0.055)   &  (0.064)   &  (0.059)   &  (0.058)   \\
\addlinespace
R2-adj         &    0.540   &    0.577   &    0.614   &    0.640   &    0.641   \\
\addlinespace

\midrule
\multicolumn{6}{l}{\textbf{Panel B. Reduced-form estimates}} \\
GIV Net Export Growth (99-05, An) x Dum99t05&   12.270***&   12.156***&    9.709***&   11.070***&   11.015***\\
               &  (3.663)   &  (3.292)   &  (3.454)   &  (3.425)   &  (3.622)   \\
\addlinespace
GIV Net Export Growth (99-05, An) x Dum08t14&  -17.318***&  -16.356***&  -14.377***&  -14.752***&  -15.258***\\
               &  (2.708)   &  (2.919)   &  (2.693)   &  (2.594)   &  (2.662)   \\
\addlinespace
R2-adj         &    0.534   &    0.568   &    0.613   &    0.642   &    0.642   \\
\addlinespace

\midrule
\multicolumn{6}{l}{\textbf{Panel C . 2SLS estimates}} \\
\addlinespace
PLMNJ Growth (99-05, An) x Dum99t05&    0.779***&    0.796***&    0.793** &    0.912***&    0.940***\\
               &  (0.246)   &  (0.258)   &  (0.312)   &  (0.305)   &  (0.342)   \\
\addlinespace
PLMNJ Growth (99-05, An) x Dum08t14&   -1.099***&   -1.071***&   -1.174***&   -1.216***&   -1.302***\\
               &  (0.289)   &  (0.323)   &  (0.410)   &  (0.401)   &  (0.433)   \\
               
\addlinespace
\addlinespace

Dep Var (Panel D): &\multicolumn{5}{c}{PLMNJ Growth (99-05, An)} \\ 
\midrule 
\multicolumn{6}{l}{\textbf{Panel D . First-stage estimates only for 99-05 (Non-stack sample)}} \\
\addlinespace
GIV Net Export Growth (99-05, An)&   15.753***&   15.278***&   12.243***&   12.134***&   11.716***\\
               &  (3.255)   &  (3.331)   &  (3.276)   &  (3.541)   &  (3.465)   \\
\addlinespace
KP F-Stat &    23.43   &    21.04   &    13.97   &    11.74   &    11.43   \\
MOP F-Stat     &    22.30   &    20.29   &    14.03   &    11.80   &    11.37   \\
\addlinespace

\midrule
\multicolumn{6}{l}{\textbf{Controls (for all Panels)}} \\
DumPeriod  &    Y        &  Y   &   Y    & Y    &  Y    \\
Basic Controls x DumPeriod &            &  Y   &   Y    & Y    &  Y    \\
Housing Controls x DumPeriod &           &      & Y       & Y    &  Y    \\
Demographic Controls x DumPeriod &            &      &        &  Y    &  Y    \\
Industry Controls x DumPeriod &            &      &        &      &  Y    \\

\midrule              
Obs (Panel A, B, and C)          &     1584   &     1584   &     1402   &     1402   &     1402   \\
Obs (Panel D)            &      792   &      792   &      701   &      701   &      701   \\
Cluster SE     &     CBSA   &     CBSA   &     CBSA   &     CBSA   &     CBSA   \\
Weight         & Ln(HU99)   & Ln(HU99)   & Ln(HU99)   & Ln(HU99)   & Ln(HU99)   \\
\bottomrule
\end{tabular}

} % end of resize box

\end{table}

%% file: Table_HH/table_HHDTI.D99t05vsD08t14.GSEM.2SLS.wide.tex
%-----------------------------------------------------------------------
%%%%%%%%%%%%%%%%%%%%%%%%%%%%%%%%%%%%
% table_HHDTI.D99t05vsD08t14.GSEM.2SLS.wide
%%%%%%%%%%%%%%%%%%%%%%%%%%%%%%%%%%%%

\noindent 

\begin{table}[h!]
\centering
\caption{
\textbf{2SLS Stacked Regression of Household Leverage in Boom (99-05) and Bust (08-14) Periods on GSEM Growth (99-05)} \smallskip \newline
{\scriptsize
This table reports 2SLS regression $\triangle_{99,05} \& \triangle_{08,14} HHDTI_{c} = \beta_{99,05} * \triangle_{99,05} Ln(GSEM_{c}) \times Dum_{99,05} + \beta_{08,14} * \triangle_{99,05} Ln(GSEM_{c}) \times Dum_{08,14} + \gamma_{99,05}* \bm{Controls_{c}} \times Dum_{99,05} + \gamma_{08,14}* \bm{Controls_{c}} \times Dum_{08,14} + \epsilon_{period, c}$. The left-hand-side dependent variable $\triangle_{99,05} \& \triangle_{08,14} HHDTI_{c}$ is the stacked increase in household leverage (debt-to-income ratio) at county $c$ 99-05 and 08-14. The key independent variable $\triangle_{99,05} Ln(GSEM_{c})$ is the growth rate of the dollar amount (deflated to 2007 USD) of government-sponsored enterprise mortgages (non-jumbo) at county $c$ 99-05. $Controls_{c}$ indicates control variables at county $c$ in the period start year 1999. We use the gravity model-based instrumental variable $\triangle_{99,05}\text{givNetExp}_{m}$ as the IV for $\triangle_{99,05}Ln(GSEM_{c})$. Regression is weighted by the natural logarithm of housing units in 1999.  For the first-stage F-test of two non-stacked samples, we report Kleibergen-Paap (2006) robust (clustered) statistics and Montiel Olea-Pflueger (2013) efficient statistics. Standard errors are clustered at the CBSA level. ***, **, and * indicate significance at the 1\%, 5\%, and 10\% levels, respectively.
\smallskip
} % end of small font size
} % end of caption
\label{table_HHDTI.D99t05vsD08t14.GSEM.2SLS.wide}

\vspace{-2mm}

\resizebox{\columnwidth}{!}{%
\begin{tabular}{l*{10}{c}}
\toprule
\textbf{TSLS estimates}            &\multicolumn{10}{c}{Household Leverage Increase (99-05 and 08-14, Annualized)} \\
            \cmidrule{2-11} 
            &\multicolumn{2}{c}{(1)}&\multicolumn{2}{c}{(2)}&\multicolumn{2}{c}{(3)}&\multicolumn{2}{c}{(4)}&\multicolumn{2}{c}{(5)}\\
            
\midrule
GSEM Growth (99-05, An) x Dum99t05&    5.646   &  (8.229)&    2.299*  &  (1.300)&    2.878   &  (2.509)&    3.143   &  (2.380)&    4.687   &  (5.481)\\ 
\addlinespace
GSEM Growth (99-05, An) x Dum08t14&   -7.968   & (11.439)&   -3.093*  &  (1.730)&   -4.262   &  (3.724)&   -4.189   &  (3.283)&   -6.493   &  (7.787)\\  \addlinespace
Dum99t05       &   -0.089   &  (0.336)&   -3.245** &  (1.478)&   -2.588   &  (2.242)&   -4.467   &  (3.391)&   -6.771   &  (7.831)\\  \addlinespace
Dum08t14       &    0.244   &  (0.466)&    4.042** &  (1.935)&    3.976   &  (3.314)&    6.711   &  (4.564)&   10.095   & (10.929)\\  \addlinespace
Ln(Num of HH, 99) x Dum99t05&            &         &    0.002   &  (0.021)&    0.021   &  (0.113)&   -0.198   &  (0.169)&   -0.263   &  (0.294)\\  \addlinespace
Ln(Num of HH, 99) x Dum08t14&            &         &   -0.039   &  (0.027)&    0.050   &  (0.188)&    0.323   &  (0.251)&    0.398   &  (0.432)\\  \addlinespace
Ln(HH Income, 99) x Dum99t05&            &         &    0.270***&  (0.105)&    0.187   &  (0.148)&    0.381   &  (0.275)&    0.559   &  (0.610)\\  \addlinespace
Ln(HH Income, 99) x Dum08t14&            &         &   -0.303** &  (0.135)&   -0.254   &  (0.219)&   -0.548   &  (0.366)&   -0.809   &  (0.849)\\  \addlinespace
Ratio of Labor Force (1999) x Dum99t05&            &         &    0.572   &  (0.350)&    0.901   &  (0.628)&    1.454*  &  (0.820)&    2.011   &  (1.946)\\  \addlinespace
Ratio of Labor Force (1999) x Dum08t14&            &         &   -0.490   &  (0.485)&   -1.014   &  (0.945)&   -1.626   &  (1.141)&   -2.442   &  (2.780)\\  \addlinespace
Ln(Num of HU, 99) x Dum99t05&            &         &            &         &   -0.005   &  (0.123)&    0.205   &  (0.190)&    0.289   &  (0.353)\\  \addlinespace
Ln(Num of HU, 99) x Dum08t14&            &         &            &         &   -0.107   &  (0.202)&   -0.367   &  (0.277)&   -0.474   &  (0.513)\\  \addlinespace
Housing supply elasticity x Dum99t05&            &         &            &         &   -0.023** &  (0.011)&   -0.020*  &  (0.010)&   -0.022   &  (0.016)\\  \addlinespace
Housing supply elasticity x Dum08t14&            &         &            &         &    0.019   &  (0.016)&    0.016   &  (0.014)&    0.017   &  (0.023)\\  \addlinespace
House Vacancy Rate (1999) x Dum99t05&            &         &            &         &   -0.339   &  (0.654)&   -0.552   &  (0.701)&   -0.865   &  (1.396)\\  \addlinespace
House Vacancy Rate (1999) x Dum08t14&            &         &            &         &    0.730   &  (0.977)&    0.890   &  (0.987)&    1.380   &  (2.021)\\  \addlinespace
Ratio of Renters (1999) x Dum99t05&            &         &            &         &   -0.254   &  (0.178)&   -0.629***&  (0.191)&   -0.409   &  (0.295)\\  \addlinespace
Ratio of Renters (1999) x Dum08t14&            &         &            &         &    0.123   &  (0.291)&    0.521*  &  (0.276)&    0.208   &  (0.444)\\  \addlinespace
Ratio of Bachelor Educated (1999) x Dum99t05&            &         &            &         &            &         &   -0.880   &  (0.582)&   -0.623   &  (0.601)\\  \addlinespace
Ratio of Bachelor Educated (1999) x Dum08t14&            &         &            &         &            &         &    1.273*  &  (0.773)&    0.825   &  (0.829)\\  \addlinespace
Ratio of White Race (1999) x Dum99t05&            &         &            &         &            &         &   -0.294   &  (0.222)&   -0.318   &  (0.372)\\  \addlinespace
Ratio of White Race (1999) x Dum08t14&            &         &            &         &            &         &    0.423   &  (0.301)&    0.453   &  (0.530)\\  \addlinespace
Ratio of Immigration (90-00) x Dum99t05&            &         &            &         &            &         &    2.351** &  (1.118)&    2.445   &  (1.738)\\  \addlinespace
Ratio of Immigration (90-00) x Dum08t14&            &         &            &         &            &         &   -2.389   &  (1.542)&   -2.426   &  (2.401)\\  \addlinespace
Ratio of Art, Enter, and Recre Emp (1999) x Dum99t05&            &         &            &         &            &         &            &         &   -6.075   &  (7.106)\\  \addlinespace
Ratio of Art, Enter, and Recre Emp (1999) x Dum08t14&            &         &            &         &            &         &            &         &    8.300   & (10.041)\\  \addlinespace
Ratio of Health Emp (1999) x Dum99t05&            &         &            &         &            &         &            &         &   -3.180   &  (3.967)\\  \addlinespace
Ratio of Health Emp (1999) x Dum08t14&            &         &            &         &            &         &            &         &    5.576   &  (5.872)\\  \addlinespace
Ratio of Tradable Service Emp (1999) x Dum99t05&            &         &            &         &            &         &            &         &   -1.566   &  (2.765)\\  \addlinespace
Ratio of Tradable Service Emp (1999) x Dum08t14&            &         &            &         &            &         &            &         &    2.565   &  (4.041)\\  \addlinespace
Ratio of College Students (1999) x Dum99t05&            &         &            &         &            &         &            &         &   -0.933   &  (0.941)\\  \addlinespace
Ratio of College Students (1999) x Dum08t14&            &         &            &         &            &         &            &         &    0.943   &  (1.402)\\  \addlinespace
\midrule
Obs            &     1584   &         &     1584   &         &     1402   &         &     1402   &         &     1402   &         \\
Cluster SE     &     CBSA   &         &     CBSA   &         &     CBSA   &         &     CBSA   &         &     CBSA   &         \\ 
Weight         & Ln(HU99)   &         & Ln(HU99)   &         & Ln(HU99)   &         & Ln(HU99)   &         & Ln(HU99)   &         \\ 
KP F-Stat (99-05, non-stack sample)       &    0.530   &         &    4.246   &         &    1.643   &         &    1.962   &         &    0.773   &         \\
MOP F-Stat (99-05, non-stack sample)  &    0.534   &         &    4.247   &         &    1.639   &         &    1.981   &         &    0.777   &         \\
CoefEqual\_Chi2 &    0.484   &         &    3.324   &         &    1.366   &         &    1.745   &         &    0.722   &         \\
CoefEqual\_PValue &    0.487   &         &    0.068   &         &    0.242   &         &    0.187   &         &    0.395   &         \\
\bottomrule

\end{tabular}

} % end of resize box

\end{table}

%% file: Table_HH/table_HHDTI.D99t05vsD08t14.GSEM.4Reg.tex
%---------------------------------------------------------------

%%%%%%%%%%%%%%%%%%%%%%%%%%%%%%%%%%%%%%%%%%%%%%%%
% table_HHDTI.D99t05vsD08t14.GSEM.4Reg
%%%%%%%%%%%%%%%%%%%%%%%%%%%%%%%%%%%%%%%%%%%%%%%%

\noindent 

\begin{table}[h!]
\centering
\caption{
\textbf{Four Stacked Regressions of Household Leverage Increase in Boom (99-05) and Bust (08-14) Periods on GSEM Growth (99-05)} \smallskip \newline
{\scriptsize
This table reports OLS, reduced-form, first stage, and second stages of stacked 2SLS regression $\triangle_{99,05} \& \triangle_{08,14} HHDTI_{c} = \beta_{99,05} * \triangle_{99,05} Ln(GSEM_{c}) \times Dum_{99,05} + \beta_{08,14} * \triangle_{99,05} Ln(GSEM_{c}) \times Dum_{08,14} + \gamma_{99,05}* \bm{Controls_{c}} \times Dum_{99,05} + \gamma_{08,14}* \bm{Controls_{c}} \times Dum_{08,14} + \epsilon_{period, c}$. The left-hand-side dependent variable $\triangle_{99,05} \& \triangle_{08,14} HHDTI_{c}$ is the stacked increase in household leverage (debt-to-income ratio) at county $c$ 99-05 and 08-14. The key independent variable $\triangle_{99,05} Ln(GSEM_{c})$ is the growth rate of the dollar amount (deflated to 2007 USD) of government-sponsored enterprise mortgages (non-jumbo) at county $c$ 99-05. $Controls_{c}$ indicates control variables at county $c$ in the period start year 1999. We use the gravity model-based instrumental variable $\triangle_{99,05}\text{givNetExp}_{m}$ as the IV for $\triangle_{99,05}Ln(GSEM_{c})$. Regression is weighted by the natural logarithm of housing units in 1999.  For the first-stage F-test of two non-stacked samples, we report Kleibergen-Paap (2006) robust (clustered) statistics and Montiel Olea-Pflueger (2013) efficient statistics. Standard errors are clustered at the CBSA level. ***, **, and * indicate significance at the 1\%, 5\%, and 10\% levels, respectively.
} % end of small font size
} % end of caption
\label{table_HHDTI.D99t05vsD08t14.GSEM.4Reg}
\resizebox{\columnwidth}{!}{%
\begin{tabular}{l*{5}{c}}
\toprule
Dep Var (Panel A, B, and C)                      &\multicolumn{5}{c}{Household Leverage Increase (99-05 \& 08-14, annualized)} \\
            \cmidrule{2-6} 
            &\multicolumn{1}{c}{(1)}&\multicolumn{1}{c}{(2)}&\multicolumn{1}{c}{(3)}&\multicolumn{1}{c}{(4)}&\multicolumn{1}{c}{(5)}\\

\midrule
\multicolumn{6}{l}{\textbf{Panel A. OLS estimates}} \\
GSEM Growth (99-05, An) x Dum99t05&   -0.042   &   -0.047   &   -0.131   &   -0.040   &   -0.041   \\
               &  (0.070)   &  (0.084)   &  (0.082)   &  (0.086)   &  (0.082)   \\
\addlinespace
GSEM Growth (99-05, An) x Dum08t14&    0.128*  &    0.085   &    0.144** &    0.044   &    0.029   \\
               &  (0.066)   &  (0.069)   &  (0.072)   &  (0.067)   &  (0.068)   \\
\addlinespace
R2-adj         &    0.527   &    0.561   &    0.609   &    0.635   &    0.635   \\
\addlinespace

\midrule
\multicolumn{6}{l}{\textbf{Panel B. Reduced-form estimates}} \\
GIV Net Export Growth (99-05, An) x Dum99t05&   12.270***&   12.156***&    9.709***&   11.070***&   11.015***\\
               &  (3.663)   &  (3.292)   &  (3.454)   &  (3.425)   &  (3.622)   \\
\addlinespace
GIV Net Export Growth (99-05, An) x Dum08t14&  -17.318***&  -16.356***&  -14.377***&  -14.752***&  -15.258***\\
               &  (2.708)   &  (2.919)   &  (2.693)   &  (2.594)   &  (2.662)   \\
\addlinespace
R2-adj         &    0.534   &    0.568   &    0.613   &    0.642   &    0.642   \\
\addlinespace

\midrule
\multicolumn{6}{l}{\textbf{Panel C . 2SLS estimates}} \\
\addlinespace
GSEM Growth (99-05, An) x Dum99t05&    5.646   &    2.299*  &    2.878   &    3.143   &    4.687   \\
               &  (8.229)   &  (1.300)   &  (2.509)   &  (2.380)   &  (5.481)   \\
\addlinespace
GSEM Growth (99-05, An) x Dum08t14&   -7.968   &   -3.093*  &   -4.262   &   -4.189   &   -6.493   \\
               & (11.439)   &  (1.730)   &  (3.724)   &  (3.283)   &  (7.787)   \\
               
\addlinespace
\addlinespace

Dep Var (Panel D): &\multicolumn{5}{c}{GSEM Growth (99-05, An)} \\ 
\midrule 
\multicolumn{6}{l}{\textbf{Panel D . First-stage estimates only for 99-05 (Non-stack sample)}} \\
\addlinespace
GIV Net Export Growth (99-05, An)&    2.173   &    5.288** &    3.373   &    3.522   &    2.350   \\
               &  (2.986)   &  (2.566)   &  (2.632)   &  (2.514)   &  (2.672)   \\
\addlinespace
KP F-Stat       &    0.530   &    4.246   &    1.643   &    1.962   &    0.773   \\
MOP F-Stat     &    0.534   &    4.247   &    1.639   &    1.981   &    0.777   \\
\addlinespace

\midrule
\multicolumn{6}{l}{\textbf{Controls (for all Panels)}} \\
DumPeriod  &    Y        &  Y   &   Y    & Y    &  Y    \\
Basic Controls x DumPeriod &            &  Y   &   Y    & Y    &  Y    \\
Housing Controls x DumPeriod &           &      & Y       & Y    &  Y    \\
Demographic Controls x DumPeriod &            &      &        &  Y    &  Y    \\
Industry Controls x DumPeriod &            &      &        &      &  Y    \\

\midrule              
Obs (Panel A, B, and C)          &     1584   &     1584   &     1402   &     1402   &     1402   \\
Obs (Panel D)            &      792   &      792   &      701   &      701   &      701   \\
Cluster SE     &     CBSA   &     CBSA   &     CBSA   &     CBSA   &     CBSA   \\
Weight         & Ln(HU99)   & Ln(HU99)   & Ln(HU99)   & Ln(HU99)   & Ln(HU99)   \\
\bottomrule
\end{tabular}

} % end of resize box

\end{table}

%% file: Table_HH/table_HHDTI.D05t08.PLMNJ.D99t05.4Reg.tex
%---------------------------------------------------------------

%%%%%%%%%%%%%%%%%%%%%%%%%%%%%%%%%%%%%%%%%%%%%%%%
% table_HHDTI.D05t08.PLMNJ.D99t05.4Reg
%%%%%%%%%%%%%%%%%%%%%%%%%%%%%%%%%%%%%%%%%%%%%%%%

\noindent 

\begin{table}[h!]
\centering
\caption{
\textbf{Four Regressions of Household Leverage Increase (05-08) on PLMNJ Growth in Boom Period (99-05)} \smallskip \newline
{\footnotesize 
This table reports OLS, reduced-form, first stage, and second stage results of 2SLS regression $\triangle_{05,08} HHDTI_{c} = \beta * \triangle_{99,05} Ln(PLMNJ_{c}) + \gamma* \bm{Controls_{c}} + \alpha + \epsilon_{c}$. The left-hand-side dependent variable $\triangle_{05,08} HHDTI_{c}$ is the household leverage (debt-to-income ratio) increase at county $c$ 05-08, and the key independent variable $\triangle_{99,05} Ln(PLMNJ_{c})$ is the growth rate of the dollar amount (deflated to 2007 USD) of private-label mortgage (non-jumbo) (PLMNJ) at county $c$ 99-05. $Controls_{c}$ indicates control variables at county $c$ in 1999. We use the gravity model-based instrumental variable ($\triangle_{99,05}\text{givNetExp}_{m}$) as IV for $\triangle_{99,05} Ln(PLMNJ_{c})$. For the first-stage F-test, we report kleibergen-Paap (2006) robust (clustered) statistics and Montiel Olea-Pflueger (2013) efficient statistics. Each regression is weighted by the natural logarithm of housing units in 1999. Standard errors are clustered at the CBSA level. ***, **, and * indicate significance at the 1\%, 5\%, and 10\% levels, respectively.
} % end of small font size
} % end of caption
\label{table_HHDTI.D05t08.PLMNJ.D99t05.4Reg}

\resizebox{\columnwidth}{!}{%

\begin{tabular}{l*{6}{c}}
\toprule
Dep Var (Panel A, B, and C)                     &\multicolumn{5}{c}{Household Leverage Increase (2005-2008, annualized)} \\
            \cmidrule{2-6} 
            &\multicolumn{1}{c}{(1)}&\multicolumn{1}{c}{(2)}&\multicolumn{1}{c}{(3)}&\multicolumn{1}{c}{(4)}&\multicolumn{1}{c}{(5)}\\
            
\midrule
\multicolumn{6}{l}{\textbf{Panel A. OLS estimates}} \\
\addlinespace
PLMNJ Growth (99-05, An)&    0.534***&    0.552***&    0.468***&    0.464***&    0.454***\\
               &  (0.074)   &  (0.071)   &  (0.078)   &  (0.073)   &  (0.072)   \\
\addlinespace
R2-adj         &   0.0841   &    0.124   &    0.186   &    0.212   &    0.217   \\
\addlinespace

\midrule
\multicolumn{6}{l}{\textbf{Panel B. Reduced-form estimates}} \\
\addlinespace
GIV Net Export Growth (99-05, An)&   14.906** &   15.732** &   11.606** &   12.498** &   11.257*  \\
               &  (6.414)   &  (6.787)   &  (5.745)   &  (5.802)   &  (5.968)   \\
\addlinespace
R2-adj         &  0.00997   &   0.0462   &    0.136   &    0.164   &    0.171   \\
\addlinespace

\midrule
\multicolumn{6}{l}{\textbf{Panel C . 2SLS estimates}} \\
\addlinespace
PLMNJ Growth (99-05, An)&    0.946** &    1.030** &    0.948*  &    1.030** &    0.961*  \\
               &  (0.397)   &  (0.434)   &  (0.506)   &  (0.521)   &  (0.552)   \\
\addlinespace

\addlinespace
\addlinespace

Dep Var (Panel D): &\multicolumn{5}{c}{PLMNJ Growth (99-05, An)} \\ 
\midrule 
\multicolumn{6}{l}{\textbf{Panel D . First-stage estimates}} \\
\addlinespace
GIV Net Export Growth (99-05, An)&   15.753***&   15.278***&   12.243***&   12.134***&   11.716***\\
               &  (3.255)   &  (3.331)   &  (3.276)   &  (3.541)   &  (3.465)   \\
\addlinespace
KP F-Stat      &    23.43   &    21.04   &    13.97   &    11.74   &    11.43   \\
MOP F-Stat     &    22.30   &    20.29   &    14.03   &    11.80   &    11.37   \\
\addlinespace
\midrule
\multicolumn{6}{l}{\textbf{Controls (for all Panels)}} \\
Basic Controls &            &  Y   &   Y    & Y    &  Y    \\
Housing Controls &           &      & Y       & Y    &  Y    \\
Demographic Controls &            &      &        &  Y    &  Y    \\
Industry Controls &            &      &        &      &  Y    \\
\midrule              
Obs            &      792   &      792   &      701   &      701   &      701   \\
Cluster SE     &     CBSA   &     CBSA   &     CBSA   &     CBSA   &     CBSA   \\
Weight         & Ln(HU99)   & Ln(HU99)   & Ln(HU99)   & Ln(HU99)   & Ln(HU99)   \\
\bottomrule

\end{tabular}

} % end of resize box

\end{table}

%% file: Table_HH/table_HousingNetWorth.D07t09.PLMNJ.D99t05.2SLS.tex
%-----------------------------------------------------------------

%%%%%%%%%%%%%%%%%%%%%%%%%%%%%%%%%%%%
% table_HousingNetWorth.D07t09.PLMNJ.D99t05.2SLS
%%%%%%%%%%%%%%%%%%%%%%%%%%%%%%%%%%%%

\noindent 

\begin{table}[h!]
\centering
\caption{
\textbf{2SLS Regression of Housing Net Worth Change (07-09) on PLMNJ Growth (99-05)} \smallskip \newline
{\footnotesize 
This table reports the first-stage and the second-stage results of 2SLS regression $\triangle_{07,09} \text{Housing Net Worth}_{c} = \beta * \triangle_{99,05} Ln(PLMNJ_{c})  + \gamma* \bm{Controls_{c}} + \alpha + \epsilon_{c}$. The left-hand-side dependent variable $\triangle_{07,09} \text{Housing Net Worth}_{c}$ is the housing net worth change at county $c$ 07-09, and the key independent variable $\triangle_{99,05} Ln(PLMNJ_{c})$ is the growth rate of the dollar amount (deflated to 2007 USD) of private-label mortgage (non-jumbo) (PLMNJ) at county $c$ 99-05. $Controls_{c}$ indicates control variables at county $c$ in 1999. We use the gravity model-based instrumental variable ($\triangle_{99,05}\text{givNetExp}_{m}$) as IV for $\triangle_{99,05} Ln(PLMNJ_{c})$. For the first-stage F-test, we report Kleibergen-Paap (2006) robust (clustered) statistics and Montiel Olea-Pflueger (2013) efficient statistics. Regression is weighted by the natural logarithm of housing units in 1999. Standard errors are clustered at the CBSA level. ***, **, and * indicate significance at the 1\%, 5\%, and 10\% levels, respectively.
} % end of small font size
} % end of caption
\label{table_HousingNetWorth.D07t09.PLMNJ.D99t05.2SLS}

\resizebox{0.95\columnwidth}{!}{%

\begin{tabular}{l*{6}{c}}
\toprule
        &\multicolumn{1}{p{3cm}}{\centering PLMNJ Growth \\ (99-05, An)}  &\multicolumn{5}{c}{\centering Housing Net Worth Change (2007-2009, annualized)} \\
            \cmidrule{2-7} 
            &\multicolumn{1}{c}{(1)}&\multicolumn{1}{c}{(2)}&\multicolumn{1}{c}{(3)}&\multicolumn{1}{c}{(4)}&\multicolumn{1}{c}{(5)}&\multicolumn{1}{c}{(6)}\\
            
\midrule
GIV Net Export Growth (99-05, An)&   12.079***&            &            &            &            &            \\
               &  (3.552)   &            &            &            &            &            \\
\addlinespace
PLMNJ Growth (07USD, 99-05, An)&            &   -0.680***&   -0.648***&   -0.687** &   -0.682** &   -0.817** \\
               &            &  (0.189)   &  (0.178)   &  (0.295)   &  (0.296)   &  (0.403)   \\
\addlinespace
Ln(Num of households, 99)&    0.006   &            &   -0.008***&   -0.078***&   -0.035   &   -0.042   \\
               &  (0.033)   &            &  (0.003)   &  (0.028)   &  (0.030)   &  (0.041)   \\
\addlinespace
Ln(household Income, 99)&    0.011   &            &   -0.011   &   -0.018   &   -0.043*  &   -0.051*  \\
               &  (0.036)   &            &  (0.014)   &  (0.024)   &  (0.024)   &  (0.029)   \\
\addlinespace
Ratio of Labor Force (1999)&    0.110   &            &    0.033   &    0.139** &    0.053   &    0.029   \\
               &  (0.100)   &            &  (0.065)   &  (0.064)   &  (0.066)   &  (0.080)   \\
\addlinespace
Ln(Num of House Units, 99)&    0.005   &            &            &    0.074** &    0.034   &    0.041   \\
               &  (0.033)   &            &            &  (0.030)   &  (0.032)   &  (0.043)   \\
\addlinespace
Housing supply elasticity&   -0.013***&            &            &   -0.003   &   -0.003   &   -0.005   \\
               &  (0.003)   &            &            &  (0.006)   &  (0.006)   &  (0.007)   \\
\addlinespace
House Vacancy Rate (1999)&    0.203** &            &            &    0.012   &    0.040   &    0.022   \\
               &  (0.082)   &            &            &  (0.075)   &  (0.076)   &  (0.076)   \\
\addlinespace
Ratio of Renters (1999)&   -0.067   &            &            &   -0.073   &   -0.009   &   -0.058   \\
               &  (0.058)   &            &            &  (0.062)   &  (0.055)   &  (0.084)   \\
\addlinespace
Ratio of Bachelor Educated (1999)&   -0.145*  &            &            &            &    0.117   &    0.023   \\
               &  (0.088)   &            &            &            &  (0.085)   &  (0.120)   \\
\addlinespace
Ratio of White Race (1999)&    0.017   &            &            &            &    0.057*  &    0.048   \\
               &  (0.034)   &            &            &            &  (0.032)   &  (0.035)   \\
\addlinespace
Ratio of Immigration (90-00)&    0.135   &            &            &            &   -0.296*  &   -0.256   \\
               &  (0.194)   &            &            &            &  (0.175)   &  (0.196)   \\
\addlinespace
Ratio of Art, Enter, and Recre Emp (1999)&            &            &            &            &            &    1.042   \\
               &            &            &            &            &            &  (0.995)   \\
\addlinespace
Ratio of Health Emp (1999)&            &            &            &            &            &    0.201   \\
               &            &            &            &            &            &  (0.333)   \\
\addlinespace
Ratio of Tradable Service Emp (1999)&            &            &            &            &            &    0.275   \\
               &            &            &            &            &            &  (0.264)   \\
\addlinespace
Ratio of College Students (1999)&            &            &            &            &            &    0.133   \\
               &            &            &            &            &            &  (0.197)   \\
\addlinespace
Constant       &   -0.088   &    0.087***&    0.275*  &    0.257   &    0.465*  &    0.603*  \\
               &  (0.365)   &  (0.033)   &  (0.157)   &  (0.269)   &  (0.254)   &  (0.320)   \\
\midrule
Obs            &      700   &      597   &      597   &      530   &      530   &      530   \\
Cluster SE     &     CBSA   &     CBSA   &     CBSA   &     CBSA   &     CBSA   &     CBSA   \\
Weight         & Ln(HU99)   & Ln(HU99)   & Ln(HU99)   & Ln(HU99)   & Ln(HU99)   & Ln(HU99)   \\
KP F-Stat      &    4.001         &    12.61   &    11.71   &    6.536   &    4.953   &    4.001   \\
MOP F-Stat    &    3.996          &    11.98   &    11.17   &    6.574   &    4.968   &    3.996    \\
\bottomrule

\end{tabular}

} % end of resize box

\end{table}

%% file: Table_HH/table_HousingNetWorth.D07t09.PLMNJ.D99t05.4Reg.tex
%---------------------------------------------------------------

%%%%%%%%%%%%%%%%%%%%%%%%%%%%%%%%%%%%%%%%%%%%%%%%
% table_HousingNetWorth.D07t09.PLMNJ.D99t05.4Reg
%%%%%%%%%%%%%%%%%%%%%%%%%%%%%%%%%%%%%%%%%%%%%%%%

\noindent 

\begin{table}[h!]
\centering
\caption{
\textbf{Four Regressions of Housing Net Worth Change (07-09) on PLMNJ Growth (99-05)} \smallskip \newline
{\footnotesize 
This table reports OLS, reduced-form, first stage, and second stage results of 2SLS regression $\triangle_{07,09} \text{Housing Net Worth}_{c} = \beta * \triangle_{99,05} Ln(PLMNJ_{c})  + \gamma* \bm{Controls_{c}} + \alpha + \epsilon_{c}$. The left-hand-side dependent variable $\triangle_{07,09} \text{Housing Net Worth}_{c}$ is the housing net worth change at county $c$ 07-09, and the key independent variable $\triangle_{99,05} Ln(PLMNJ_{c})$ is the growth rate of the dollar amount (deflated to 2007 USD) of private-label mortgage (non-jumbo) (PLMNJ) at county $c$ 99-05. $Controls_{c}$ indicates control variables at county $c$ in 1999. We use the gravity model-based instrumental variable ($\triangle_{99,05}\text{givNetExp}_{m}$) as IV for $\triangle_{99,05} Ln(PLMNJ_{c})$. For the first-stage F-test, we report Kleibergen-Paap (2006) robust (clustered) statistics and Montiel Olea-Pflueger (2013) efficient statistics. Regression is weighted by the natural logarithm of housing units in 1999. Standard errors are clustered at the CBSA level. ***, **, and * indicate significance at the 1\%, 5\%, and 10\% levels, respectively.
} % end of small font size
} % end of caption
\label{table_HousingNetWorth.D07t09.PLMNJ.D99t05.4Reg}

\resizebox{\columnwidth}{!}{%

\begin{tabular}{l*{6}{c}}
\toprule
Dep Var (Panel A, B, and C)                     &\multicolumn{5}{c}{Housing Net Worth Change (2007-2009, annualized)} \\
            \cmidrule{2-6} 
            &\multicolumn{1}{c}{(1)}&\multicolumn{1}{c}{(2)}&\multicolumn{1}{c}{(3)}&\multicolumn{1}{c}{(4)}&\multicolumn{1}{c}{(5)}\\
            
\midrule
\multicolumn{6}{l}{\textbf{Panel A. OLS estimates}} \\
\addlinespace
PLMNJ Growth (07USD, 99-05, An)&   -0.244***&   -0.237***&   -0.171***&   -0.150***&   -0.142***\\
               &  (0.038)   &  (0.040)   &  (0.034)   &  (0.029)   &  (0.028)   \\
\addlinespace
R2-adj         &    0.183   &    0.241   &    0.278   &    0.391   &    0.400   \\
\addlinespace

\midrule
\multicolumn{6}{l}{\textbf{Panel B. Reduced-form estimates}} \\
\addlinespace
GIV Net Export Growth (99-05, An)&   -9.093***&   -8.656***&   -6.124***&   -5.910***&   -6.363***\\
               &  (2.228)   &  (2.357)   &  (1.742)   &  (1.828)   &  (1.876)   \\
\addlinespace
R2-adj         &   0.0365   &    0.104   &    0.212   &    0.344   &    0.363   \\
\addlinespace

\midrule
\multicolumn{6}{l}{\textbf{Panel C . 2SLS estimates}} \\
\addlinespace
PLMNJ Growth (07USD, 99-05, An)&   -0.680***&   -0.648***&   -0.687** &   -0.682** &   -0.817** \\
               &  (0.189)   &  (0.178)   &  (0.295)   &  (0.296)   &  (0.403)   \\
\addlinespace

\addlinespace
\addlinespace

Dep Var (Panel D): &\multicolumn{5}{c}{PLMNJ Growth (99-05, An)} \\ 
\midrule 
\multicolumn{6}{l}{\textbf{Panel D . First-stage estimates}} \\
\addlinespace
GIV Net Export Growth (99-05, An)&   15.599***&   15.025***&   12.173***&   12.079***&   11.844***\\
               &  (3.219)   &  (3.265)   &  (3.288)   &  (3.552)   &  (3.477)   \\
\addlinespace
KP F-Stat      &    12.61   &    11.71   &    6.536   &    4.953   &    4.001   \\
MOP F-Stat     &    11.98   &    11.17   &    6.574   &    4.968   &    3.996   \\
\addlinespace
\midrule
\multicolumn{6}{l}{\textbf{Controls (for all Panels)}} \\
Basic Controls &            &  Y   &   Y    & Y    &  Y    \\
Housing Controls &           &      & Y       & Y    &  Y    \\
Demographic Controls &            &      &        &  Y    &  Y    \\
Industry Controls &            &      &        &      &  Y    \\
\midrule              
Obs            &      597   &      597   &      530   &      530   &      530   \\
Cluster SE     &     CBSA   &     CBSA   &     CBSA   &     CBSA   &     CBSA   \\
Weight         & Ln(HU99)   & Ln(HU99)   & Ln(HU99)   & Ln(HU99)   & Ln(HU99)   \\
\bottomrule

\end{tabular}

} % end of resize box

\end{table}

%% file: Table_HH/table_HPI.D99t05vsD07t09.HHDTI.2SLS.wide.tex
%-----------------------------------------------------------------------
%%%%%%%%%%%%%%%%%%%%%%%%%%%%%%%%%%%%
% table_HPI.D99t05vsD07t09.HHDTI.2SLS.wide
%%%%%%%%%%%%%%%%%%%%%%%%%%%%%%%%%%%%

\noindent 

\begin{table}[h!]
\centering
\caption{
\textbf{2SLS Stacked Regression of Housing Price Growth in Boom (99-05) and Bust (07-09) Periods on Household Leverage Increase (99-05)} \smallskip \newline
{\scriptsize
This table reports 2SLS regression $\triangle_{99,05} \& \triangle_{07,09} Ln(HPI_{c}) = \beta_{99,05} * \triangle_{99,05} HHDTI_{c} \times Dum_{99,05} + \beta_{07,09} * \triangle_{99,05} HHDTI_{c} \times Dum_{07,09} + \gamma_{99,05}* \bm{Controls_{c}} \times Dum_{99,05} + \gamma_{07,09}* \bm{Controls_{c}} \times Dum_{07,09} + \epsilon_{period, c}$. The left-hand-side dependent variable $\triangle_{99,05} \& \triangle_{07,09} Ln(HPI_{c})$ is the stacked growth rate of the house price index (deflated to 2007) at county $c$ 99-05 and 07-09. The key independent variable $\triangle_{99,05} HHDTI_{c}$ is the rise in household leverage (debt-to-income ratio) at county $c$ 99-05. $Controls_{c}$ indicates control variables at county $c$ in the period start year 1999. We use the gravity model-based instrumental variable $\triangle_{99,05}\text{givNetExp}_{m}$ as the IV for $\triangle_{99,05}HHDTI_{c}$. Regression is weighted by the natural logarithm of housing units in 1999. For the first-stage F-test of two non-stacked samples, we report Kleibergen-Paap (2006) robust (clustered) statistics and Montiel Olea-Pflueger (2013) efficient statistics. Standard errors are clustered at the CBSA level. ***, **, and * indicate significance at the 1\%, 5\%, and 10\% levels, respectively.
\smallskip
} % end of small font size
} % end of caption
\label{table_HPI.D99t05vsD07t09.HHDTI.2SLS.wide}

\vspace{-2mm}

\resizebox{\columnwidth}{!}{%
\begin{tabular}{l*{10}{c}}
\toprule
\textbf{TSLS estimates}            &\multicolumn{10}{c}{Housing Price Growth (99-05 and 07-09, Annualized)} \\
            \cmidrule{2-11} 
            &\multicolumn{2}{c}{(1)}&\multicolumn{2}{c}{(2)}&\multicolumn{2}{c}{(3)}&\multicolumn{2}{c}{(4)}&\multicolumn{2}{c}{(5)}\\
            
\midrule
HH Debt-to-Income Rise (99-05, An) x Dum99t05&    0.630***&  (0.174)&    0.534***&  (0.173)&    0.479** &  (0.195)&    0.455***&  (0.161)&    0.445** &  (0.175)\\ 
\addlinespace
HH Debt-to-Income Rise (99-05, An) x Dum07t09&   -0.378***&  (0.104)&   -0.337***&  (0.091)&   -0.329***&  (0.116)&   -0.296***&  (0.091)&   -0.304***&  (0.103)\\  
\addlinespace
Dum99t05       &   -0.053** &  (0.024)&   -0.044   &  (0.177)&   -0.316** &  (0.140)&   -0.104   &  (0.208)&   -0.141   &  (0.198)\\   \addlinespace
Dum07t09       &    0.037** &  (0.015)&   -0.119   &  (0.084)&   -0.002   &  (0.083)&    0.143   &  (0.153)&    0.161   &  (0.143)\\   \addlinespace
Ln(Num of HH, 99) x DumD99t05&            &         &    0.024***&  (0.006)&    0.003   &  (0.034)&    0.031   &  (0.029)&    0.032   &  (0.029)\\   \addlinespace
Ln(Num of HH, 99) x DumD07t09&            &         &   -0.016***&  (0.003)&   -0.004   &  (0.024)&   -0.012   &  (0.020)&   -0.014   &  (0.021)\\   \addlinespace
Ln(HH Income, 99) x DumD99t05&            &         &   -0.011   &  (0.024)&    0.023   &  (0.019)&    0.003   &  (0.025)&    0.006   &  (0.024)\\   \addlinespace
Ln(HH Income, 99) x DumD07t09&            &         &    0.023** &  (0.011)&    0.008   &  (0.011)&   -0.006   &  (0.017)&   -0.007   &  (0.016)\\   \addlinespace
Ratio of Labor Force (1999) x DumD99t05&            &         &   -0.214***&  (0.062)&   -0.243***&  (0.079)&   -0.296***&  (0.086)&   -0.283***&  (0.088)\\   \addlinespace
Ratio of Labor Force (1999) x DumD07t09&            &         &    0.122***&  (0.040)&    0.161** &  (0.067)&    0.165***&  (0.063)&    0.160** &  (0.063)\\   \addlinespace
Ln(Num of HU, 99) x DumD99t05&            &         &            &         &    0.008   &  (0.038)&   -0.019   &  (0.031)&   -0.020   &  (0.031)\\   \addlinespace
Ln(Num of HU, 99) x DumD07t09&            &         &            &         &   -0.008   &  (0.026)&   -0.001   &  (0.021)&    0.000   &  (0.022)\\   \addlinespace
Housing supply elasticity x DumD99t05&            &         &            &         &   -0.002   &  (0.004)&   -0.003   &  (0.003)&   -0.003   &  (0.003)\\   \addlinespace
Housing supply elasticity x DumD07t09&            &         &            &         &   -0.001   &  (0.002)&   -0.001   &  (0.002)&   -0.001   &  (0.002)\\   \addlinespace
House Vacancy Rate (1999) x DumD99t05&            &         &            &         &    0.008   &  (0.112)&    0.033   &  (0.091)&    0.020   &  (0.097)\\   \addlinespace
House Vacancy Rate (1999) x DumD07t09&            &         &            &         &    0.053   &  (0.071)&    0.032   &  (0.058)&    0.067   &  (0.062)\\   \addlinespace
Ratio of Renters (1999) x DumD99t05&            &         &            &         &    0.258***&  (0.088)&    0.282***&  (0.098)&    0.235** &  (0.100)\\   \addlinespace
Ratio of Renters (1999) x DumD07t09&            &         &            &         &   -0.101*  &  (0.057)&   -0.124** &  (0.058)&   -0.110*  &  (0.058)\\   \addlinespace
Ratio of Bachelor Educated (1999) x DumD99t05&            &         &            &         &            &         &    0.114*  &  (0.065)&    0.091   &  (0.079)\\   \addlinespace
Ratio of Bachelor Educated (1999) x DumD07t09&            &         &            &         &            &         &    0.023   &  (0.047)&   -0.024   &  (0.067)\\   \addlinespace
Ratio of White Race (1999) x DumD99t05&            &         &            &         &            &         &    0.015   &  (0.031)&    0.001   &  (0.031)\\   \addlinespace
Ratio of White Race (1999) x DumD07t09&            &         &            &         &            &         &   -0.003   &  (0.018)&   -0.001   &  (0.018)\\   \addlinespace
Ratio of Immigration (90-00) x DumD99t05&            &         &            &         &            &         &   -0.323   &  (0.236)&   -0.235   &  (0.259)\\   \addlinespace
Ratio of Immigration (90-00) x DumD07t09&            &         &            &         &            &         &    0.165   &  (0.131)&    0.172   &  (0.155)\\   \addlinespace
Ratio of Art, Enter, and Recre Emp (1999) x DumD99t05&            &         &            &         &            &         &            &         &    0.714***&  (0.200)\\   \addlinespace
Ratio of Art, Enter, and Recre Emp (1999) x DumD07t09&            &         &            &         &            &         &            &         &   -0.689***&  (0.171)\\   \addlinespace
Ratio of Health Emp (1999) x DumD99t05&            &         &            &         &            &         &            &         &    0.377   &  (0.235)\\   \addlinespace
Ratio of Health Emp (1999) x DumD07t09&            &         &            &         &            &         &            &         &    0.101   &  (0.194)\\   \addlinespace
Ratio of Tradable Service Emp (1999) x DumD99t05&            &         &            &         &            &         &            &         &   -0.090   &  (0.172)\\   \addlinespace
Ratio of Tradable Service Emp (1999) x DumD07t09&            &         &            &         &            &         &            &         &    0.258*  &  (0.132)\\   \addlinespace
Ratio of College Students (1999) x DumD99t05&            &         &            &         &            &         &            &         &    0.145   &  (0.122)\\   \addlinespace
Ratio of College Students (1999) x DumD07t09&            &         &            &         &            &         &            &         &   -0.098   &  (0.068)\\   \addlinespace

\addlinespace
\midrule
Obs            &     1572   &         &     1572   &         &     1390   &         &     1390   &         &     1390   &         \\
Cluster SE     &     CBSA   &         &     CBSA   &         &     CBSA   &         &     CBSA   &         &     CBSA   &         \\
Weight         & Ln(HU99)   &         & Ln(HU99)   &         & Ln(HU99)   &         & Ln(HU99)   &         & Ln(HU99)   &         \\
KP F-Stat (99-05, non-stack sample)       &    23.08   &         &    20.62   &         &    14.24   &         &    11.79   &         &    11.22   &         \\
MOP F-Stat (99-05, non-stack sample)  &    22.02   &         &    19.84   &         &    14.40   &         &    11.90   &         &    11.20   &         \\
CoefEqual\_Chi2 &   14.521   &         &   12.057   &         &    7.399   &         &    9.896   &         &    8.301   &         \\
CoefEqual\_PValue &    0.000   &         &    0.001   &         &    0.007   &         &    0.002   &         &    0.004   &         \\
\bottomrule

\end{tabular}

} % end of resize box

\end{table}

%% file: Table_HH/table_HPI.D99t05vsD07t09.HHDTI.4Reg.tex
%---------------------------------------------------------------

%%%%%%%%%%%%%%%%%%%%%%%%%%%%%%%%%%%%%%%%%%%%%%%%
% table_HPI.D99t05vsD07t09.HHDTI.4Reg
%%%%%%%%%%%%%%%%%%%%%%%%%%%%%%%%%%%%%%%%%%%%%%%%

\noindent 

\begin{table}[h!]
\centering
\caption{
\textbf{Four Stacked Regressions of Housing Price Growth in Boom (99-05) and Bust (07-09) Periods on Household Leverage Increase (99-05)} \smallskip \newline
{\scriptsize
This table reports OLS, reduced-form, first stage, and second stages of stacked 2SLS regression $\triangle_{99,05} \& \triangle_{07,09} Ln(HPI_{c}) = \beta_{99,05} * \triangle_{99,05} HHDTI_{c} \times Dum_{99,05} + \beta_{07,09} * \triangle_{99,05} HHDTI_{c} \times Dum_{07,09} + \gamma_{99,05}* \bm{Controls_{c}} \times Dum_{99,05} + \gamma_{07,09}* \bm{Controls_{c}} \times Dum_{07,09} + \epsilon_{period, c}$. The left-hand-side dependent variable $\triangle_{99,05} \& \triangle_{07,09} Ln(HPI_{c})$ is the stacked growth rate of the house price index (deflated to 2007) at county $c$ 99-05 and 07-09. The key independent variable $\triangle_{99,05} HHDTI_{c}$ is the rise in household leverage (debt-to-income ratio) at county $c$ 99-05. $Controls_{c}$ indicates control variables at county $c$ in the period start year 1999. We use the gravity model-based instrumental variable $\triangle_{99,05}\text{givNetExp}_{m}$ as the IV for $\triangle_{99,05}HHDTI_{c}$. Regression is weighted by the natural logarithm of housing units in 1999. For the first-stage F-test of two non-stacked samples, we report Kleibergen-Paap (2006) robust (clustered) statistics and Montiel Olea-Pflueger (2013) efficient statistics. Standard errors are clustered at the CBSA level. ***, **, and * indicate significance at the 1\%, 5\%, and 10\% levels, respectively.
} % end of small font size
} % end of caption
\label{table_HPI.D99t05vsD07t09.HHDTI.4Reg}
\resizebox{\columnwidth}{!}{%
\begin{tabular}{l*{6}{c}}
\toprule
Dep Var (Panel A, B, and C)                      &\multicolumn{5}{c}{House Price Growth (99-05 \& 07-09, annualized)} \\
            \cmidrule{2-6} 
            &\multicolumn{1}{c}{(1)}&\multicolumn{1}{c}{(2)}&\multicolumn{1}{c}{(3)}&\multicolumn{1}{c}{(4)}&\multicolumn{1}{c}{(5)}\\

\midrule
\multicolumn{6}{l}{\textbf{Panel A. OLS estimates}} \\
HH Debt-to-Income Rise (99-05, An) x Dum99t05&    0.054***&    0.071***&    0.064***&    0.056***&    0.057***\\
               &  (0.014)   &  (0.015)   &  (0.014)   &  (0.013)   &  (0.013)   \\
\addlinespace
HH Debt-to-Income Rise (99-05, An) x Dum07t09&   -0.042***&   -0.060***&   -0.046***&   -0.035***&   -0.036***\\
               &  (0.010)   &  (0.011)   &  (0.011)   &  (0.010)   &  (0.009)   \\
\addlinespace
R2-adj         &    0.553   &    0.655   &    0.741   &    0.751   &    0.764   \\
\addlinespace

\midrule
\multicolumn{6}{l}{\textbf{Panel B. Reduced-form estimates}} \\
GIV Net Export Growth (99-05, An) x Dum99t05&    7.639***&    6.530***&    4.655***&    5.119***&    4.971***\\
               &  (1.784)   &  (1.853)   &  (1.070)   &  (1.133)   &  (1.267)   \\
\addlinespace
GIV Net Export Growth (99-05, An) x Dum07t09&   -4.591***&   -4.128***&   -3.198***&   -3.333***&   -3.391***\\
               &  (0.923)   &  (0.946)   &  (0.717)   &  (0.697)   &  (0.694)   \\
\addlinespace
R2-adj         &    0.557   &    0.640   &    0.728   &    0.748   &    0.759   \\
\addlinespace

\midrule
\multicolumn{6}{l}{\textbf{Panel C . 2SLS estimates}} \\
\addlinespace
HH Debt-to-Income Rise (99-05, An) x Dum99t05&    0.630***&    0.534***&    0.479** &    0.455***&    0.445** \\
               &  (0.174)   &  (0.173)   &  (0.195)   &  (0.161)   &  (0.175)   \\
\addlinespace
HH Debt-to-Income Rise (99-05, An) x Dum07t09&   -0.378***&   -0.337***&   -0.329***&   -0.296***&   -0.304***\\
               &  (0.104)   &  (0.091)   &  (0.116)   &  (0.091)   &  (0.103)   \\
               
\addlinespace
\addlinespace

Dep Var (Panel D): &\multicolumn{5}{c}{Household Leverage Increase (99-05, An)} \\ 
\midrule 
\multicolumn{6}{l}{\textbf{Panel D . First-stage estimates only for 99-05 (Non-stack sample)}} \\
\addlinespace
GIV NEG (99-05, An)&   12.130***&   12.237***&    9.725***&   11.243***&   11.172***\\
               &  (3.677)   &  (3.364)   &  (3.487)   &  (3.521)   &  (3.780)   \\
\addlinespace
KP F-Stat      &    10.88   &    13.23   &    7.776   &    10.19   &    8.736   \\
MOP F-Stat     &    9.760   &    12.36   &    7.059   &    9.542   &    7.959   \\
\addlinespace

\midrule
\multicolumn{6}{l}{\textbf{Controls (for all Panels)}} \\
DumPeriod  &    Y        &  Y   &   Y    & Y    &  Y    \\
Basic Controls x DumPeriod &            &  Y   &   Y    & Y    &  Y    \\
Housing Controls x DumPeriod &           &      & Y       & Y    &  Y    \\
Demographic Controls x DumPeriod &            &      &        &  Y    &  Y    \\
Industry Controls x DumPeriod &            &      &        &      &  Y    \\

\midrule              
Obs (Panel A, B, and C)          &     1572   &     1572   &     1390   &     1390   &     1390   \\
Obs (Panel D)            &      786   &      786   &      695   &      695   &      695   \\
Cluster SE     &     CBSA   &     CBSA   &     CBSA   &     CBSA   &     CBSA   \\
Weight         & Ln(HU99)   & Ln(HU99)   & Ln(HU99)   & Ln(HU99)   & Ln(HU99)   \\
\bottomrule
\end{tabular}

} % end of resize box

\end{table}

%% file: Table_HH/table_Permit.D99t05vsD05t09.HHDTI.2SLS.wide.tex
%-----------------------------------------------------------------------
%%%%%%%%%%%%%%%%%%%%%%%%%%%%%%%%%%%%
% table_Permit.D99t05vsD05t09.HHDTI.2SLS.wide
%%%%%%%%%%%%%%%%%%%%%%%%%%%%%%%%%%%%

\noindent 

\begin{table}[h!]
\centering
\caption{
\textbf{2SLS Stacked Regression of Residential Construction Investment Growth in Boom (99-05) and Bust (05-09) Periods on Household Leverage Rise (99-05)} \smallskip \newline
{\scriptsize
This table reports 2SLS regression $\triangle_{99,05} \& \triangle_{05,09} Ln(\text{PermitValue}_{c}) = \beta_{99,05} * \triangle_{99,05} HHDTI_{c} \times Dum_{99,05} + \beta_{05,09} * \triangle_{99,05} HHDTI_{c} \times Dum_{05,09} + \gamma_{99,05}* \bm{Controls_{c}} \times Dum_{99,05} + \gamma_{05,09}* \bm{Controls_{c}} \times Dum_{05,09} + \epsilon_{period, c}$. The left-hand-side dependent variable $\triangle_{99,05} \& \triangle_{05,09} Ln(\text{PermitValue}_{c})$ is the stacked growth rate of the Residential Construction Investment (building permit value) (deflated to 2007) at county $c$ 99-05 and 05-09. The key independent variable $\triangle_{99,05} HHDTI_{c}$ is the rise in household leverage (debt-to-income ratio) at county $c$ 99-05. $Controls_{c}$ indicates control variables at county $c$ in the period start year 1999. We use the gravity model-based instrumental variable $\triangle_{99,05}\text{givNetExp}_{m}$ as the IV for $\triangle_{99,05}HHDTI_{c}$. Regression is weighted by the natural logarithm of housing units in 1999. For the first-stage F-test of two non-stacked samples, we report Kleibergen-Paap (2006) robust (clustered) statistics and Montiel Olea-Pflueger (2013) efficient statistics. Standard errors are clustered at the CBSA level. ***, **, and * indicate significance at the 1\%, 5\%, and 10\% levels, respectively.
\smallskip
} % end of small font size
} % end of caption
\label{table_Permit.D99t05vsD05t09.HHDTI.2SLS.wide}

\vspace{-2mm}

\resizebox{\columnwidth}{!}{%
\begin{tabular}{l*{10}{c}}
\toprule
\textbf{TSLS estimates}            &\multicolumn{10}{c}{Residential Permit Value Growth (99-05 and 05-09, Annualized)} \\
            \cmidrule{2-11} 
            &\multicolumn{2}{c}{(1)}&\multicolumn{2}{c}{(2)}&\multicolumn{2}{c}{(3)}&\multicolumn{2}{c}{(4)}&\multicolumn{2}{c}{(5)}\\
            
\midrule
HH Debt-to-Income Rise (99-05, An) x Dum99t05&    0.919** &  (0.372)&    1.080***&  (0.390)&    0.853*  &  (0.449)&    0.798** &  (0.389)&    0.749*  &  (0.391)\\
\addlinespace
HH Debt-to-Income Rise (99-05, An) x Dum05t09&   -2.061***&  (0.718)&   -1.844***&  (0.625)&   -2.098** &  (0.884)&   -1.808***&  (0.693)&   -1.839** &  (0.738)\\ \addlinespace
Dum99t05       &   -0.067   &  (0.053)&    1.753***&  (0.402)&    0.919***&  (0.307)&    1.255** &  (0.516)&    1.335***&  (0.490)\\ \addlinespace
Dum05t09       &    0.002   &  (0.102)&   -0.604   &  (0.719)&   -0.146   &  (0.640)&    1.636   &  (1.124)&    1.519   &  (1.134)\\ \addlinespace
Ln(Num of HH, 99) x DumD99t05&            &         &    0.033***&  (0.013)&   -0.027   &  (0.069)&   -0.001   &  (0.061)&    0.008   &  (0.063)\\ \addlinespace
Ln(Num of HH, 99) x DumD05t09&            &         &   -0.069***&  (0.022)&    0.103   &  (0.128)&    0.061   &  (0.108)&    0.033   &  (0.111)\\ \addlinespace
Ln(HH Income, 99) x DumD99t05&            &         &   -0.188***&  (0.054)&   -0.098** &  (0.039)&   -0.124** &  (0.058)&   -0.128** &  (0.055)\\ \addlinespace
Ln(HH Income, 99) x DumD05t09&            &         &    0.104   &  (0.095)&    0.062   &  (0.084)&   -0.105   &  (0.123)&   -0.094   &  (0.122)\\ \addlinespace
Ratio of Labor Force (1999) x DumD99t05&            &         &   -0.287** &  (0.137)&   -0.312*  &  (0.189)&   -0.355*  &  (0.210)&   -0.362*  &  (0.200)\\ \addlinespace
Ratio of Labor Force (1999) x DumD05t09&            &         &    0.327   &  (0.243)&    0.491   &  (0.455)&    0.470   &  (0.444)&    0.515   &  (0.437)\\ \addlinespace
Ln(Num of HU, 99) x DumD99t05&            &         &            &         &    0.041   &  (0.074)&    0.015   &  (0.063)&    0.006   &  (0.063)\\ \addlinespace
Ln(Num of HU, 99) x DumD05t09&            &         &            &         &   -0.167   &  (0.138)&   -0.128   &  (0.109)&   -0.108   &  (0.110)\\ \addlinespace
Housing supply elasticity x DumD99t05&            &         &            &         &    0.006   &  (0.009)&    0.004   &  (0.008)&    0.004   &  (0.008)\\ \addlinespace
Housing supply elasticity x DumD05t09&            &         &            &         &   -0.019   &  (0.019)&   -0.013   &  (0.014)&   -0.018   &  (0.014)\\ \addlinespace
House Vacancy Rate (1999) x DumD99t05&            &         &            &         &    0.037   &  (0.237)&    0.059   &  (0.205)&    0.063   &  (0.205)\\ \addlinespace
House Vacancy Rate (1999) x DumD05t09&            &         &            &         &    0.583   &  (0.471)&    0.414   &  (0.377)&    0.572   &  (0.394)\\ \addlinespace
Ratio of Renters (1999) x DumD99t05&            &         &            &         &    0.364** &  (0.166)&    0.350*  &  (0.205)&    0.326*  &  (0.187)\\ \addlinespace
Ratio of Renters (1999) x DumD05t09&            &         &            &         &   -0.424   &  (0.358)&   -0.647*  &  (0.372)&   -0.704*  &  (0.367)\\ \addlinespace
Ratio of Bachelor Educated (1999) x DumD99t05&            &         &            &         &            &         &    0.178   &  (0.145)&    0.162   &  (0.167)\\ \addlinespace
Ratio of Bachelor Educated (1999) x DumD05t09&            &         &            &         &            &         &    0.395   &  (0.318)&    0.020   &  (0.387)\\ \addlinespace
Ratio of White Race (1999) x DumD99t05&            &         &            &         &            &         &   -0.045   &  (0.059)&   -0.041   &  (0.058)\\ \addlinespace
Ratio of White Race (1999) x DumD05t09&            &         &            &         &            &         &   -0.035   &  (0.100)&   -0.078   &  (0.098)\\ \addlinespace
Ratio of Immigration (90-00) x DumD99t05&            &         &            &         &            &         &   -0.555   &  (0.515)&   -0.573   &  (0.503)\\ \addlinespace
Ratio of Immigration (90-00) x DumD05t09&            &         &            &         &            &         &    1.268   &  (0.965)&    1.655   &  (1.050)\\ \addlinespace
Ratio of Art, Enter, and Recre Emp (1999) x DumD99t05&            &         &            &         &            &         &            &         &    0.491   &  (0.525)\\ \addlinespace
Ratio of Art, Enter, and Recre Emp (1999) x DumD05t09&            &         &            &         &            &         &            &         &   -2.276***&  (0.840)\\ \addlinespace
Ratio of Health Emp (1999) x DumD99t05&            &         &            &         &            &         &            &         &   -0.434   &  (0.547)\\ \addlinespace
Ratio of Health Emp (1999) x DumD05t09&            &         &            &         &            &         &            &         &    2.496** &  (1.184)\\ \addlinespace
Ratio of Tradable Service Emp (1999) x DumD99t05&            &         &            &         &            &         &            &         &    0.127   &  (0.373)\\ \addlinespace
Ratio of Tradable Service Emp (1999) x DumD05t09&            &         &            &         &            &         &            &         &    1.188   &  (0.756)\\ \addlinespace
Ratio of College Students (1999) x DumD99t05&            &         &            &         &            &         &            &         &    0.165   &  (0.452)\\ \addlinespace
Ratio of College Students (1999) x DumD05t09&            &         &            &         &            &         &            &         &   -0.290   &  (0.738)\\ 
\addlinespace
\midrule
Obs            &     1580   &         &     1580   &         &     1400   &         &     1400   &         &     1400   &         \\
Cluster SE     &     CBSA   &         &     CBSA   &         &     CBSA   &         &     CBSA   &         &     CBSA   &         \\
Weight         & Ln(HU99)   &         & Ln(HU99)   &         & Ln(HU99)   &         & Ln(HU99)   &         & Ln(HU99)   &         \\
KP F-Stat (99-05, non-stack sample)       &    10.99   &         &    13.02   &         &    7.889   &         &    10.46   &         &    9.267   &         \\
MOP F-Stat (99-05, non-stack sample)  &    9.896   &         &    12.26   &         &    7.322   &         &    10.04   &         &    8.726   &         \\
CoefEqual\_Chi2 &    8.974   &         &   10.377   &         &    5.999   &         &    7.162   &         &    6.538   &         \\
CoefEqual\_PValue &    0.003   &         &    0.001   &         &    0.014   &         &    0.007   &         &    0.011   &         \\
\bottomrule

\end{tabular}

} % end of resize box

\end{table}

%% file: Table_HH/table_Permit.D99t05vsD05t09.HHDTI.4Reg.tex
%---------------------------------------------------------------

%%%%%%%%%%%%%%%%%%%%%%%%%%%%%%%%%%%%%%%%%%%%%%%%
% table_Permit.D99t05vsD05t09.HHDTI.4Reg
%%%%%%%%%%%%%%%%%%%%%%%%%%%%%%%%%%%%%%%%%%%%%%%%

\noindent 

\begin{table}[h!]
\centering
\caption{
\textbf{Four Stacked Regressions of Residential Construction Permit Value Growth in Boom (99-05) and Bust (05-09) Periods on Household Leverage Rise (99-05)} \smallskip \newline
{\scriptsize
This table reports OLS, reduced-form, first stage, and second stages of stacked 2SLS regression $\triangle_{99,05} \& \triangle_{05,09} Ln(\text{PermitValue}_{c}) = \beta_{99,05} * \triangle_{99,05} HHDTI_{c} \times Dum_{99,05} + \beta_{05,09} * \triangle_{99,05} HHDTI_{c} \times Dum_{05,09} + \gamma_{99,05}* \bm{Controls_{c}} \times Dum_{99,05} + \gamma_{05,09}* \bm{Controls_{c}} \times Dum_{05,09} + \epsilon_{period, c}$. The left-hand-side dependent variable $\triangle_{99,05} \& \triangle_{05,09} Ln(\text{PermitValue}_{c})$ is the stacked growth rate of the residential construction permit value (deflated to 2007) at county $c$ 99-05 and 05-09. The key independent variable $\triangle_{99,05} HHDTI_{c}$ is the rise in household leverage (debt-to-income ratio) at county $c$ 99-05. $Controls_{c}$ indicates control variables at county $c$ in the period start year 1999. We use the gravity model-based instrumental variable $\triangle_{99,05}\text{givNetExp}_{m}$ as the IV for $\triangle_{99,05}HHDTI_{c}$. Regression is weighted by the natural logarithm of housing units in 1999. For the first-stage F-test of two non-stacked samples, we report Kleibergen-Paap (2006) robust (clustered) statistics and Montiel Olea-Pflueger (2013) efficient statistics. Standard errors are clustered at the CBSA level. ***, **, and * indicate significance at the 1\%, 5\%, and 10\% levels, respectively.
} % end of small font size
} % end of caption
\label{table_Permit.D99t05vsD05t09.HHDTI.4Reg}
\resizebox{\columnwidth}{!}{%
\begin{tabular}{l*{6}{c}}
\toprule
Dep Var (Panel A, B, and C)                      &\multicolumn{5}{c}{Residential Permit Value Growth (99-05 \& 05-09, annualized)} \\
            \cmidrule{2-6} 
            &\multicolumn{1}{c}{(1)}&\multicolumn{1}{c}{(2)}&\multicolumn{1}{c}{(3)}&\multicolumn{1}{c}{(4)}&\multicolumn{1}{c}{(5)}\\

\midrule
\multicolumn{6}{l}{\textbf{Panel A. OLS estimates}} \\
HH Debt-to-Income Rise (99-05, An) x Dum99t05&    0.044*  &    0.082***&    0.065** &    0.041   &    0.039   \\
               &  (0.025)   &  (0.029)   &  (0.029)   &  (0.029)   &  (0.029)   \\
\addlinespace
HH Debt-to-Income Rise (99-05, An) x Dum05t09&   -0.381***&   -0.424***&   -0.339***&   -0.280***&   -0.286***\\
               &  (0.077)   &  (0.083)   &  (0.078)   &  (0.065)   &  (0.064)   \\
\addlinespace
R2-adj         &    0.736   &    0.749   &    0.761   &    0.769   &    0.774   \\
\addlinespace

\midrule
\multicolumn{6}{l}{\textbf{Panel B. Reduced-form estimates}} \\
GIV Net Export Growth (99-05, An) x Dum99t05&   11.050***&   12.899***&    8.282** &    8.856***&    8.263** \\
               &  (3.113)   &  (3.428)   &  (3.334)   &  (3.388)   &  (3.463)   \\
\addlinespace
GIV Net Export Growth (99-05, An) x Dum05t09&  -24.777***&  -22.025***&  -20.374***&  -20.050***&  -20.300***\\
               &  (6.732)   &  (6.424)   &  (6.864)   &  (6.806)   &  (7.156)   \\
\addlinespace
R2-adj         &    0.725   &    0.735   &    0.754   &    0.766   &    0.770   \\
\addlinespace

\midrule
\multicolumn{6}{l}{\textbf{Panel C . 2SLS estimates}} \\
\addlinespace
HH Debt-to-Income Rise (99-05, An) x Dum99t05&    0.919** &    1.080***&    0.853*  &    0.798** &    0.749*  \\
               &  (0.372)   &  (0.390)   &  (0.449)   &  (0.389)   &  (0.391)   \\
\addlinespace
HH Debt-to-Income Rise (99-05, An) x Dum05t09&   -2.061***&   -1.844***&   -2.098** &   -1.808***&   -1.839** \\
               &  (0.718)   &  (0.625)   &  (0.884)   &  (0.693)   &  (0.738)   \\
               
\addlinespace
\addlinespace

Dep Var (Panel D): &\multicolumn{5}{c}{Household Leverage Rise (99-05, An)} \\ 
\midrule 
\multicolumn{6}{l}{\textbf{Panel D . First-stage estimates only for 99-05 (Non-stack sample)}} \\
\addlinespace
GIV NEG (99-05, An)&   12.023***&   11.942***&    9.711***&   11.092***&   11.039***\\
               &  (3.627)   &  (3.310)   &  (3.457)   &  (3.429)   &  (3.626)   \\
\addlinespace
KP F-Stat      &    10.99   &    13.02   &    7.889   &    10.46   &    9.267   \\
MOP F-Stat     &    9.896   &    12.26   &    7.322   &    10.04   &    8.726   \\
\addlinespace

\midrule
\multicolumn{6}{l}{\textbf{Controls (for all Panels)}} \\
DumPeriod  &    Y        &  Y   &   Y    & Y    &  Y    \\
Basic Controls x DumPeriod &            &  Y   &   Y    & Y    &  Y    \\
Housing Controls x DumPeriod &           &      & Y       & Y    &  Y    \\
Demographic Controls x DumPeriod &            &      &        &  Y    &  Y    \\
Industry Controls x DumPeriod &            &      &        &      &  Y    \\

\midrule              
Obs (Panel A, B, and C)          &     1580   &     1580   &     1400   &     1400   &     1400   \\
Obs (Panel D)            &      790   &      790   &      700   &      700   &      700   \\
Cluster SE     &     CBSA   &     CBSA   &     CBSA   &     CBSA   &     CBSA   \\
Weight         & Ln(HU99)   & Ln(HU99)   & Ln(HU99)   & Ln(HU99)   & Ln(HU99)   \\
\bottomrule
\end{tabular}

} % end of resize box

\end{table}

%% file: Table_HH/table_RefineHouse.D00t06vsD07t10.HHDTI.2SLS.wide.tex
%-----------------------------------------------------------------------
%%%%%%%%%%%%%%%%%%%%%%%%%%%%%%%%%%%%
% table_RefineHouse.D00t06vsD07t10.HHDTI.2SLS.wide
%%%%%%%%%%%%%%%%%%%%%%%%%%%%%%%%%%%%

\noindent 

\begin{table}[h!]
\centering
\caption{
\textbf{2SLS Stacked Regression of Refined House Employment Growth in Boom (00-06) and Bust (07-10) Periods on Household Leverage Rise (99-05)} \smallskip \newline
{\scriptsize
This table reports 2SLS regression $\triangle_{00,06} \& \triangle_{07,10} RefinedHouseEmpShr_{c} = \beta_{00,06} * \triangle_{99,05} HHDTI_{c} \times Dum_{00,06} + \beta_{07,10} * \triangle_{99,05} HHDTI_{c} \times Dum_{07,10} + \gamma_{00,06}* \bm{Controls_{c}} \times Dum_{00,06} + \gamma_{07,10}* \bm{Controls_{c}} \times Dum_{07,10} + \epsilon_{period, c}$. The left-hand-side dependent variable $\triangle_{00,06} \& \triangle_{07,10} RefinedHouseEmpShr_{c}$ is the change of the refined house employment share in working-age population at county $c$ 00-06 and 07-10. To reduce the impact of outliers, the dependent variable is winsorized at 5\% and 95\% levels in each period. The key independent variable $\triangle_{99,05} HHDTI_{c}$ is the rise in household leverage (debt-to-income ratio) at county $c$ 99-05. $Controls_{c}$ indicates control variables at county $c$ in the period start year 1999. We use the gravity model-based instrumental variable $\triangle_{99,05}\text{givNetExp}_{m}$ as the IV for $\triangle_{99,05}HHDTI_{c}$. Regression is weighted by the natural logarithm of housing units in 1999.  For the first-stage F-test of two non-stacked samples, we report Kleibergen-Paap (2006) robust (clustered) statistics and Montiel Olea-Pflueger (2013) efficient statistics. Standard errors are clustered at the CBSA level. ***, **, and * indicate significance at the 1\%, 5\%, and 10\% levels, respectively.
\smallskip
} % end of small font size
} % end of caption
\label{table_RefineHouse.D00t06vsD07t10.HHDTI.2SLS.wide}

\vspace{-2mm}

\resizebox{\columnwidth}{!}{%
\begin{tabular}{l*{8}{c}}
\toprule
\textbf{TSLS estimates}            &\multicolumn{8}{c}{Refined House Employment Growth x 100 (00-06 and 07-10, An)} \\
            \cmidrule{2-9} 
            &\multicolumn{2}{c}{(1)}&\multicolumn{2}{c}{(2)}&\multicolumn{2}{c}{(3)}&\multicolumn{2}{c}{(4)}\\
            
\midrule
HH Debt-to-Income Rise (99-05, An) x Dum00t06&    0.446** &  (0.180)&    0.391** &  (0.164)&    0.422*  &  (0.230)&    0.421** &  (0.198)\\  \addlinespace 
HH Debt-to-Income Rise (99-05, An) x Dum07t10&   -0.785***&  (0.254)&   -0.599***&  (0.221)&   -0.680** &  (0.336)&   -0.696** &  (0.318)\\  \addlinespace
Dum00t06       &   -0.025   &  (0.026)&    0.027   &  (0.159)&   -0.162   &  (0.142)&    0.012   &  (0.266)\\  \addlinespace
Dum07t10       &    0.028   &  (0.037)&    0.377   &  (0.238)&    0.680***&  (0.238)&   -0.104   &  (0.419)\\  \addlinespace
Ln(Num of HH, 99) x DumD00t06&            &         &    0.015***&  (0.005)&   -0.046   &  (0.039)&   -0.033   &  (0.030)\\  \addlinespace
Ln(Num of HH, 99) x DumD07t10&            &         &   -0.030***&  (0.008)&    0.075   &  (0.056)&    0.044   &  (0.048)\\  \addlinespace
Ln(HH Income, 99) x DumD00t06&            &         &   -0.016   &  (0.021)&   -0.002   &  (0.019)&   -0.019   &  (0.030)\\  \addlinespace
Ln(HH Income, 99) x DumD07t10&            &         &   -0.009   &  (0.032)&   -0.026   &  (0.032)&    0.046   &  (0.047)\\  \addlinespace
Ratio of Labor Force (1999) x DumD00t06&            &         &   -0.060   &  (0.062)&   -0.044   &  (0.101)&   -0.073   &  (0.110)\\  \addlinespace
Ratio of Labor Force (1999) x DumD07t10&            &         &    0.077   &  (0.101)&    0.017   &  (0.162)&    0.096   &  (0.186)\\  \addlinespace
Ln(Num of HU, 99) x DumD00t06&            &         &            &         &    0.059   &  (0.043)&    0.046   &  (0.033)\\  \addlinespace
Ln(Num of HU, 99) x DumD07t10&            &         &            &         &   -0.106*  &  (0.062)&   -0.076   &  (0.052)\\  \addlinespace
Housing supply elasticity x DumD00t06&            &         &            &         &    0.004   &  (0.004)&    0.003   &  (0.003)\\  \addlinespace
Housing supply elasticity x DumD07t10&            &         &            &         &   -0.006   &  (0.007)&   -0.006   &  (0.006)\\  \addlinespace
House Vacancy Rate (1999) x DumD00t06&            &         &            &         &   -0.029   &  (0.119)&   -0.022   &  (0.103)\\  \addlinespace
House Vacancy Rate (1999) x DumD07t10&            &         &            &         &    0.049   &  (0.166)&    0.053   &  (0.148)\\  \addlinespace
Ratio of Renters (1999) x DumD00t06&            &         &            &         &    0.113   &  (0.093)&    0.120   &  (0.111)\\  \addlinespace
Ratio of Renters (1999) x DumD07t10&            &         &            &         &   -0.148   &  (0.147)&   -0.115   &  (0.181)\\  \addlinespace
Ratio of Bachelor Educated (1999) x DumD00t06&            &         &            &         &            &         &    0.075   &  (0.072)\\  \addlinespace
Ratio of Bachelor Educated (1999) x DumD07t10&            &         &            &         &            &         &   -0.293** &  (0.126)\\  \addlinespace
Ratio of White Race (1999) x DumD00t06&            &         &            &         &            &         &    0.008   &  (0.036)\\  \addlinespace
Ratio of White Race (1999) x DumD07t10&            &         &            &         &            &         &    0.016   &  (0.044)\\  \addlinespace
Ratio of Immigration (90-00) x DumD00t06&            &         &            &         &            &         &   -0.127   &  (0.302)\\  \addlinespace
Ratio of Immigration (90-00) x DumD07t10&            &         &            &         &            &         &    0.322   &  (0.464)\\  \addlinespace
\midrule
Obs            &     1578   &         &     1578   &         &     1396   &         &     1396   &         \\
Cluster SE     &     CBSA   &         &     CBSA   &         &     CBSA   &         &     CBSA   &         \\
Weight         & Ln(HU99)   &         & Ln(HU99)   &         & Ln(HU99)   &         & Ln(HU99)   &         \\
KP F-Stat (99-05, non-stack sample)       &    11.21   &         &    13.66   &         &    7.745   &         &    10.34   &         \\
MOP F-Stat (99-05, non-stack sample)  &    10.11   &         &    12.95   &         &    7.178   &         &    9.921   &         \\
CoefEqual\_Chi2 &    8.508   &         &    7.159   &         &    4.042   &         &    4.957   &         \\
CoefEqual\_PValue &    0.004   &         &    0.007   &         &    0.044   &         &    0.026   &         \\
\bottomrule

\end{tabular}

} % end of resize box

\end{table}

%% file: Table_HH/table_RefineHouse.D00t06vsD07t10.HHDTI.4Reg.tex
%---------------------------------------------------------------

%%%%%%%%%%%%%%%%%%%%%%%%%%%%%%%%%%%%%%%%%%%%%%%%
% table_RefineHouse.D00t06vsD07t10.HHDTI.4Reg
%%%%%%%%%%%%%%%%%%%%%%%%%%%%%%%%%%%%%%%%%%%%%%%%

\noindent 

\begin{table}[h!]
\centering
\caption{
\textbf{Four Stacked Regressions of Refined House Employment Growth in Boom (00-06) and Bust (07-10) Periods on Household Leverage Rise (99-05)} \smallskip \newline
{\scriptsize
This table reports OLS, reduced-form, first stage, and second stages of stacked 2SLS regression $\triangle_{00,06} \& \triangle_{07,10} RefinedHouseEmpShr_{c} = \beta_{00,06} * \triangle_{99,05} HHDTI_{c} \times Dum_{00,06} + \beta_{07,10} * \triangle_{99,05} HHDTI_{c} \times Dum_{07,10} + \gamma_{00,06}* \bm{Controls_{c}} \times Dum_{00,06} + \gamma_{07,10}* \bm{Controls_{c}} \times Dum_{07,10} + \epsilon_{period, c}$. The left-hand-side dependent variable $\triangle_{00,06} \& \triangle_{07,10} RefinedHouseEmpShr_{c}$ is the change of the refined house employment share in working-age population at county $c$ 00-06 and 07-10. To reduce the impact of outliers, the dependent variable is winsorized at 5\% and 95\% levels in each period. The key independent variable $\triangle_{99,05} HHDTI_{c}$ is the rise in household leverage (debt-to-income ratio) at county $c$ 99-05. $Controls_{c}$ indicates control variables at county $c$ in the period start year 1999. We use the gravity model-based instrumental variable $\triangle_{99,05}\text{givNetExp}_{m}$ as the IV for $\triangle_{99,05}HHDTI_{c}$. Regression is weighted by the natural logarithm of housing units in 1999.  For the first-stage F-test of two non-stacked samples, we report Kleibergen-Paap (2006) robust (clustered) statistics and Montiel Olea-Pflueger (2013) efficient statistics. Standard errors are clustered at the CBSA level. ***, **, and * indicate significance at the 1\%, 5\%, and 10\% levels, respectively.
} % end of small font size
} % end of caption
\label{table_RefineHouse.D00t06vsD07t10.HHDTI.4Reg}
\resizebox{0.95\columnwidth}{!}{%
\begin{tabular}{l*{4}{c}}
\toprule
Dep Var (Panel A, B, and C)                      &\multicolumn{4}{c}{Refined House Employment Growth x 100 (00-06 \& 07-10, An)} \\
            \cmidrule{2-5} 
            &\multicolumn{1}{c}{(1)}&\multicolumn{1}{c}{(2)}&\multicolumn{1}{c}{(3)}&\multicolumn{1}{c}{(4)}\\

\midrule
\multicolumn{5}{l}{\textbf{Panel A. OLS estimates}} \\
HH Debt-to-Income Rise (99-05, An) x Dum00t06&    0.046***&    0.049***&    0.030** &    0.012   \\
               &  (0.014)   &  (0.015)   &  (0.015)   &  (0.015)   \\
\addlinespace
HH Debt-to-Income Rise (99-05, An) x Dum07t10&   -0.076***&   -0.087***&   -0.054** &   -0.033   \\
               &  (0.023)   &  (0.024)   &  (0.023)   &  (0.024)   \\
\addlinespace
R2-adj         &    0.561   &    0.609   &    0.624   &    0.635   \\
\addlinespace

\midrule
\multicolumn{5}{l}{\textbf{Panel B. Reduced-form estimates}} \\
GIV Net Export Growth (99-05, An) x Dum00t06&    5.476***&    4.760***&    4.056***&    4.647***\\
               &  (1.426)   &  (1.381)   &  (1.479)   &  (1.446)   \\
\addlinespace
GIV Net Export Growth (99-05, An) x Dum07t10&   -9.654***&   -7.286***&   -6.538***&   -7.686***\\
               &  (2.087)   &  (2.066)   &  (2.142)   &  (2.220)   \\
\addlinespace
R2-adj         &    0.562   &    0.605   &    0.625   &    0.640   \\
\addlinespace

\midrule
\multicolumn{5}{l}{\textbf{Panel C . 2SLS estimates}} \\
\addlinespace
HH Debt-to-Income Rise (99-05, An) x Dum00t06&    0.446** &    0.391** &    0.422*  &    0.421** \\
               &  (0.180)   &  (0.164)   &  (0.230)   &  (0.198)   \\
\addlinespace
HH Debt-to-Income Rise (99-05, An) x Dum07t10&   -0.785***&   -0.599***&   -0.680** &   -0.696** \\
               &  (0.254)   &  (0.221)   &  (0.336)   &  (0.318)   \\
               
\addlinespace
\addlinespace

Dep Var (Panel D): &\multicolumn{4}{c}{Household Leverage Rise (99-05, An)} \\ 
\midrule 
\multicolumn{5}{l}{\textbf{Panel D . First-stage estimates only for 99-05 (Non-stack sample)}} \\
\addlinespace
GIV NEG (99-05, An)&   12.291***&   12.166***&    9.612***&   11.037***\\
               &  (3.671)   &  (3.291)   &  (3.454)   &  (3.432)   \\
\addlinespace
KP F-Stat      &    11.21   &    13.66   &    7.745   &    10.34   \\
MOP F-Stat     &    10.11   &    12.95   &    7.178   &    9.921   \\
\addlinespace

\midrule
\multicolumn{5}{l}{\textbf{Controls (for all Panels)}} \\
DumPeriod  &    Y        &  Y   &   Y    & Y        \\
Basic Controls x DumPeriod &            &  Y   &   Y    & Y    \\
Housing Controls x DumPeriod &           &      & Y       & Y   \\
Demographic Controls x DumPeriod &            &      &      &  Y \\

\midrule              
Obs (Panel A, B, and C)          &     1578   &     1578   &     1396   &     1396   \\
Obs (Panel D)          &      789   &      789   &      698   &      698   \\
Cluster SE     &     CBSA   &     CBSA   &     CBSA   &     CBSA   \\
Weight         & Ln(HU99)   & Ln(HU99)   & Ln(HU99)   & Ln(HU99)   \\
\bottomrule
\end{tabular}

} % end of resize box

\end{table}

%% file: Table_HH/table_BEAConstEmp.D00t06vsD07t10.HHDTI.2SLS.wide.tex
%-----------------------------------------------------------------------
%%%%%%%%%%%%%%%%%%%%%%%%%%%%%%%%%%%%
% table_BEAConstEmp.D00t06vsD07t10.HHDTI.2SLS.wide
%%%%%%%%%%%%%%%%%%%%%%%%%%%%%%%%%%%%

\noindent 

\begin{table}[h!]
\centering
\caption{
\textbf{2SLS Stacked Regression of BEA Construction Employment Growth in Boom (00-06) and Bust (07-10) Periods on Household Leverage Rise (99-05)} \smallskip \newline
{\scriptsize
This table reports 2SLS regression $\triangle_{00,06} \& \triangle_{07,10} BEAConstEmpShr_{c} = \beta_{00,06} * \triangle_{99,05} HHDTI_{c} \times Dum_{00,06} + \beta_{07,10} * \triangle_{99,05} HHDTI_{c} \times Dum_{07,10} + \gamma_{00,06}* \bm{Controls_{c}} \times Dum_{00,06} + \gamma_{07,10}* \bm{Controls_{c}} \times Dum_{07,10} + \epsilon_{period, c}$. The left-hand-side dependent variable $\triangle_{00,06} \& \triangle_{07,10} BEAConstEmpShr_{c}$ is the change of the BEA construction employment share in working-age population at county $c$ 00-06 and 07-10. To reduce the impact of outliers, the dependent variable is winsorized at 2\% and 98\% levels in each period. The key independent variable $\triangle_{99,05} HHDTI_{c}$ is the rise in household leverage (debt-to-income ratio) at county $c$ 99-05. $Controls_{c}$ indicates control variables at county $c$ in the period start year 1999. We use the gravity model-based instrumental variable $\triangle_{99,05}\text{givNetExp}_{m}$ as the IV for $\triangle_{99,05}HHDTI_{c}$. Regression is weighted by the natural logarithm of housing units in 1999. For the first-stage F-test of two non-stacked samples, we report Kleibergen-Paap (2006) robust (clustered) statistics and Montiel Olea-Pflueger (2013) efficient statistics. Standard errors are clustered at the CBSA level. ***, **, and * indicate significance at the 1\%, 5\%, and 10\% levels, respectively.
\smallskip
} % end of small font size
} % end of caption
\label{table_BEAConstEmp.D00t06vsD07t10.HHDTI.2SLS.wide}

\vspace{-2mm}

\resizebox{\columnwidth}{!}{%
\begin{tabular}{l*{8}{c}}
\toprule
\textbf{TSLS estimates}            &\multicolumn{8}{c}{BEA Construction Employment Growth (00-06 and 07-10, An)} \\
            \cmidrule{2-9} 
            &\multicolumn{2}{c}{(1)}&\multicolumn{2}{c}{(2)}&\multicolumn{2}{c}{(3)}&\multicolumn{2}{c}{(4)}\\
            
\midrule
HH Debt-to-Income Rise (99-05, An) x Dum00t06&    0.022** &  (0.011)&    0.020** &  (0.009)&    0.018   &  (0.011)&    0.019** &  (0.010)\\ 
\addlinespace
HH Debt-to-Income Rise (99-05, An) x Dum07t10&   -0.030***&  (0.012)&   -0.023** &  (0.010)&   -0.024*  &  (0.014)&   -0.025** &  (0.011)\\  \addlinespace
Dum00t06       &   -0.002   &  (0.002)&    0.009   &  (0.010)&   -0.005   &  (0.006)&    0.012   &  (0.013)\\  \addlinespace
Dum07t10       &    0.000   &  (0.002)&    0.006   &  (0.011)&    0.012   &  (0.009)&   -0.015   &  (0.018)\\  \addlinespace
Ln(Num of HH, 99) x DumD00t06&            &         &    0.001*  &  (0.000)&   -0.003*  &  (0.002)&   -0.001   &  (0.002)\\  \addlinespace
Ln(Num of HH, 99) x DumD07t10&            &         &   -0.001***&  (0.000)&    0.001   &  (0.002)&   -0.001   &  (0.002)\\  \addlinespace
Ln(HH Income, 99) x DumD00t06&            &         &   -0.001   &  (0.001)&   -0.000   &  (0.001)&   -0.002   &  (0.001)\\  \addlinespace
Ln(HH Income, 99) x DumD07t10&            &         &    0.000   &  (0.001)&    0.000   &  (0.001)&    0.003   &  (0.002)\\  \addlinespace
Ratio of Labor Force (1999) x DumD00t06&            &         &   -0.006** &  (0.003)&   -0.005   &  (0.004)&   -0.009*  &  (0.005)\\  \addlinespace
Ratio of Labor Force (1999) x DumD07t10&            &         &    0.000   &  (0.004)&   -0.003   &  (0.007)&    0.000   &  (0.007)\\  \addlinespace
Ln(Num of HU, 99) x DumD00t06&            &         &            &         &    0.003** &  (0.002)&    0.002   &  (0.002)\\  \addlinespace
Ln(Num of HU, 99) x DumD07t10&            &         &            &         &   -0.002   &  (0.002)&   -0.001   &  (0.002)\\  \addlinespace
Housing supply elasticity x DumD00t06&            &         &            &         &    0.000   &  (0.000)&    0.000   &  (0.000)\\  \addlinespace
Housing supply elasticity x DumD07t10&            &         &            &         &   -0.000   &  (0.000)&   -0.000   &  (0.000)\\  \addlinespace
House Vacancy Rate (1999) x DumD00t06&            &         &            &         &   -0.002   &  (0.005)&   -0.001   &  (0.004)\\  \addlinespace
House Vacancy Rate (1999) x DumD07t10&            &         &            &         &   -0.003   &  (0.006)&   -0.003   &  (0.005)\\  \addlinespace
Ratio of Renters (1999) x DumD00t06&            &         &            &         &    0.002   &  (0.004)&    0.004   &  (0.005)\\  \addlinespace
Ratio of Renters (1999) x DumD07t10&            &         &            &         &   -0.000   &  (0.005)&    0.001   &  (0.006)\\  \addlinespace
Ratio of Bachelor Educated (1999) x DumD00t06&            &         &            &         &            &         &    0.007*  &  (0.004)\\  \addlinespace
Ratio of Bachelor Educated (1999) x DumD07t10&            &         &            &         &            &         &   -0.010** &  (0.005)\\  \addlinespace
Ratio of White Race (1999) x DumD00t06&            &         &            &         &            &         &    0.002   &  (0.001)\\  \addlinespace
Ratio of White Race (1999) x DumD07t10&            &         &            &         &            &         &    0.001   &  (0.002)\\  \addlinespace
Ratio of Immigration (90-00) x DumD00t06&            &         &            &         &            &         &   -0.011   &  (0.012)\\  \addlinespace
Ratio of Immigration (90-00) x DumD07t10&            &         &            &         &            &         &    0.011   &  (0.015)\\  \addlinespace
\midrule
Obs            &     1530   &         &     1530   &         &     1368   &         &     1368   &         \\
Cluster SE     &     CBSA   &         &     CBSA   &         &     CBSA   &         &     CBSA   &         \\
Weight         & Ln(HU99)   &         & Ln(HU99)   &         & Ln(HU99)   &         & Ln(HU99)   &         \\
KP F-Stat (99-05, non-stack sample)       &    8.888   &         &    11.69   &         &    5.655   &         &    7.386   &         \\
MOP F-Stat (99-05, non-stack sample)  &    8.363   &         &    11.32   &         &    5.309   &         &    7.208   &         \\
CoefEqual\_Chi2 &    6.148   &         &    6.467   &         &    3.477   &         &    5.416   &         \\
CoefEqual\_PValue &    0.013   &         &    0.011   &         &    0.062   &         &    0.020   &         \\
\bottomrule

\end{tabular}

} % end of resize box

\end{table}

%% file: Table_HH/table_BEAConstEmp.D00t06vsD07t10.HHDTI.4Reg.tex
%---------------------------------------------------------------

%%%%%%%%%%%%%%%%%%%%%%%%%%%%%%%%%%%%%%%%%%%%%%%%
% table_BEAConstEmp.D00t06vsD07t10.HHDTI.4Reg
%%%%%%%%%%%%%%%%%%%%%%%%%%%%%%%%%%%%%%%%%%%%%%%%

\noindent 

\begin{table}[h!]
\centering
\caption{
\textbf{Four Stacked Regressions of BEA Construction Employment Growth in Boom (00-06) and Bust (07-10) Periods on Household Leverage Rise (99-05)} \smallskip \newline
{\scriptsize
This table reports OLS, reduced-form, first stage, and second stages of stacked 2SLS regression $\triangle_{00,06} \& \triangle_{07,10} BEAConstEmpShr_{c} = \beta_{00,06} * \triangle_{99,05} HHDTI_{c} \times Dum_{00,06} + \beta_{07,10} * \triangle_{99,05} HHDTI_{c} \times Dum_{07,10} + \gamma_{00,06}* \bm{Controls_{c}} \times Dum_{00,06} + \gamma_{07,10}* \bm{Controls_{c}} \times Dum_{07,10} + \epsilon_{period, c}$. The left-hand-side dependent variable $\triangle_{00,06} \& \triangle_{07,10} BEAConstEmpShr_{c}$ is the change of the BEA construction employment share in working-age population at county $c$ 00-06 and 07-10. To reduce the impact of outliers, the dependent variable is winsorized at 2\% and 98\% levels in each period. The key independent variable $\triangle_{99,05} HHDTI_{c}$ is the rise in household leverage (debt-to-income ratio) at county $c$ 99-05. $Controls_{c}$ indicates control variables at county $c$ in the period start year 1999. We use the gravity model-based instrumental variable $\triangle_{99,05}\text{givNetExp}_{m}$ as the IV for $\triangle_{99,05}HHDTI_{c}$. Regression is weighted by the natural logarithm of housing units in 1999. For the first-stage F-test of two non-stacked samples, we report Kleibergen-Paap (2006) robust (clustered) statistics and Montiel Olea-Pflueger (2013) efficient statistics. Standard errors are clustered at the CBSA level. ***, **, and * indicate significance at the 1\%, 5\%, and 10\% levels, respectively.
} % end of small font size
} % end of caption
\label{table_BEAConstEmp.D00t06vsD07t10.HHDTI.4Reg}
\resizebox{0.95\columnwidth}{!}{%
\begin{tabular}{l*{4}{c}}
\toprule
Dep Var (Panel A, B, and C)                      &\multicolumn{4}{c}{BEA Construction Employment Growth (00-06 \& 07-10, An)} \\
            \cmidrule{2-5} 
            &\multicolumn{1}{c}{(1)}&\multicolumn{1}{c}{(2)}&\multicolumn{1}{c}{(3)}&\multicolumn{1}{c}{(4)}\\

\midrule
\multicolumn{5}{l}{\textbf{Panel A. OLS estimates}} \\
HH Debt-to-Income Rise (99-05, An) x Dum00t06&    0.002***&    0.002** &    0.001   &    0.001   \\
               &  (0.001)   &  (0.001)   &  (0.001)   &  (0.001)   \\
\addlinespace
HH Debt-to-Income Rise (99-05, An) x Dum07t10&   -0.006***&   -0.006***&   -0.004***&   -0.003** \\
               &  (0.001)   &  (0.001)   &  (0.001)   &  (0.001)   \\
\addlinespace
R2-adj         &    0.596   &    0.614   &    0.629   &    0.634   \\
\addlinespace

\midrule
\multicolumn{5}{l}{\textbf{Panel B. Reduced-form estimates}} \\
GIV Net Export Growth (99-05, An) x Dum00t06&    0.223***&    0.217***&    0.148*  &    0.175** \\
               &  (0.073)   &  (0.070)   &  (0.076)   &  (0.077)   \\
\addlinespace
GIV Net Export Growth (99-05, An) x Dum07t10&   -0.307***&   -0.254** &   -0.190   &   -0.232*  \\
               &  (0.108)   &  (0.099)   &  (0.119)   &  (0.119)   \\
\addlinespace
R2-adj         &    0.585   &    0.603   &    0.625   &    0.633   \\
\addlinespace

\midrule
\multicolumn{5}{l}{\textbf{Panel C . 2SLS estimates}} \\
\addlinespace
HH Debt-to-Income Rise (99-05, An) x Dum00t06&    0.022** &    0.020** &    0.018   &    0.019** \\
               &  (0.011)   &  (0.009)   &  (0.011)   &  (0.010)   \\
\addlinespace
HH Debt-to-Income Rise (99-05, An) x Dum07t10&   -0.030***&   -0.023** &   -0.024*  &   -0.025** \\
               &  (0.012)   &  (0.010)   &  (0.014)   &  (0.011)   \\
               
\addlinespace
\addlinespace

Dep Var (Panel D): &\multicolumn{4}{c}{Household Leverage Rise (99-05, An)} \\ 
\midrule 
\multicolumn{5}{l}{\textbf{Panel D . First-stage estimates only for 99-05 (Non-stack sample)}} \\
\addlinespace
GIV NEG (99-05, An)&   10.158***&   10.879***&    8.070** &    9.290***\\
               &  (3.407)   &  (3.182)   &  (3.393)   &  (3.418)   \\
\addlinespace
KP F-Stat      &    8.888   &    11.69   &    5.655   &    7.386   \\
MOP F-Stat     &    8.363   &    11.32   &    5.309   &    7.208   \\
\addlinespace

\midrule
\multicolumn{5}{l}{\textbf{Controls (for all Panels)}} \\
DumPeriod  &    Y        &  Y   &   Y    & Y        \\
Basic Controls x DumPeriod &            &  Y   &   Y    & Y    \\
Housing Controls x DumPeriod &           &      & Y       & Y   \\
Demographic Controls x DumPeriod &            &      &      &  Y \\

\midrule              
Obs (Panel A, B, and C)          &     1530   &     1530   &     1368   &     1368   \\
Obs (Panel D)          &      765   &      765   &      684   &      684   \\
Cluster SE     &     CBSA   &     CBSA   &     CBSA   &     CBSA   \\
Weight         & Ln(HU99)   & Ln(HU99)   & Ln(HU99)   & Ln(HU99)   \\
\bottomrule
\end{tabular}

} % end of resize box

\end{table}

%% file: Table_HH/table_Tradable.D00t06vsD07t10.NEP.D99t05.2SLS.wide.tex
%-----------------------------------------------------------------------
%%%%%%%%%%%%%%%%%%%%%%%%%%%%%%%%%%%%
% table_Tradable.D00t06vsD07t10.NEP.D99t05.2SLS.wide
%%%%%%%%%%%%%%%%%%%%%%%%%%%%%%%%%%%%

\noindent 

\begin{table}[h!]
\centering
\caption{
\textbf{2SLS Stacked Regression of Tradable Employment Growth in Boom (00-06) and Bust (07-10) Periods on Net Export Growth (99-05)} \smallskip \newline
{\scriptsize
This table reports 2SLS regression $\triangle_{00,06} \& \triangle_{07,10} TradableEmpShr_{c} = \beta_{00,06} * \triangle_{99,05} \text{NetExp}_{m} \times Dum_{00,06} + \beta_{07,10} * \triangle_{99,05} \text{NetExp}_{m} \times Dum_{07,10} + \gamma_{00,06}* \bm{Controls_{c}} \times Dum_{00,06} + \gamma_{07,10}* \bm{Controls_{c}} \times Dum_{07,10} + \epsilon_{period, c}$. The left-hand-side dependent variable $\triangle_{00,06} \& \triangle_{07,10} TradableEmpShr_{c}$ is the change of the Tradable employment share in working-age population at county $c$ 00-06 and 07-10. The key independent variable $\triangle_{99,05} \text{NetExp}_{m}$ is the growth rate of net export growth at county $c$ 99-05. $Controls_{c}$ indicates control variables at county $c$ in the period start year 1999. We use the gravity model-based instrumental variable $\triangle_{99,05}\text{givNetExp}_{m}$ as the IV for $\triangle_{99,05}\text{NetExp}_{m}$. Regression is weighted by the natural logarithm of housing units in 1999.  For the first-stage F-test of two non-stacked samples, we report Kleibergen-Paap (2006) robust (clustered) statistics and Montiel Olea-Pflueger (2013) efficient statistics. Standard errors are clustered at the CBSA level. ***, **, and * indicate significance at the 1\%, 5\%, and 10\% levels, respectively.
\smallskip
} % end of small font size
} % end of caption
\label{table_Tradable.D00t06vsD07t10.NEP.D99t05.2SLS.wide}

\vspace{-2mm}

\resizebox{\columnwidth}{!}{%
\begin{tabular}{l*{8}{c}}
\toprule
\textbf{TSLS estimates}            &\multicolumn{8}{c}{Tradable Employment Growth (00-06 and 07-10, An)} \\
            \cmidrule{2-9} 
            &\multicolumn{2}{c}{(1)}&\multicolumn{2}{c}{(2)}&\multicolumn{2}{c}{(3)}&\multicolumn{2}{c}{(4)} \\
            
\midrule
Net Export Growth (99-05, An) x Dum00t06&    0.790***&  (0.191)&    0.804***&  (0.191)&    0.701***&  (0.173)&    0.742***&  (0.174)\\ 
\addlinespace
Net Export Growth (99-05, An) x Dum07t10&    0.783***&  (0.215)&    0.814***&  (0.228)&    0.776***&  (0.249)&    0.835***&  (0.259)\\  \addlinespace
Dum00t06       &   -0.001***&  (0.000)&    0.004   &  (0.006)&    0.007   &  (0.007)&    0.045***&  (0.016)\\  \addlinespace
Dum07t10       &   -0.002***&  (0.000)&    0.009   &  (0.007)&    0.004   &  (0.009)&    0.057***&  (0.019)\\  \addlinespace
Ln(Num of HH, 99) x DumD00t06&            &         &   -0.000   &  (0.000)&    0.006*  &  (0.004)&    0.007*  &  (0.004)\\  \addlinespace
Ln(Num of HH, 99) x DumD07t10&            &         &   -0.000   &  (0.000)&    0.000   &  (0.002)&    0.000   &  (0.002)\\  \addlinespace
Ln(HH Income, 99) x DumD00t06&            &         &   -0.000   &  (0.001)&   -0.001   &  (0.001)&   -0.004***&  (0.002)\\  \addlinespace
Ln(HH Income, 99) x DumD07t10&            &         &   -0.000   &  (0.001)&   -0.000   &  (0.001)&   -0.005***&  (0.002)\\  \addlinespace
Ratio of Labor Force (1999) x DumD00t06&            &         &    0.000   &  (0.004)&   -0.004   &  (0.004)&   -0.005   &  (0.004)\\  \addlinespace
Ratio of Labor Force (1999) x DumD07t10&            &         &   -0.009** &  (0.004)&   -0.006   &  (0.004)&   -0.006   &  (0.005)\\  \addlinespace
Ln(Num of HU, 99) x DumD00t06&            &         &            &         &   -0.006*  &  (0.004)&   -0.007*  &  (0.004)\\  \addlinespace
Ln(Num of HU, 99) x DumD07t10&            &         &            &         &   -0.000   &  (0.002)&   -0.001   &  (0.002)\\  \addlinespace
Housing supply elasticity x DumD00t06&            &         &            &         &    0.000   &  (0.000)&    0.000   &  (0.000)\\  \addlinespace
Housing supply elasticity x DumD07t10&            &         &            &         &   -0.000   &  (0.000)&   -0.000   &  (0.000)\\  \addlinespace
House Vacancy Rate (1999) x DumD00t06&            &         &            &         &    0.012** &  (0.006)&    0.011** &  (0.006)\\  \addlinespace
House Vacancy Rate (1999) x DumD07t10&            &         &            &         &    0.009** &  (0.004)&    0.007*  &  (0.004)\\  \addlinespace
Ratio of Renters (1999) x DumD00t06&            &         &            &         &   -0.002   &  (0.002)&   -0.006** &  (0.003)\\  \addlinespace
Ratio of Renters (1999) x DumD07t10&            &         &            &         &    0.001   &  (0.002)&   -0.006** &  (0.003)\\  \addlinespace
Ratio of Bachelor Educated (1999) x DumD00t06&            &         &            &         &            &         &    0.011** &  (0.005)\\  \addlinespace
Ratio of Bachelor Educated (1999) x DumD07t10&            &         &            &         &            &         &    0.014***&  (0.005)\\  \addlinespace
Ratio of White Race (1999) x DumD00t06&            &         &            &         &            &         &    0.001   &  (0.001)\\  \addlinespace
Ratio of White Race (1999) x DumD07t10&            &         &            &         &            &         &   -0.001   &  (0.002)\\  \addlinespace
Ratio of Immigration (90-00) x DumD00t06&            &         &            &         &            &         &    0.012   &  (0.008)\\  \addlinespace
Ratio of Immigration (90-00) x DumD07t10&            &         &            &         &            &         &    0.022** &  (0.009)\\
\midrule
Obs             &     1578   &         &     1578   &         &     1396   &         &     1396   &         \\
Cluster SE      &     CBSA   &         &     CBSA   &         &     CBSA   &         &     CBSA   &         \\
Weight         & Ln(HU99)   &         & Ln(HU99)   &         & Ln(HU99)   &         & Ln(HU99)   &         \\
KP F-Stat (99-05, non-stack sample)      &    23.44   &         &    23.20   &         &    16.52   &         &    17.83   &         \\
MOP F-Stat (99-05, non-stack sample)  &    22.98   &         &    22.71   &         &    16.16   &         &    17.49   &         \\
CoefEqual\_Chi2  &    0.001   &         &    0.002   &         &    0.117   &         &    0.183   &         \\
CoefEqual\_PValue &    0.973   &         &    0.961   &         &    0.732   &         &    0.669   &         \\
\bottomrule

\end{tabular}

} % end of resize box

\end{table}

%% file: Table_HH/table_Tradable.D00t06vsD07t10.NEP.D99t05.4Reg.tex
%---------------------------------------------------------------

%%%%%%%%%%%%%%%%%%%%%%%%%%%%%%%%%%%%%%%%%%%%%%%%
% table_Tradable.D00t06vsD07t10.NEP.D99t05.4Reg
%%%%%%%%%%%%%%%%%%%%%%%%%%%%%%%%%%%%%%%%%%%%%%%%

\noindent 

\begin{table}[h!]
\centering
\caption{
\textbf{Four Stacked Regressions of Tradable Employment Growth in Boom (00-06) and Bust (07-10) Periods on Net Export Growth (99-05)} \smallskip \newline
{\scriptsize
This table reports OLS, reduced-form, first stage, and second stages of stacked 2SLS regression $\triangle_{00,06} \& \triangle_{07,10} TradableEmpShr_{c} = \beta_{00,06} * \triangle_{99,05} \text{NetExp}_{m} \times Dum_{00,06} + \beta_{07,10} * \triangle_{99,05} \text{NetExp}_{m} \times Dum_{07,10} + \gamma_{00,06}* \bm{Controls_{c}} \times Dum_{00,06} + \gamma_{07,10}* \bm{Controls_{c}} \times Dum_{07,10} + \epsilon_{period, c}$. The left-hand-side dependent variable $\triangle_{00,06} \& \triangle_{07,10} TradableEmpShr_{c}$ is the change of the Tradable employment share in working-age population at county $c$ 00-06 and 07-10. The key independent variable $\triangle_{99,05} \text{NetExp}_{m}$ is the growth rate of net export growth in metropolitan area $m$ 99-05. $Controls_{c}$ indicates control variables at county $c$ in the period start year 1999. We use the gravity model-based instrumental variable $\triangle_{99,05}\text{givNetExp}_{m}$ as the IV for $\triangle_{99,05}\text{NetExp}_{m}$. Regression is weighted by the natural logarithm of housing units in 1999.  For the first-stage F-test of two non-stacked samples, we report Kleibergen-Paap (2006) robust (clustered) statistics and Montiel Olea-Pflueger (2013) efficient statistics. Standard errors are clustered at the CBSA level. ***, **, and * indicate significance at the 1\%, 5\%, and 10\% levels, respectively.
} % end of small font size
} % end of caption
\label{table_Tradable.D00t06vsD07t10.NEP.D99t05.4Reg}
\resizebox{0.95\columnwidth}{!}{%
\begin{tabular}{l*{4}{c}}
\toprule
Dep Var (Panel A, B, and C)                      &\multicolumn{4}{c}{Tradable Employment Growth (00-06 \& 07-10, An)} \\
            \cmidrule{2-5} 
            &\multicolumn{1}{c}{(1)}&\multicolumn{1}{c}{(2)}&\multicolumn{1}{c}{(3)}&\multicolumn{1}{c}{(4)} \\

\midrule
\multicolumn{5}{l}{\textbf{Panel A. OLS estimates}} \\
Net Export Growth (99-05, An) x Dum00t06&    0.537***&    0.544***&    0.472***&    0.451***\\
               &  (0.104)   &  (0.103)   &  (0.099)   &  (0.097)   \\
\addlinespace
Net Export Growth (99-05, An) x Dum07t10&    0.625***&    0.638***&    0.599***&    0.576***\\
               &  (0.149)   &  (0.155)   &  (0.163)   &  (0.154)   \\
\addlinespace
R2-adj         &    0.427   &    0.430   &    0.449   &    0.459   \\
\addlinespace

\midrule
\multicolumn{5}{l}{\textbf{Panel B. Reduced-form estimates}} \\
GIV Net Export Growth (99-05, An) x Dum00t06&    0.910***&    0.921***&    0.772***&    0.831***\\
               &  (0.187)   &  (0.183)   &  (0.180)   &  (0.168)   \\
\addlinespace
GIV Net Export Growth (99-05, An) x Dum07t10&    0.903***&    0.932***&    0.854***&    0.936***\\
               &  (0.176)   &  (0.179)   &  (0.188)   &  (0.190)   \\
\addlinespace
R2-adj         &    0.413   &    0.416   &    0.437   &    0.454   \\
\addlinespace

\midrule
\multicolumn{5}{l}{\textbf{Panel C . 2SLS estimates}} \\
\addlinespace
Net Export Growth (99-05, An) x Dum00t06&    0.790***&    0.804***&    0.701***&    0.742***\\
               &  (0.191)   &  (0.191)   &  (0.173)   &  (0.174)   \\
\addlinespace
Net Export Growth (99-05, An) x Dum07t10&    0.783***&    0.814***&    0.776***&    0.835***\\
               &  (0.215)   &  (0.228)   &  (0.249)   &  (0.259)   \\
               
\addlinespace
\addlinespace

Dep Var (Panel D): &\multicolumn{4}{c}{Net Export Growth (99-05, An)} \\ 
\midrule 
\multicolumn{5}{l}{\textbf{Panel D . First-stage estimates only for 99-05 (Non-stack sample)}} \\
\addlinespace
GIV NEG (99-05, An)&    1.153***&    1.145***&    1.101***&    1.121***\\
               &  (0.238)   &  (0.238)   &  (0.271)   &  (0.265)   \\
\addlinespace
KP F-Stat      &    23.44   &    23.20   &    16.52   &    17.83   \\
MOP F-Stat     &    22.98   &    22.71   &    16.16   &    17.49   \\
\addlinespace

\midrule
\multicolumn{5}{l}{\textbf{Controls (for all Panels)}} \\
DumPeriod  &    Y        &  Y   &   Y    & Y        \\
Basic Controls x DumPeriod &            &  Y   &   Y    & Y    \\
Housing Controls x DumPeriod &           &      & Y       & Y   \\
Demographic Controls x DumPeriod &            &      &      &  Y \\

\midrule              
Obs (Panel A, B, and C)          &     1578   &     1578   &     1396   &     1396   \\
Obs (Panel D)          &      789   &      789   &      698   &      698   \\
Cluster SE     &     CBSA   &     CBSA   &     CBSA   &     CBSA    \\
Weight         & Ln(HU99)   & Ln(HU99)   & Ln(HU99)   & Ln(HU99)    \\
\bottomrule
\end{tabular}

} % end of resize box

\end{table}

%% file: Table_HH/table_ComConstEmp.D00t06vsD07t10.NEP.D99t05.2SLS.wide.tex
%-----------------------------------------------------------------------
%%%%%%%%%%%%%%%%%%%%%%%%%%%%%%%%%%%%
% table_ComConstEmp.D00t06vsD07t10.NEP.D99t05.2SLS.wide
%%%%%%%%%%%%%%%%%%%%%%%%%%%%%%%%%%%%

\noindent 

\begin{table}[h!]
\centering
\caption{
\textbf{2SLS Stacked Regression of Commercial Construction Employment Growth in Boom (00-06) and Bust (07-10) Periods on Net Export Growth (99-05)} \smallskip \newline
{\scriptsize
This table reports 2SLS regression $\triangle_{00,06} \& \triangle_{07,10} ComConstEmpShr_{c} = \beta_{00,06} * \triangle_{99,05} \text{NetExp}_{m} \times Dum_{00,06} + \beta_{07,10} * \triangle_{99,05} \text{NetExp}_{m} \times Dum_{07,10} + \gamma_{00,06}* \bm{Controls_{c}} \times Dum_{00,06} + \gamma_{07,10}* \bm{Controls_{c}} \times Dum_{07,10} + \epsilon_{period, c}$. The left-hand-side dependent variable $\triangle_{00,06} \& \triangle_{07,10} ComConstEmpShr_{c}$ is the change of the commercial construction employment share in working-age population at county $c$ 00-06 and 07-10. The key independent variable $\triangle_{99,05} \text{NetExp}_{m}$ is the growth rate of net export growth at county $c$ 99-05. $Controls_{c}$ indicates control variables at county $c$ in the period start year 1999. We use the gravity model-based instrumental variable $\triangle_{99,05}\text{givNetExp}_{m}$ as the IV for $\triangle_{99,05}\text{NetExp}_{m}$. Regression is weighted by the natural logarithm of housing units in 1999.  For the first-stage F-test of two non-stacked samples, we report Kleibergen-Paap (2006) robust (clustered) statistics and Montiel Olea-Pflueger (2013) efficient statistics. Standard errors are clustered at the CBSA level. ***, **, and * indicate significance at the 1\%, 5\%, and 10\% levels, respectively.
\smallskip
} % end of small font size
} % end of caption
\label{table_ComConstEmp.D00t06vsD07t10.NEP.D99t05.2SLS.wide}

\vspace{-2mm}

\resizebox{\columnwidth}{!}{%
\begin{tabular}{l*{8}{c}}
\toprule
\textbf{TSLS estimates}            &\multicolumn{8}{c}{Commercial Construction Employment Growth (00-06 and 07-10, An)} \\
            \cmidrule{2-9} 
            &\multicolumn{2}{c}{(1)}&\multicolumn{2}{c}{(2)}&\multicolumn{2}{c}{(3)}&\multicolumn{2}{c}{(4)} \\
            
\midrule
Net Export Growth (99-05, An) x Dum00t06&    0.684   &  (1.519)&    0.605   &  (1.564)&    0.036   &  (1.579)&    0.011   &  (1.529)\\ 
\addlinespace
Net Export Growth (99-05, An) x Dum07t10&    0.076   &  (1.633)&    0.592   &  (1.723)&    0.124   &  (2.046)&   -0.161   &  (2.028)\\  \addlinespace
Dum00t06       &    0.004   &  (0.004)&   -0.155** &  (0.068)&   -0.160*  &  (0.085)&   -0.230   &  (0.143)\\  \addlinespace
Dum07t10       &   -0.015***&  (0.005)&    0.492***&  (0.130)&    0.671***&  (0.148)&    0.619** &  (0.289)\\  \addlinespace
Ln(Num of HH, 99) x DumD00t06&            &         &   -0.002   &  (0.002)&   -0.023   &  (0.028)&   -0.031   &  (0.029)\\  \addlinespace
Ln(Num of HH, 99) x DumD07t10&            &         &    0.004   &  (0.003)&    0.055** &  (0.028)&    0.054*  &  (0.032)\\  \addlinespace
Ln(HH Income, 99) x DumD00t06&            &         &    0.019** &  (0.008)&    0.014   &  (0.009)&    0.022   &  (0.014)\\  \addlinespace
Ln(HH Income, 99) x DumD07t10&            &         &   -0.054***&  (0.014)&   -0.066***&  (0.016)&   -0.059** &  (0.028)\\  \addlinespace
Ratio of Labor Force (1999) x DumD00t06&            &         &   -0.038   &  (0.028)&    0.004   &  (0.034)&    0.020   &  (0.037)\\  \addlinespace
Ratio of Labor Force (1999) x DumD07t10&            &         &    0.049   &  (0.065)&   -0.011   &  (0.063)&   -0.011   &  (0.065)\\  \addlinespace
Ln(Num of HU, 99) x DumD00t06&            &         &            &         &    0.023   &  (0.028)&    0.031   &  (0.029)\\  \addlinespace
Ln(Num of HU, 99) x DumD07t10&            &         &            &         &   -0.052*  &  (0.027)&   -0.051*  &  (0.031)\\  \addlinespace
Housing supply elasticity x DumD00t06&            &         &            &         &   -0.001   &  (0.002)&   -0.001   &  (0.002)\\  \addlinespace
Housing supply elasticity x DumD07t10&            &         &            &         &    0.000   &  (0.003)&   -0.000   &  (0.003)\\  \addlinespace
House Vacancy Rate (1999) x DumD00t06&            &         &            &         &    0.057   &  (0.047)&    0.051   &  (0.049)\\  \addlinespace
House Vacancy Rate (1999) x DumD07t10&            &         &            &         &   -0.027   &  (0.075)&   -0.028   &  (0.077)\\  \addlinespace
Ratio of Renters (1999) x DumD00t06&            &         &            &         &   -0.022   &  (0.022)&   -0.030   &  (0.029)\\  \addlinespace
Ratio of Renters (1999) x DumD07t10&            &         &            &         &   -0.018   &  (0.035)&   -0.015   &  (0.048)\\  \addlinespace
Ratio of Bachelor Educated (1999) x DumD00t06&            &         &            &         &            &         &   -0.031   &  (0.037)\\  \addlinespace
Ratio of Bachelor Educated (1999) x DumD07t10&            &         &            &         &            &         &    0.000   &  (0.073)\\  \addlinespace
Ratio of White Race (1999) x DumD00t06&            &         &            &         &            &         &   -0.011   &  (0.013)\\  \addlinespace
Ratio of White Race (1999) x DumD07t10&            &         &            &         &            &         &   -0.027   &  (0.025)\\  \addlinespace
Ratio of Immigration (90-00) x DumD00t06&            &         &            &         &            &         &    0.048   &  (0.066)\\  \addlinespace
Ratio of Immigration (90-00) x DumD07t10&            &         &            &         &            &         &   -0.160   &  (0.119)\\  
\midrule
Obs            &     1520   &         &     1520   &         &     1340   &         &     1340   &         \\
Cluster SE      &     CBSA   &         &     CBSA   &         &     CBSA   &         &     CBSA   &         \\
Weight         & Ln(HU99)   &         & Ln(HU99)   &         & Ln(HU99)   &         & Ln(HU99)   &         \\
KP F-Stat (99-05, non-stack sample)      &    22.58   &         &    22.23   &         &    15.88   &         &    17.18   &         \\
MOP F-Stat (99-05, non-stack sample)  &    22.01   &         &    21.62   &         &    15.43   &         &    16.74   &         \\
CoefEqual\_Chi2  &    0.048   &         &    0.000   &         &    0.001   &         &    0.003   &         \\
CoefEqual\_PValue &    0.826   &         &    0.997   &         &    0.979   &         &    0.958   &         \\
\bottomrule

\end{tabular}

} % end of resize box

\end{table}

%% file: Table_HH/table_ComConstEmp.D00t06vsD07t10.NEP.D99t05.4Reg.tex
%---------------------------------------------------------------

%%%%%%%%%%%%%%%%%%%%%%%%%%%%%%%%%%%%%%%%%%%%%%%%
% table_ComConstEmp.D00t06vsD07t10.NEP.D99t05.4Reg
%%%%%%%%%%%%%%%%%%%%%%%%%%%%%%%%%%%%%%%%%%%%%%%%

\noindent 

\begin{table}[h!]
\centering
\caption{
\textbf{Four Stacked Regressions of Commercial Construction Employment Growth in Boom (00-06) and Bust (07-10) Periods on Net Export Growth (99-05)} \smallskip \newline
{\scriptsize
This table reports OLS, reduced-form, first stage, and second stages of stacked 2SLS regression $\triangle_{00,06} \& \triangle_{07,10} ComConstEmpShr_{c} = \beta_{00,06} * \triangle_{99,05} \text{NetExp}_{m} \times Dum_{00,06} + \beta_{07,10} * \triangle_{99,05} \text{NetExp}_{m} \times Dum_{07,10} + \gamma_{00,06}* \bm{Controls_{c}} \times Dum_{00,06} + \gamma_{07,10}* \bm{Controls_{c}} \times Dum_{07,10} + \epsilon_{period, c}$. The left-hand-side dependent variable $\triangle_{00,06} \& \triangle_{07,10} ComConstEmpShr_{c}$ is the change of the commercial construction employment share in working-age population at county $c$ 00-06 and 07-10. The key independent variable $\triangle_{99,05} \text{NetExp}_{m}$ is the growth rate of net export growth at county $c$ 99-05. $Controls_{c}$ indicates control variables at county $c$ in the period start year 1999. We use the gravity model-based instrumental variable $\triangle_{99,05}\text{givNetExp}_{m}$ as the IV for $\triangle_{99,05}\text{NetExp}_{m}$. Regression is weighted by the natural logarithm of housing units in 1999.  For the first-stage F-test of two non-stacked samples, we report Kleibergen-Paap (2006) robust (clustered) statistics and Montiel Olea-Pflueger (2013) efficient statistics. Standard errors are clustered at the CBSA level. ***, **, and * indicate significance at the 1\%, 5\%, and 10\% levels, respectively.
} % end of small font size
} % end of caption
\label{table_ComConstEmp.D00t06vsD07t10.NEP.D99t05.4Reg}
\resizebox{0.95\columnwidth}{!}{%
\begin{tabular}{l*{4}{c}}
\toprule
Dep Var (Panel A, B, and C)                      &\multicolumn{4}{c}{Commercial Construction Employment Growth (00-06 \& 07-10, An)} \\
            \cmidrule{2-5} 
            &\multicolumn{1}{c}{(1)}&\multicolumn{1}{c}{(2)}&\multicolumn{1}{c}{(3)}&\multicolumn{1}{c}{(4)} \\

\midrule
\multicolumn{5}{l}{\textbf{Panel A. OLS estimates}} \\
Net Export Growth (99-05, An) x Dum00t06&   -0.373   &   -0.463   &   -0.796   &   -0.726   \\
               &  (0.853)   &  (0.890)   &  (0.921)   &  (0.914)   \\
\addlinespace
Net Export Growth (99-05, An) x Dum07t10&    0.401   &    0.788   &    1.099   &    1.000   \\
               &  (1.016)   &  (1.089)   &  (1.205)   &  (1.220)   \\
\addlinespace
R2-adj         &   0.0348   &   0.0507   &   0.0723   &   0.0708   \\
\addlinespace

\midrule
\multicolumn{5}{l}{\textbf{Panel B. Reduced-form estimates}} \\
GIV Net Export Growth (99-05, An) x Dum00t06&    0.787   &    0.691   &    0.039   &    0.013   \\
               &  (1.675)   &  (1.725)   &  (1.749)   &  (1.732)   \\
\addlinespace
GIV Net Export Growth (99-05, An) x Dum07t10&    0.087   &    0.676   &    0.136   &   -0.181   \\
               &  (1.888)   &  (2.016)   &  (2.284)   &  (2.284)   \\
\addlinespace
R2-adj         &   0.0347   &   0.0504   &   0.0712   &   0.0699   \\
\addlinespace

\midrule
\multicolumn{5}{l}{\textbf{Panel C . 2SLS estimates}} \\
\addlinespace
Net Export Growth (99-05, An) x Dum00t06&    0.684   &    0.605   &    0.036   &    0.011   \\
               &  (1.519)   &  (1.564)   &  (1.579)   &  (1.529)   \\
\addlinespace
Net Export Growth (99-05, An) x Dum07t10&    0.076   &    0.592   &    0.124   &   -0.161   \\
               &  (1.633)   &  (1.723)   &  (2.046)   &  (2.028)   \\
               
\addlinespace
\addlinespace

Dep Var (Panel D): &\multicolumn{4}{c}{Net Export Growth (99-05, An)} \\ 
\midrule 
\multicolumn{5}{l}{\textbf{Panel D . First-stage estimates only for 99-05 (Non-stack sample)}} \\
\addlinespace
GIV Net Export Growth (99-05, An)&    1.150***&    1.142***&    1.102***&    1.122***\\
               &  (0.242)   &  (0.242)   &  (0.276)   &  (0.271)   \\
\addlinespace
KP F-Stat      &    22.58   &    22.23   &    15.88   &    17.18   \\
MOP F-Stat     &    22.01   &    21.62   &    15.43   &    16.74   \\
\addlinespace

\midrule
\multicolumn{5}{l}{\textbf{Controls (for all Panels)}} \\
DumPeriod  &    Y        &  Y   &   Y    & Y        \\
Basic Controls x DumPeriod &            &  Y   &   Y    & Y    \\
Housing Controls x DumPeriod &           &      & Y       & Y   \\
Demographic Controls x DumPeriod &            &      &      &  Y \\

\midrule              
Obs (Panel A, B, and C)          &     1520   &     1520   &     1340   &     1340   \\
Obs (Panel D)          &      760   &      760   &      670   &      670   \\
Cluster SE     &     CBSA   &     CBSA   &     CBSA   &     CBSA    \\
Weight         & Ln(HU99)   & Ln(HU99)   & Ln(HU99)   & Ln(HU99)    \\
\bottomrule
\end{tabular}

} % end of resize box

\end{table}

%% file: Appendix.tex
%------------------------------------------------------------
%------------------------------------------------------------
\clearpage
%------------------------------------------------------------

%------------------------------------------------------------
\section{Appendix}

%--------------------------------------------------------------------------------------
%\subsect{Appendix for Data Details}
%--------------------------------------------------------------------------------------

%\input{App_DataDetails}

%--------------------------------------------------------------------------------------
%\subsection{Appendix: GIV for Imports}
%--------------------------------------------------------------------------------------

\input{App_GIVImport}

%--------------------------------------------------------------------------------------
%\subsection{Empirical.Robustness} This subsection is moved to the main context
%--------------------------------------------------------------------------------------

%\input{Empirical.Robustness}

%--------------------------------------------------------------------------------------
%\subsection{Appendixe: Empirical for Credit Expansion}
%--------------------------------------------------------------------------------------

%\input{App_EmpCreditExpansion}

%---------------------------------------------------------------
%---------------------------------------------------------------
%---------------------------------------------------------------
%---------------------------------------------------------------
% Appendix Figures and Tables
%---------------------------------------------------------------
%---------------------------------------------------------------
%---------------------------------------------------------------
%---------------------------------------------------------------

\clearpage 
%-----------------------------------------------------
\subsection{Appendix Figures and Tables}

%%%%%%%%%%%%%%%%%%%%%%%%%%%%%%%%%%%%%%%%%%%%%%%%%%%%%%%%%%%%%%%%%%%%%%%%%%%
%%%%%%%%%%%%%%%%%%%%%%%%%%%%%%%%%%%%%%%%%%%%%%%%%%%%%%%%%%%%%%%%%%%%%%%%%%%
% 
% Robustness Tests: State-level differences
% 
%%%%%%%%%%%%%%%%%%%%%%%%%%%%%%%%%%%%%%%%%%%%%%%%%%%%%%%%%%%%%%%%%%%%%%%%%%%
%%%%%%%%%%%%%%%%%%%%%%%%%%%%%%%%%%%%%%%%%%%%%%%%%%%%%%%%%%%%%%%%%%%%%%%%%%%

%\pagebreak 
%---------------------------------------------------------------

%%%%%%%%%%%%%%%%%%%%%%%%%%%%%%%%%%%%%%%%%%%%%%%%
% table_Robust.APLvsNone.HPI.D99t05.D07t09.HHDTI
%%%%%%%%%%%%%%%%%%%%%%%%%%%%%%%%%%%%%%%%%%%%%%%%

\input{Table_HH/table_Robust.APLvsNone.HPI.D99t05.D07t09.HHDTI}

%\pagebreak 
%---------------------------------------------------------------

%%%%%%%%%%%%%%%%%%%%%%%%%%%%%%%%%%%%%%%%%%%%%%%%
% table_Robust.NRCvsRC.HPI.D99t05.D07t09.HHDTI
%%%%%%%%%%%%%%%%%%%%%%%%%%%%%%%%%%%%%%%%%%%%%%%%

\input{Table_HH/table_Robust.NRCvsRC.HPI.D99t05.D07t09.HHDTI}

\pagebreak 
%---------------------------------------------------------------

%%%%%%%%%%%%%%%%%%%%%%%%%%%%%%%%%%%%%%%%%%%%%%%%
% table_Robust.NJDvsJD.HPI.D99t05.D07t09.HHDTI
%%%%%%%%%%%%%%%%%%%%%%%%%%%%%%%%%%%%%%%%%%%%%%%%

\input{Table_HH/table_Robust.NJDvsJD.HPI.D99t05.D07t09.HHDTI}

\pagebreak 
%---------------------------------------------------------------

%%%%%%%%%%%%%%%%%%%%%%%%%%%%%%%%%%%%%%%%%%%%%%%%
% table_Robust.SandvsNone.HPI.D99t05.D07t09.HHDTI 
%%%%%%%%%%%%%%%%%%%%%%%%%%%%%%%%%%%%%%%%%%%%%%%%

\input{Table_HH/table_Robust.SandvsNone.HPI.D99t05.D07t09.HHDTI}

\pagebreak 
%---------------------------------------------------------------

%%%%%%%%%%%%%%%%%%%%%%%%%%%%%%%%%%%%%%%%%%%%%%%%
% table_Robust.StCapGainTax.HPI.D99t05.D07t09.HHDTI
%%%%%%%%%%%%%%%%%%%%%%%%%%%%%%%%%%%%%%%%%%%%%%%%

\input{Table_HH/table_Robust.StCapGainTax.HPI.D99t05.D07t09.HHDTI}

%---------------------------------------------------------------
%---------------------------------------------------------------
% Empirical: Placebo Tests 
%
%---------------------------------------------------------------
%---------------------------------------------------------------

\pagebreak 
%---------------------------------------------------------------

%%%%%%%%%%%%%%%%%%%%%%%%%%%%%%%%%%%%%%%%%%%%%%%%
% 
%%%%%%%%%%%%%%%%%%%%%%%%%%%%%%%%%%%%%%%%%%%%%%%%

%\input{Table_HH/}

%################################################################################
%################################################################################
% This is the end of the entire section (tex file)
%################################################################################
%################################################################################

%% file: App_GIVImport.tex
%-----------------------------------------------------
%-----------------------------------------------------
%-----------------------------------------------------
%-----------------------------------------------------

\subsection{Gravity Model-based IV: US Imports}\label{subsec:GIV_imports}

We have illustrated the central idea of the gravity model-based instrument for US exports in Section \ref{subsec:GIV_exports}. For completeness, we also show how \cite{feenstra2019us} construct IV for US imports here. The gravity-based IV for US imports begins with a simple symmetric constant-elasticity equation in \cite{romalis2007nafta}:
\vspace{-1mm}
\begin{equation}{\label{eq:imp_gravity}}
    \frac{X^{j,US}_{s,v,t}}{X^{j,i}_{s,v,t}} = \Bigg( \frac{w^{j}_{s,t}d^{j,US}\tau^{j,US}_{s,t}}{w^{j}_{s,t}d^{j,i}\tau^{j,i}_{s,t}} \Bigg) ^{1-\sigma} \frac{(P^{US}_{s,t})^{\sigma-1}E^{US}_{s,t}}{(P^{i}_{s,t})^{\sigma-1}E^{i}_{s,t}} = \Bigg(\frac{d^{j,US}\tau^{j,US}_{s,t}}{d^{j,i}\tau^{j,i}_{s,t}} \Bigg) ^{1-\sigma} \frac{(P^{US}_{s,t})^{\sigma-1}E^{US}_{s,t}}{(P^{i}_{s,t})^{\sigma-1}E^{i}_{s,t}}
\end{equation}
$X^{j,US}_{s,v,t}$ is country $j$'s export to US in industry $s$ in product variant $v$ in year $t$. By the similar notation, $X^{j,i}_{s,v,t}$ represents country $j$'s export to country $i$. $w^{j}_{s,t}$ denotes the relative marginal cost of production in industry $s$ in country $j$, which is canceled out in the above equation. $\tau^{j,US}_{s,t}$ and $\tau^{j,i}_{s,t}$ are the \textit{ad valorem} import tariff on country $j$'s export to the US and country $i$, respectively. $d^{j,US}$ and $d^{j,i}$ are the pre-determined bilateral distance and other fixed trade costs from country $j$ to the US and to country $i$, respectively. $P^{US}_{s,t}$ and $P^{i}_{s,t}$ represent the aggregate price index in the US and country $i$. $E^{US}_{s,t}$ and $E^{i}_{s,t}$ denote the total expenditure in the US and country $i$. Lastly, $\sigma$ is the constant elasticity of substitution ($\sigma>1$). 

Like before, the intuition of this gravity-style model is straightforward. The ratio of country $i$'s export to the US relative to country $j$ is decreasing with the ratio of bilateral distance and the ratio of \textit{ad valorem} total import tariff, but increasing with the ratio of aggregate price index and total expenditure.

Suppose that there are $N^{j}_{s,t}$ identical product varieties exported by country $j$ in year $t$ and industry $s$, one can re-arrange the above equation, multiply both sides by $N^{j}_{s,t}$, and sum over countries $i \neq US$:
\vspace{-1mm}
\begin{equation*}
    N^{j}_{s,t}X^{j,US}_{s,v,t}*\sum_{i\neq US} \big[ ( d^{j,i})^{1-\sigma} (P^{i}_{s,t})^{\sigma-1}E^{i}_{s,t} \big] = (d^{j,US}\tau^{j,US}_{s,t})^{1-\sigma} (P^{US}_{s,t})^{\sigma-1}E^{US}_{s,t}* \sum_{i \neq US} \big[ N^{j}_{s,t}X^{j,i}_{s,v,t} (\tau^{j,i}_{s,t})^{\sigma-1} \big ]
\end{equation*}

Since the above equation holds for any countries $i \neq US$, one can choose the set of countries that have similar economic conditions with the US (so that they are market substitution of US market when country $j$ considers its export) to make my prediction more accurate. \cite{feenstra2019us} use the same eight high-income countries by \cite{autor2013china}. 

We denote the sectoral export from country $j$ to the US and to country $i$  as $X^{j,US}_{s,t} \equiv X^{j,US}_{s,v,t}*N^{j}_{s,t}$ and $X^{j,i}_{s,t} \equiv X^{j,i}_{s,v,t}*N^{j}_{s,t}$. Consequently, we can get 
\vspace{-1mm}
\begin{equation*}
    X^{j,US}_{s,t}*\sum_{i\neq US} \big[ ( d^{j,i})^{1-\sigma} (P^{i}_{s,t})^{\sigma-1}E^{i}_{s,t} \big] = (d^{j,US}\tau^{j,US}_{s,t})^{1-\sigma} (P^{US}_{s,t})^{\sigma-1}E^{US}_{s,t}* \sum_{i \neq US} \big[ X^{j,i}_{s,t} (\tau^{j,i}_{s,t})^{\sigma-1} \big ]
\end{equation*}

With a few re-arrangement, we can get the formula for $ X^{j,US}_{s,t}$:
\vspace{-1mm}
\begin{equation}
    X^{j,US}_{s,t} =   \frac{(d^{j,US}\tau^{j,US}_{s,t})^{1-\sigma}(P^{US}_{s,t})^{\sigma-1}E^{US}_{s,t}}{\sum_{i\neq US} \big[ (d^{j,i})^{1-\sigma}(P^{i}_{s,t})^{\sigma-1}E^{i}_{s,t} \big]}  
    * \bigg( \sum_{k\neq US} X^{j,k}_{s,t} \bigg)  * \Bigg\{  \sum_{i\neq US} \bigg[ \frac{ X^{j,i}_{s,t} }{\sum_{k\neq US} X^{j,k}_{s,t}} (\tau^{j,i}_{s,t})^{\sigma -1} \bigg] \Bigg\}
\end{equation}
 
Note in the above formula, we multiply and divide by $\sum_{k\neq US} X^{j,k}_{s,t}$ to prepare for the regression setup in the next step. Now we can take logs of both sides and move the term $\lnb{\sum_{k\neq US} X^{j,k}_{s,t}}$ to the left-hand side of the equation to derive the regression-style formula:
\vspace{-1mm}
\begin{equation} \label{eq:imp_gravityRegression}
\resizebox{0.92\textwidth}{!}{%
\begin{math}
\begin{aligned}
\lnb{X^{j,US}_{s,t}} & = \underbrace{ \lnb{\sum_{k\neq US}X^{j,k}_{s,t}} }_{\text{Term 0}} + \underbrace{ \lnb{(P^{US}_{s,t})^{\sigma-1}E^{US}_{s,t}} }_{\text{Ind-Year FE: } \gamma^{US}_{s,t}} + \underbrace{(1-\sigma)\lnb{d^{j,US}}}_{\text{Exporting-country FE: } \delta^{j,US}} \\
& + \underbrace{(1-\sigma)\lnb{\tau^{j,US}_{s,t}}}_{\text{Term 1}} + \underbrace{(\sigma-1) \lnb{ \Bigg\{  \sum_{i\neq US} \bigg[ \frac{ X^{j,i}_{s,t} }{\sum_{k\neq US} X^{j,k}_{s,t}} (\tau^{j,i}_{s,t})^{\sigma -1} \bigg] \Bigg\}^{\frac{1}{\sigma-1}} }}_{\text{Term 2: } (\sigma-1) \lnb{T^{j}_{s,t}}} + \epsilon^{j}_{s,t} \\
\end{aligned}
\end{math}
} %end of \scalemath \resizebos
\end{equation}
We can see that the US import from country $j$ in the industry $s$ year $t$ can be divided into six terms. ``Term 0'' is the other eight high-income countries' import from country $j$, which represents the world supply. The second term $\gamma^{US}_{s,t}$ is the US demand shocks, which is potentially endogenous. We remove this term by the US industry-by-year fixed effects. The third term $\delta^{j,US}$ is the distance from country $j$ to the US and all other industry- and year-invariant trade costs. Since this term is predetermined rather than a shock, we remove it by the exporting-country fixed effects. ``Term 1" is the tariff on country $j$'s exports imposed by the US, which is by definition out of the control of exporting firms. I keep this term to capture the shock from tariffs. ``Term 2" represents the weighted average tariffs on country $j$'s exports charged by other eight high-income countries. Intuitively, when this weighted average tariffs on country $j$'s exports increase, destination country $j$ will export to the US as a substitution. We keep this term to capture this substitution effect. The last term $\epsilon^{j}_{s,t} = - \lnb{ \sum_{i\neq US} [ (d^{j,i})^{1-\sigma} (P^{i}_{s,t})^{\sigma-1}E^{i}_{s,t} ] } $ is unobserved and only shows up in the regression error term. 

After the above regression, we can construct predicted US imports that are presumably exogenous:
\begin{equation} \label{eq:imp_gravityPreUSImp}
 \lnb{ \widehat{X^{j,US}_{s,t}} } = \lnb{\sum_{k\neq US}X^{j,k}_{s,t}} + \hat{\beta_1} *\lnb{\tau^{j,US}_{s,t}} + \hat{\beta_2}* \lnb{T^{j}_{s,t}}
\end{equation}

%################################################################################
%################################################################################
% This is the end of the entire section (tex file)
%################################################################################
%################################################################################

%% file: Table_HH/table_Robust.APLvsNone.HPI.D99t05.D07t09.HHDTI.tex
%---------------------------------------------------------------

%%%%%%%%%%%%%%%%%%%%%%%%%%%%%%%%%%%%%%%%%%%%%%%%
% table_Robust.APLvsNone.HPI.D99t05.D07t09.HHDTI
%%%%%%%%%%%%%%%%%%%%%%%%%%%%%%%%%%%%%%%%%%%%%%%%

\noindent 

\begin{table}[h!]
\centering
\caption{
\textbf{Robustness Test for Anti-Predatory Lending States vs. Non-Anti-Predatory Lending States.} \\
Stacked 2SLS Regressions of House Price Growth in Boom (99-05) and Bust (07-09) Periods on Household Leverage Increase (99-05)  \smallskip \newline
{\scriptsize
This table reports OLS, reduced-form, first stage, and second stages of stacked 2SLS regression $\triangle_{99,05} \& \triangle_{07,09} Ln(HPI_{c})  = \beta_{Boom} * \triangle_{99,05} HHDTI_{c} \times Dum_{99,05} + \beta_{Bust} * \triangle_{99,05} HHDTI_{c} \times Dum_{07,09} + \beta_{APL, Boom} * \triangle_{99,05} HHDTI_{c} \times Dum_{99,05} \times Dum_{APL} + \beta_{APL, Bust} * \triangle_{99,05} HHDTI_{c} \times Dum_{07,09} \times Dum_{APL} + \gamma_{Boom} * \bm{Controls_{c}} \times Dum_{99,05} + \gamma_{Bust} * \bm{Controls_{c}} \times Dum_{07,09} + \epsilon_{c}$. The left-hand-side dependent variable $\triangle_{99,05} \& \triangle_{07,09} Ln(HPI_{c})$ is the stacked growth rate of the house price index (deflated to 2007) at county $c$ 99-05 and 07-09. The key independent variable $\triangle_{99,05} HHDTI_{c}$ is the rise in household leverage (debt-to-income ratio) at county $c$ 99-05. $Dum_{APL}$ is the dummy variable for counties in states with anti-predatory lending laws. $Controls_{c}$ indicates control variables at county $c$ in the period start year 1999. In either boom or bust period, we have two endogenous variables here: $\triangle_{99,05} HHDTI_{c}$ is instrumented by $\triangle_{99,05}\text{givNetExp}_{m}$ and $\triangle_{99,05} HHDTI_{c} \times Dum_{APL}$ is instrumented by $\triangle_{99,05}\text{givNetExp}_{m} \times Dum_{APL}$. For each of the first-stage F-tests of two endogenous variables, we report Sanderson-Windmeijer (2016) robust (clustered) statistics. To evaluate the overall strength of instruments, we report the p-value of robust (clustered) Kleibergen-Paap test statistics calculated by Windmeijer (2021). Each regression is weighted by the natural logarithm of housing units in 1999. Standard errors are clustered at the CBSA level. ***, **, and * indicate significance at the 1\%, 5\%, and 10\% levels, respectively.
} % end of small font size
} % end of caption
\label{table_Robust.APLvsNone.HPI.D99t05.D07t09.HHDTI}

\resizebox{\columnwidth}{!}{%

\begin{tabular}{l*{6}{c}}
\toprule
\textbf{TSLS Estimates}              &\multicolumn{5}{c}{House Price Growth (07USD, 99-05 or 07-09, An)} \\
            \cmidrule{2-6} 
            &\multicolumn{1}{c}{(1)}&\multicolumn{1}{c}{(2)}&\multicolumn{1}{c}{(3)}&\multicolumn{1}{c}{(4)}&\multicolumn{1}{c}{(5)}\\
            
\midrule
HH Debt-to-Income Rise (99-05, An) x Dum99t05&    0.661***&    0.562***&    0.486***&    0.468***&    0.483***\\
               &  (0.172)   &  (0.170)   &  (0.172)   &  (0.152)   &  (0.179)   \\
\addlinespace
HH Debt-to-Income Rise (99-05, An) x Dum07t09&   -0.399***&   -0.356***&   -0.333***&   -0.305***&   -0.326***\\
               &  (0.108)   &  (0.095)   &  (0.113)   &  (0.092)   &  (0.112)   \\
\addlinespace
HH Debt-to-Income Rise (99-05, An) x Dum99t05 x DumAPL&   -0.097   &   -0.110** &   -0.127** &   -0.105** &   -0.112*  \\
               &  (0.060)   &  (0.048)   &  (0.056)   &  (0.050)   &  (0.058)   \\
\addlinespace
HH Debt-to-Income Rise (99-05, An) x Dum07t09 x DumAPL&    0.062*  &    0.072***&    0.077** &    0.068***&    0.066** \\
               &  (0.033)   &  (0.027)   &  (0.033)   &  (0.026)   &  (0.029)   \\
\addlinespace
\midrule
DumPeriod  &    Y        &  Y   &   Y    & Y    &  Y    \\
Basic Controls x DumPeriod &            &  Y   &   Y    & Y    &  Y    \\
Housing Controls x DumPeriod &           &      & Y       & Y    &  Y    \\
Demographic Controls x DumPeriod &            &      &        &  Y    &  Y    \\
Industry Controls x DumPeriod &            &      &        &      &  Y    \\
\midrule
Obs          &     1572   &     1572   &     1390   &     1390   &     1390   \\
Cluster SE     &     CBSA   &     CBSA   &     CBSA   &     CBSA   &     CBSA   \\
Weight         & Ln(HU99)   & Ln(HU99)   & Ln(HU99)   & Ln(HU99)   & Ln(HU99)   \\
SW F-Stat: HHDTI (99 to 05) 1st-Stage &    13.09   &    17.61   &    10.51   &    11.59   &    9.283   \\
SW F-Stat: HHDTIxDumAPL (99 to 05) 1st-Stage &    15.25   &    18.53   &    14.68   &    25.90   &    22.78   \\
KP Robust (99 to 05) UnderID P-Value &   0.0006   &   0.0005   &   0.0063   &   0.0052   &   0.0110   \\
CoefEqual\_Chi2 &   16.050   &   13.564   &    9.275   &   11.311   &    8.819   \\
CoefEqual\_PValue &    0.000   &    0.000   &    0.002   &    0.001   &    0.003   \\

\bottomrule

\end{tabular}

} % end of resize box

\end{table}

%% file: Table_HH/table_Robust.NRCvsRC.HPI.D99t05.D07t09.HHDTI.tex
%---------------------------------------------------------------

%%%%%%%%%%%%%%%%%%%%%%%%%%%%%%%%%%%%%%%%%%%%%%%%
% table_Robust.NRCvsRC.HPI.D99t05.D07t09.HHDTI
%%%%%%%%%%%%%%%%%%%%%%%%%%%%%%%%%%%%%%%%%%%%%%%%

\noindent 

\begin{table}[h!]
\centering
\caption{
\textbf{Robustness Test for Non-Recourse States vs. Recourse States.} \\
Stacked 2SLS Regressions of House Price Growth in Boom (99-05) and Bust (07-09) Periods on Household Leverage Increase (99-05)  \smallskip \newline
{\scriptsize
This table reports stacked 2SLS regression $\triangle_{99,05} \& \triangle_{07,09} Ln(HPI_{c}) = \beta_{Boom} * \triangle_{99,05} HHDTI_{c} \times Dum_{99,05} + \beta_{Bust} * \triangle_{99,05} HHDTI_{c} \times Dum_{07,09}  + \beta_{NRC, Boom} * \triangle_{99,05} HHDTI_{c} \times Dum_{99,05} \times Dum_{NRC} + \beta_{NRC, Bust} * \triangle_{99,05} HHDTI_{c} \times Dum_{07,09} \times Dum_{NRC} + \gamma_{Boom} * \bm{Controls_{c}} \times Dum_{99,05} + \gamma_{Bust} * \bm{Controls_{c}} \times Dum_{07,09} + \epsilon_{c}$. The left-hand-side dependent variable $\triangle_{99,05} \& \triangle_{07,09} Ln(HPI_{c})$ is the stacked growth rate of the house price index (deflated to 2007) at county $c$ 99-05 and 07-09 in either non-recourse state or recourse state. The key independent variable $\triangle_{99,05} HHDTI_{c}$ is the rise in household leverage (debt-to-income ratio) at the county $c$ 99-05. $Dum_{NRC}$ is the dummy variable for counties in non-recourse states. $Controls_{c}$ indicates control variables at county $c$ in the start year 1999. In either boom or bust period, we have two endogenous variables here: $\triangle_{99,05} HHDTI_{c}$ is instrumented by $\triangle_{99,05}\text{givNetExp}_{m}$ and $\triangle_{99,05} HHDTI_{c} \times Dum_{APL}$ is instrumented by $\triangle_{99,05}\text{givNetExp}_{m} \times Dum_{APL}$. For each of the first-stage F-tests of two endogenous variables, we report Sanderson-Windmeijer (2016) robust (clustered) statistics. To evaluate the overall strength of instruments, we report the p-value of robust (clustered) Kleibergen-Paap test statistics calculated by Windmeijer (2021). Each regression is weighted by the natural logarithm of housing units in 1999. Standard errors are clustered at the CBSA level. ***, **, and * indicate significance at the 1\%, 5\%, and 10\% levels, respectively.
} % end of small font size
} % end of caption
\label{table_Robust.NRCvsRC.HPI.D99t05.D07t09.HHDTI}

\resizebox{\columnwidth}{!}{%

\begin{tabular}{l*{5}{c}}
\toprule
\textbf{TSLS Estimates}              &\multicolumn{5}{c}{House Price Growth (07USD, 99-05 or 07-09, An)} \\
            \cmidrule{2-6} 
            &\multicolumn{1}{c}{(1)}&\multicolumn{1}{c}{(2)}&\multicolumn{1}{c}{(3)}&\multicolumn{1}{c}{(4)}&\multicolumn{1}{c}{(5)}\\
            
\midrule
HH Debt-to-Income Rise (99-05, An) x Dum99t05&    0.626***&    0.530***&    0.452***&    0.442***&    0.447***\\
               &  (0.164)   &  (0.165)   &  (0.161)   &  (0.143)   &  (0.166)   \\
\addlinespace
HH Debt-to-Income Rise (99-05, An) x Dum07t09&   -0.378***&   -0.338***&   -0.327***&   -0.299***&   -0.304***\\
               &  (0.101)   &  (0.090)   &  (0.107)   &  (0.089)   &  (0.103)   \\
\addlinespace
HH Debt-to-Income Rise (99-05, An) x Dum99t05 x DumNRC&   -0.037   &   -0.026   &   -0.086   &   -0.061   &   -0.089   \\
               &  (0.068)   &  (0.053)   &  (0.065)   &  (0.061)   &  (0.071)   \\
\addlinespace
HH Debt-to-Income Rise (99-05, An) x Dum07t09 x DumNRC&    0.002   &   -0.007   &    0.005   &   -0.012   &    0.002   \\
               &  (0.040)   &  (0.031)   &  (0.042)   &  (0.035)   &  (0.036)   \\
\addlinespace
\midrule
DumPeriod  &    Y        &  Y   &   Y    & Y    &  Y    \\
Basic Controls x DumPeriod &            &  Y   &   Y    & Y    &  Y    \\
Housing Controls x DumPeriod &           &      & Y       & Y    &  Y    \\
Demographic Controls x DumPeriod &            &      &        &  Y    &  Y    \\
Industry Controls x DumPeriod &            &      &        &      &  Y    \\
\midrule
Obs           &     1572   &     1572   &     1390   &     1390   &     1390   \\
Cluster SE     &     CBSA   &     CBSA   &     CBSA   &     CBSA   &     CBSA   \\
Weight         & Ln(HU99)   & Ln(HU99)   & Ln(HU99)   & Ln(HU99)   & Ln(HU99)   \\
SW F-Stat: HHDTI (99 to 05) 1st-Stage &    17.31   &    21.93   &    12.32   &    12.90   &    10.25   \\
SW F-Stat: HHDTIxDumNRC (99 to 05) 1st-Stage &    20.52   &    25.44   &    20.83   &    25.91   &    21.15   \\
KP Robust (99 to 05) UnderID P-Value &   0.0004   &   0.0003   &   0.0051   &   0.0044   &   0.0100   \\
CoefEqual\_Chi2 &   15.939   &   12.922   &    9.445   &   11.302   &    8.833   \\
CoefEqual\_PValue &    0.000   &    0.000   &    0.002   &    0.001   &    0.003   \\

\bottomrule

\end{tabular}

} % end of resize box

\end{table}

%% file: Table_HH/table_Robust.NJDvsJD.HPI.D99t05.D07t09.HHDTI.tex
%---------------------------------------------------------------

%%%%%%%%%%%%%%%%%%%%%%%%%%%%%%%%%%%%%%%%%%%%%%%%
% table_Robust.NJDvsJD.HPI.D99t05.D07t09.HHDTI
%%%%%%%%%%%%%%%%%%%%%%%%%%%%%%%%%%%%%%%%%%%%%%%%

\noindent 

\begin{table}[h!]
\centering
\caption{
\textbf{Robustness Test for Non-Judicial States vs. Judicial States.} \\
Stacked 2SLS Regressions of House Price Growth in Boom (99-05) and Bust (07-09) Periods on Household Leverage Increase (99-05)  \smallskip \newline
{\scriptsize
This table reports stacked 2SLS regression $\triangle_{99,05} \& \triangle_{07,09} Ln(HPI_{c}) = \beta_{Boom} * \triangle_{99,05} HHDTI_{c} \times Dum_{99,05} + \beta_{Bust} * \triangle_{99,05} HHDTI_{c} \times Dum_{07,09}  + \beta_{NJD, Boom} * \triangle_{99,05} HHDTI_{c} \times Dum_{99,05} \times Dum_{NJD} + \beta_{NJD, Bust} * \triangle_{99,05} HHDTI_{c} \times Dum_{07,09} \times Dum_{NJD} + \gamma_{Boom} * \bm{Controls_{c}} \times Dum_{99,05} + \gamma_{Bust} * \bm{Controls_{c}} \times Dum_{07,09} + \epsilon_{c}$. The left-hand-side dependent variable $\triangle_{99,05} \& \triangle_{07,09} Ln(HPI_{c})$ is the stacked growth rate of the house price index (deflated to 2007) at county $c$ 99-05 and 07-09 in either judicial state or non-judicial state. The key independent variable $\triangle_{99,05} HHDTI_{c}$ is the rise in household leverage (debt-to-income ratio) at the county $c$ 99-05. $Dum_{NJD}$ is the dummy variable for counties in states where foreclosure of a delinquent property needs judicial judgment. $Controls_{c}$ indicates control variables at county $c$ in the start year 1999. In either boom or bust period, we have two endogenous variables here: $\triangle_{99,05} HHDTI_{c}$ is instrumented by $\triangle_{99,05}\text{givNetExp}_{m}$ and $\triangle_{99,05} HHDTI_{c} \times Dum_{APL}$ is instrumented by $\triangle_{99,05}\text{givNetExp}_{m} \times Dum_{APL}$. For each of the first-stage F-tests of two endogenous variables, we report Sanderson-Windmeijer (2016) robust (clustered) statistics. To evaluate the overall strength of instruments, we report the p-value of robust (clustered) Kleibergen-Paap test statistics calculated by Windmeijer (2021). Each regression is weighted by the natural logarithm of housing units in 1999. Standard errors are clustered at the CBSA level. ***, **, and * indicate significance at the 1\%, 5\%, and 10\% levels, respectively.
} % end of small font size
} % end of caption
\label{table_Robust.NJDvsJD.HPI.D99t05.D07t09.HHDTI}

\resizebox{\columnwidth}{!}{%

\begin{tabular}{l*{6}{c}}
\toprule
\textbf{TSLS Estimates}              &\multicolumn{5}{c}{House Price Growth (07USD, 99-05 or 07-09, An)} \\
            \cmidrule{2-6} 
            &\multicolumn{1}{c}{(1)}&\multicolumn{1}{c}{(2)}&\multicolumn{1}{c}{(3)}&\multicolumn{1}{c}{(4)}&\multicolumn{1}{c}{(5)}\\
            
\midrule
HH Debt-to-Income Rise (99-05, An) x Dum99t05&    0.683***&    0.555***&    0.514** &    0.490***&    0.485***\\
               &  (0.173)   &  (0.178)   &  (0.201)   &  (0.167)   &  (0.185)   \\
\addlinespace
HH Debt-to-Income Rise (99-05, An) x Dum07t09&   -0.399***&   -0.338***&   -0.334***&   -0.295***&   -0.304***\\
               &  (0.101)   &  (0.095)   &  (0.122)   &  (0.097)   &  (0.111)   \\
\addlinespace
HH Debt-to-Income Rise (99-05, An) x Dum99t05 x DumNJD&   -0.124** &   -0.044   &   -0.081*  &   -0.068*  &   -0.062   \\
               &  (0.054)   &  (0.043)   &  (0.049)   &  (0.040)   &  (0.042)   \\
\addlinespace
HH Debt-to-Income Rise (99-05, An) x Dum07t09 x DumNJD&    0.048*  &    0.002   &    0.013   &   -0.003   &    0.001   \\
               &  (0.029)   &  (0.026)   &  (0.033)   &  (0.027)   &  (0.026)   \\
\addlinespace
\midrule
DumPeriod  &    Y        &  Y   &   Y    & Y    &  Y    \\
Basic Controls x DumPeriod &            &  Y   &   Y    & Y    &  Y    \\
Housing Controls x DumPeriod &           &      & Y       & Y    &  Y    \\
Demographic Controls x DumPeriod &            &      &        &  Y    &  Y    \\
Industry Controls x DumPeriod &            &      &        &      &  Y    \\
\midrule
Obs           &     1572   &     1572   &     1390   &     1390   &     1390   \\
Cluster SE     &     CBSA   &     CBSA   &     CBSA   &     CBSA   &     CBSA   \\
Weight         & Ln(HU99)   & Ln(HU99)   & Ln(HU99)   & Ln(HU99)   & Ln(HU99)   \\
SW F-Stat: HHDTI (99 to 05) 1st-Stage &    14.25   &    15.87   &    9.340   &    11.68   &    9.933   \\
SW F-Stat: HHDTIxDumNJD (99 to 05) 1st-Stage &    34.68   &    53.54   &    20.77   &    60.14   &    57.47   \\
KP Robust (99 to 05) UnderID P-Value &   0.0002   &   0.0002   &   0.0046   &   0.0041   &   0.0076   \\
CoefEqual\_Chi2 &   17.727   &   12.040   &    7.721   &    9.950   &    8.177   \\
CoefEqual\_PValue &    0.000   &    0.001   &    0.005   &    0.002   &    0.004   \\

\bottomrule

\end{tabular}

} % end of resize box

\end{table}

%% file: Table_HH/table_Robust.SandvsNone.HPI.D99t05.D07t09.HHDTI.tex
%---------------------------------------------------------------

%%%%%%%%%%%%%%%%%%%%%%%%%%%%%%%%%%%%%%%%%%%%%%%%
% table_Robust.SandvsNone.HPI.D99t05.D07t09.HHDTI
%%%%%%%%%%%%%%%%%%%%%%%%%%%%%%%%%%%%%%%%%%%%%%%%

\noindent 

\begin{table}[h!]
\centering
\caption{
\textbf{Robustness Test for Sand States vs. Non-Sand States.} \\
Stacked 2SLS Regressions of House Price Growth in Boom (99-05) and Bust (07-09) Periods on Household Leverage Increase (99-05)  \smallskip \newline
{\scriptsize
This table reports stacked 2SLS regression $\triangle_{99,05} \& \triangle_{07,09} Ln(HPI_{c}) = \beta_{Boom} * \triangle_{99,05} HHDTI_{c} \times Dum_{99,05} + \beta_{Bust} * \triangle_{99,05} HHDTI_{c} \times Dum_{07,09}  + \beta_{Sand, Boom} \times Dum_{99,05} \times Dum_{Sand} + \beta_{Sand, Bust} * \times Dum_{07,09} \times Dum_{Sand} + \gamma_{Boom} * \bm{Controls_{c}} \times Dum_{99,05} + \gamma_{Bust} * \bm{Controls_{c}} \times Dum_{07,09} + \epsilon_{c}$. The left-hand-side dependent variable $\triangle_{99,05} \& \triangle_{07,09} Ln(HPI_{c})$ is the stacked growth rate of the house price index (deflated to 2007) at county $c$ 99-05 and 07-09 in either sand state or non-sand state. The key independent variable $\triangle_{99,05} HHDTI_{c}$ is the rise in household leverage (debt-to-income ratio) at the county $c$ 99-05. $Dum_{Sand}$ is the dummy variable for counties in four sand states.  $Controls_{c}$ indicates control variables at county $c$ in the start year 1999. We use the gravity model-based instrumental variable ($\triangle_{99,05}\text{givNetExp}_{m}$) as IV for $\triangle_{99,05}HHDTI_{c}$. For the first-stage F-test of the non-stacked sample (99-05), we report kleibergen-Paap (2006) robust (clustered) statistics and Montiel Olea-Pflueger (2013) efficient statistics. Each regression is weighted by the natural logarithm of housing units in 1999. Standard errors are clustered at the CBSA level. ***, **, and * indicate significance at the 1\%, 5\%, and 10\% levels, respectively.
} % end of small font size
} % end of caption
\label{table_Robust.SandvsNone.HPI.D99t05.D07t09.HHDTI}

\resizebox{\columnwidth}{!}{%

\begin{tabular}{l*{6}{c}}
\toprule
\textbf{TSLS Estimates}              &\multicolumn{5}{c}{House Price Growth (07USD, 99-05 or 07-09, An)} \\
            \cmidrule{2-6} 
            &\multicolumn{1}{c}{(1)}&\multicolumn{1}{c}{(2)}&\multicolumn{1}{c}{(3)}&\multicolumn{1}{c}{(4)}&\multicolumn{1}{c}{(5)}\\
            
\midrule
HH Debt-to-Income Rise (99-05, An) x Dum99t05&    0.486***&    0.506** &    0.445*  &    0.385** &    0.381*  \\
               &  (0.186)   &  (0.245)   &  (0.252)   &  (0.174)   &  (0.196)   \\
\addlinespace
HH Debt-to-Income Rise (99-05, An) x Dum07t09&   -0.250***&   -0.269** &   -0.247** &   -0.195***&   -0.221** \\
               &  (0.084)   &  (0.109)   &  (0.118)   &  (0.071)   &  (0.090)   \\
\addlinespace
Dum\_Sand\_xD99t05&    0.045***&    0.009   &    0.011   &    0.023*  &    0.023   \\
               &  (0.013)   &  (0.027)   &  (0.023)   &  (0.012)   &  (0.015)   \\
\addlinespace
Dum\_Sand\_xD07t09&   -0.040***&   -0.023*  &   -0.027** &   -0.034***&   -0.029***\\
               &  (0.006)   &  (0.012)   &  (0.012)   &  (0.007)   &  (0.008)   \\
\addlinespace
\midrule
DumPeriod  &    Y        &  Y   &   Y    & Y    &  Y    \\
Basic Controls x DumPeriod &            &  Y   &   Y    & Y    &  Y    \\
Housing Controls x DumPeriod &           &      & Y       & Y    &  Y    \\
Demographic Controls x DumPeriod &            &      &        &  Y    &  Y    \\
Industry Controls x DumPeriod &            &      &        &      &  Y    \\
\midrule
Obs           &     1572   &     1572   &     1390   &     1390   &     1390   \\
Cluster SE     &     CBSA   &     CBSA   &     CBSA   &     CBSA   &     CBSA   \\
Weight         & Ln(HU99)   & Ln(HU99)   & Ln(HU99)   & Ln(HU99)   & Ln(HU99)   \\
KP F-Stat (99 to 05, non-stacked sample) &    10.88   &    13.23   &    7.776   &    10.19   &    8.736   \\
MOP F-Stat (99 to 05, non-stacked sample)    &    9.760   &    12.36   &    7.059   &    9.542   &    7.959   \\
CoefEqual\_Chi2 &    8.324   &    5.303   &    3.858   &    6.495   &    5.205   \\
CoefEqual\_PValue&    0.004   &    0.021   &    0.049   &    0.011   &    0.023   \\

\bottomrule

\end{tabular}

} % end of resize box

\end{table}

%% file: Table_HH/table_Robust.StCapGainTax.HPI.D99t05.D07t09.HHDTI.tex
%---------------------------------------------------------------

%%%%%%%%%%%%%%%%%%%%%%%%%%%%%%%%%%%%%%%%%%%%%%%%
% table_Robust.StCapGainTax.HPI.D99t05.D07t09.HHDTI
%%%%%%%%%%%%%%%%%%%%%%%%%%%%%%%%%%%%%%%%%%%%%%%%

\noindent 

\begin{table}[h!]
\centering
\caption{
\textbf{Robustness Test for State Capital Gain Tax Rates.} \\
Stacked 2SLS Regressions of House Price Growth in Boom (99-05) and Bust (07-09) Periods on Household Leverage Increase (99-05)  \smallskip \newline
{\scriptsize
This table reports stacked 2SLS regression $\triangle_{99,05} \& \triangle_{07,09} Ln(HPI_{c}) = \beta_{99,05} * \triangle_{99,05} HHDTI_{c} \times Dum_{99,05} + \beta_{07,09} * \triangle_{99,05} HHDTI_{c} \times Dum_{07,09} + \beta_{Tax, Boom} \times StateCapGainTax_{s} \times Dum_{00,06}  + \beta_{Tax, Bust} \times StateCapGainTax_{s} * \times Dum_{07,10} + \gamma_{99,05}* \bm{Controls_{c}} \times Dum_{99,05} + \gamma_{07,09}* \bm{Controls_{c}} \times Dum_{07,09} + \epsilon_{period, c}$. The left-hand-side dependent variable $\triangle_{99,05} \& \triangle_{07,09} Ln(HPI_{c})$ is the stacked growth rate of the house price index at county $c$ 99-05 and 07-09. The key independent variable $\triangle_{99,05} HHDTI_{c}$ is the rise in household leverage (debt-to-income ratio) at the county $c$ 99-05. $StateCapGainTax_{s}$ is the 2005 state-level capital gain tax rate. $Controls_{c}$ indicates control variables at county $c$ in the start year 1999. We add state-level capital gain tax rate interacted with period dummies. We use the gravity model-based instrumental variable ($\triangle_{99,05}\text{givNetExp}_{m}$) as IV for $\triangle_{99,05}HHDTI_{c}$. For the first-stage F-test of the non-stacked sample (99-05), we report kleibergen-Paap (2006) robust (clustered) statistics and Montiel Olea-Pflueger (2013) efficient statistics. Each regression is weighted by the natural logarithm of housing units in 1999. Standard errors are clustered at the CBSA level. ***, **, and * indicate significance at the 1\%, 5\%, and 10\% levels, respectively.
} % end of small font size
} % end of caption
\label{table_Robust.StCapGainTax.HPI.D99t05.D07t09.HHDTI}

\resizebox{\columnwidth}{!}{%

\begin{tabular}{l*{6}{c}}
\toprule
\textbf{TSLS Estimates}              &\multicolumn{5}{c}{House Price Growth (07USD, 99-05 or 07-09, An)} \\
            \cmidrule{2-6} 
            &\multicolumn{1}{c}{(1)}&\multicolumn{1}{c}{(2)}&\multicolumn{1}{c}{(3)}&\multicolumn{1}{c}{(4)}&\multicolumn{1}{c}{(5)}\\
            
\midrule
HH Debt-to-Income Rise (99-05, An) x Dum99t05&    0.578***&    0.518***&    0.440***&    0.418***&    0.414***\\
               &  (0.156)   &  (0.167)   &  (0.169)   &  (0.139)   &  (0.155)   \\
\addlinespace
HH Debt-to-Income Rise (99-05, An) x Dum07t09&   -0.340***&   -0.320***&   -0.297***&   -0.268***&   -0.282***\\
               &  (0.086)   &  (0.086)   &  (0.100)   &  (0.079)   &  (0.091)   \\
\addlinespace
State Capital Gain Tax x Dum99t05&   -0.313** &   -0.092   &   -0.199*  &   -0.203*  &   -0.209*  \\
               &  (0.123)   &  (0.100)   &  (0.120)   &  (0.113)   &  (0.125)   \\
\addlinespace
State Capital Gain Tax x Dum07t09&    0.233***&    0.105*  &    0.164** &    0.154** &    0.149** \\
               &  (0.073)   &  (0.062)   &  (0.075)   &  (0.069)   &  (0.074)   \\
\addlinespace
\midrule
DumPeriod  &    Y        &  Y   &   Y    & Y    &  Y    \\
Basic Controls x DumPeriod &            &  Y   &   Y    & Y    &  Y    \\
Housing Controls x DumPeriod &           &      & Y       & Y    &  Y    \\
Demographic Controls x DumPeriod &            &      &        &  Y    &  Y    \\
Industry Controls x DumPeriod &            &      &        &      &  Y    \\
\midrule
Obs           &     1572   &     1572   &     1390   &     1390   &     1390   \\
Cluster SE     &     CBSA   &     CBSA   &     CBSA   &     CBSA   &     CBSA   \\
Weight         & Ln(HU99)   & Ln(HU99)   & Ln(HU99)   & Ln(HU99)   & Ln(HU99)   \\
KP F-Stat (99 to 05, non-stacked sample)   &    10.88   &    13.23   &    7.776   &    10.19   &    8.736   \\
MOP F-Stat (99 to 05, non-stacked sample)   &    9.760   &    12.36   &    7.059   &    9.542   &    7.959   \\
CoefEqual\_Chi2 &   15.977   &   12.080   &    8.346   &   11.201   &    9.255   \\
CoefEqual\_PValue&    0.000   &    0.001   &    0.004   &    0.001   &    0.002   \\

\bottomrule

\end{tabular}

} % end of resize box

\end{table}